\documentclass{aa}

\usepackage{graphicx}
\usepackage{txfonts}
\usepackage{lipsum}
\usepackage{subcaption}         %
\usepackage{lscape}             %
\usepackage{placeins}           %
                                
\usepackage[breaklinks]{hyperref}
\hypersetup{
  colorlinks=True,
  citecolor=[rgb]{0.06, 0.2, 0.65},
  linkcolor=[rgb]{0.8, 0.2, 1.0},
  urlcolor=[rgb]{0.06, 0.2, 0.65},
}

\usepackage{color}
\usepackage{xcolor}
\usepackage{booktabs}
\RequirePackage[linesnumbered, ruled, vlined]{algorithm2e}
\usepackage{siunitx}

\makeatletter
\addto\extrasenglish{%
}
\makeatother

\newcommand{\fwdRelRmseMean}{8.91}
\newcommand{\fwdRelRmseMedian}{9.16}

\newcommand{\fwdRelRmseQNinetyFive}{13.11}
\newcommand{\fwdRelRmseQNineNineSeven}{13.70}

\newcommand{\fwdPerSpecMean}{6.88}
\newcommand{\fwdPerSpecQFifty}{4.84}
\newcommand{\fwdPerSpecQSixtyEight}{6.43}
\newcommand{\fwdPerSpecQNinetyFive}{18.79}
\newcommand{\fwdPerSpecQNineNineSeven}{41.20}

\newcommand{\fwdBandUniAllMean}{7.48}
\newcommand{\fwdBandUniAllMedian}{6.54}

\newcommand{\fwdBandUniOpticalMean}{6.38}
\newcommand{\fwdBandUniOpticalMedian}{6.30}

\newcommand{\fwdBandUniNirMean}{8.60}
\newcommand{\fwdBandUniNirMedian}{7.85}

\newcommand{\fwdBandUniMirMean}{11.06}
\newcommand{\fwdBandUniMirMedian}{11.21}

\newcommand{\fwdBandUniFirMean}{6.69}
\newcommand{\fwdBandUniFirMedian}{6.52}

\newcommand{\fwdBandUniSubMmMean}{4.85}
\newcommand{\fwdBandUniSubMmMedian}{4.61}

\newcommand{\fwdBandUniSilicateMean}{13.09}
\newcommand{\fwdBandUniSilicateMedian}{13.03}

\newcommand{\fwdBandUniIceMean}{13.34}
\newcommand{\fwdBandUniIceMedian}{12.62}

\newcommand{\fwdResHighMax}{198}

\newcommand{\fwdResSpitzer}{600}
\newcommand{\fwdResJwstNirspecIfu}{2700}
\newcommand{\fwdResJwstMiriMrs}{3500}

\newcommand{\EsoHaESS}{8.7\times10^{3}}        %

\newcommand{\EsoHaLogMgasMed}{-1.55^{+0.18}_{-0.22}}

\newcommand{\EsoHaMgasMed}{0.028^{+0.015}_{-0.011}}

\newcommand{\EsoHaLogRcMed}{1.45^{+0.39}_{-0.19}}

\newcommand{\EsoHaRcMed}{28^{+41}_{-10}}

\newcommand{\EsoHaLogHcMed}{-1.12^{+0.11}_{-0.12}}

\newcommand{\EsoHaLogPsiMed}{-0.92^{+0.43}_{-0.95}}

\newcommand{\EsoHaLogGammaMed}{-0.14^{+0.18}_{-0.16}}

\newcommand{\EsoHaLogOneMinusFlMed}{-1.72^{+1.10}_{-0.17}}

\newcommand{\EsoHaLogAminMed}{-1.44^{+0.57}_{-0.20}}

\newcommand{\EsoHaLogAmaxSmallMed}{0.91^{+0.77}_{-0.34}}

\newcommand{\EsoHaLogXMed}{-0.26^{+0.06}_{-0.84}}

\newcommand{\EsoHaIncMed}{79.8^{+5.2}_{-2.4}}

\newcommand{\EsoHaLogDeltaCavMed}{-2.37^{+1.24}_{-0.81}}

\newcommand{\EsoHaLogRcavMed}{0.36^{+1.54}_{-1.06}}

\newcommand{\EsoHaLogXcoMed}{-5.56^{+2.02}_{-1.63}}

\newcommand{\EsoHaLogXcoTwoMed}{-6.75^{+1.61}_{-0.66}}

\newcommand{\EsoHaLogXhTwoOMed}{-4.75^{+0.63}_{-0.72}}

\newcommand{\EsoHaLogLstarMed}{-0.59^{+0.73}_{-0.22}}

\newcommand{\EsoHaLogRstarMed}{0.15^{+0.15}_{-0.40}}

\newcommand{\EsoHaLogTstarMed}{3.521^{+0.019}_{-0.022}}

\newcommand{\EsoHaDistMed}{186^{+26}_{-20}}

\newcommand{\EsoHaAvMed}{1.01^{+0.45}_{-0.30}}

\newcommand{\EsoHaIncModeA}{79.5}
\newcommand{\EsoHaIncModeB}{85.5}

\begin{document}

   \title{The Fast \& Fourier: Using neural operators to speed up radiative modeling of protoplanetary disks}
   \titlerunning{Using neural operators to speed up radiative modeling of protoplanetary disks}

   \author{L. Welzel\inst{1}\corrauth{welzel@strw.leidenuniv.nl}        %
        \and M. K. McClure\inst{1}
        }

   \institute{Leiden Observatory, Leiden University, PO Box 9513, 2300 RA Leiden, The Netherlands
   }

   \date{Received Month day, year; accepted Month day, year}

  \abstract
  {Edge-on protoplanetary disks provide a spatially resolved view of the vertical distribution of icy planetary building blocks, but inferring their relative abundances requires radiative-transfer (RT) modeling with multiple anisotropic scattering. Each such model costs $10^{1}$-$10^{3}$ CPU hours, which precludes principled Bayesian inference and homogeneous, population-scale analyses.}
  {We develop and validate a fast and reusable surrogate for Bayesian inference with ice-bearing edge-on disk models.}  %
  {We train a Fourier Neural Operator on a grid of \texttt{RADMC-3D} spectra. The network maps disk parameters and analytic physical input fields to the spectrum, using mixed-resolution spectra. We use the surrogate in an explicit-likelihood nested sampler. Then we validate it on held-out simulations and synthetic recovery tests with TARP coverage. Finally, we use it to retrieve the disk structure and ice abundances of ESO-H$\alpha$~569 from its spectral energy distribution.}
  {The surrogate successfully reproduces the \texttt{RADMC-3D} spectra with a median error of $\fwdPerSpecQFifty\%$. We recover synthetic parameters without detectable bias and with nominal coverage. For ESO-H$\alpha$~569, the posterior favors a massive ($M_{\rm gas}=\EsoHaMgasMed\,M_\odot$), nearly edge-on ($i=\EsoHaIncMed\,^{\circ}$) disk consistent with the image-plus-SED model of \citet{wolff2017hubblespacetelescope}. Its ice abundances and cavity size remain weakly constrained by broadband data alone. The marginal spectrum-evaluation speed-up is approximately $10^{9}$--$10^{10}$.}
  {Neural-operator emulation makes full posterior inference feasible within the adopted model and training prior. Scientifically important posterior points should still be checked with high-fidelity RT models. By dramatically decreasing the evaluation time, this approach enables for the first time homogeneous population studies of ices in disks, which can be capitalized on for new JWST surveys to map the range of compositions within planetary nurseries.}


   \maketitle
   \nolinenumbers

\section{Introduction}
\label{sec:introduction}

The vertical distribution of icy solids in protoplanetary disks is a central, but poorly constrained, ingredient of planet formation models. Vertical settling regulates the mid-plane solid density and thus the timescales for grain growth, while turbulent mixing exchanges material between the irradiated surface layers and the shielded interior, and freeze-out at the snow surfaces sets the phase in which the main C, N, and O carriers are available \citep{dullemond2004effectdustsettling, oberg2011effectssnowlinesplanetary, oberg2023protoplanetarydiskchemistry}.

For most disks, the vertical disk structure is projected along the line of sight and is difficult to tie to observables. However, in edge-on systems the optically thick mid-plane occults the central star, allowing the hot infrared continuum from the inner disk to backlight the outer disk's upper layers. Infrared images trace the small-grains' vertical extent \citep{burrows1996hubblespacetelescope, wolff2017hubblespacetelescope}, while millimeter imaging probes large grains settled to the much thinner mid-plane \citep{villenave2020observationsedgeonprotoplanetary, villenave2023modestdustsettling}. Ice mantles in the outer disk absorb against the infrared continuum, even in Class II disks without a cold envelope. JWST has transformed the few earlier ice detections \citep[e.g.,][]{pontoppidan2005icesedgeondisk, terada2007detectionwaterice, honda2009detectionwaterice, aikawa2012akariobservationsice, honda2016watericesurface} into complete ice inventories, and the succession of its edge-on disk programs has built up a modest sample in which disk ice reservoirs could be characterized \citep{sturm2023jwstinventoryprotoplanetary,sturm2023edgeonprotoplanetarydisk,sturm2024jwstmirianalysis,dartois2025edgeondisctau}.

Initial analysis of edge-on-disks in Cycle 1 JWST programs (i.e. HH 48 NE in the Ice Age ERS program, \#1309, P.I.~M.~K.~McClure, and Tau 042021 in MIDAS, \#1752, P.I.~M.~K.~McClure) detected H$_2$O (3, 6 $\mu$m), CO$_2$ (4.2, 15.2 $\mu$m), and CO ice (4.7 $\mu$m), but not less abundant ice species like CH$_3$OH or NH$_3$ \citep{sturm2023jwstinventoryprotoplanetary, sturm2024jwstmirianalysis, arulanantham2024jwstmirimrs, dartois2025edgeondisctau}. These studies established that ices are more vertically extended than anticipated, with vertical optical-depth profiles indicating all three major ice species reside high in the disk upper layers, where UV radiation should desorb them. Maintaining ice at these heights requires efficient UV shielding or replenishment by vertical mixing \citep{oberg2007photodesorptioncoice, oberg2009photodesorptionicesco, oberg2009photodesorptionicesii, fayolle2011coicephotodesorption, sturm2024jwstmirianalysis, duchene2024jwstimagingedgeon, dartois2025edgeondisctau}. Moreover, the high vertical extension of CO, the most volatile of these three species, implies that a fraction of the total CO ice is trapped in less volatile ice matrices. Such trapping of CO in CO$_2$ and H$_2$O ice has been demonstrated in the laboratory \citep{collings2004laboratorysurveythermal, simon2019entrapmentcoco2, ligterink2024rapidformationmacromolecules}, and was confirmed observationally in HH 48 NE by \citet{bergner2024jwsticeband}, whose fits to the CO band profile show that CO$_2$ and CO reside in mixed, partly apolar matrices, with a fraction of the CO trapped in CO$_2$ and H$_2$O ice. The JEDIce survey (\#5299, P.I.~J.~Bergner; \citealt{bergner2026jwstedgeondisk, dartois2026nearinfraredsilhouetteflying}), together with programs targeting individual systems (MINDS GTO, \#1282, P.I.~T.~Henning, \citealt{henning2024mindsjwstmiri}; GO \#1741, P.I.~A.~Potapov, \citealt{potapov2025simplemoleculescomplex, potapov2026firstdetectionhdo}), has since completed the sample of known edge-on disks and demonstrates that elevated mixed ices are the norm rather than a peculiarity of HH 48 NE. H$_2$O, CO$_2$, and CO are detected in all six JEDIce disks and the CO band profiles appear consistent with apolar CO:CO$_2$ mixtures \citep{bergner2026jwstedgeondisk}. The degree of ice mixing directly impacts planet formation: the fraction of CO trapped in CO$_2$ or H$_2$O ice mantles remains solid inside its nominal snowline, changing the C/O ratio of the gas and solids accreted by forming planets \citep{ligterink2024mindtrapnonnegligible, williams2025lockedicehow}. Calculating observationally calibrated disk ice maps for individual disks would allow us to assess in a statistically significant way the potential building blocks of different exoplanets and Solar System bodies.

\label{sec:background_edge_on_disks}

However, ice abundances do not follow directly from these detections. Although the optical depths of the main ice features of HH 48 NE, Tau 042021, and the JEDIce sample approach unity at most, comparison of formal column densities calculated from the $^{12}$CO$_2$ and $^{13}$CO$_2$ band depths reveals that the $^{12}$CO$_2$ band depth is saturated and a factor of $\sim$~10 smaller than its true value \citep{sturm2023edgeonprotoplanetarydisk}. Radiative transfer modeling shows that the flux at each wavelength samples many photon paths, each tracing different amounts of disk ice. The apparent ice band depth grows sublinearly with ice column and saturates at values of order unity, even where the true column keeps increasing. Therefore column densities along the line of sight cannot be calculated directly from spatially resolved optical depths in a manner similar to molecular cloud ices \citep{sturm2023edgeonprotoplanetarydisk, sturm2023edgeonprotoplanetarydiska, sturm2024jwstmirianalysis,mcclure2023iceagejwst, dartois2024spectroscopicsizinginterstellar, smith2025cospatialicemapping}.

The ice optical depths also depend on the disk structure. Inclination, disk mass, settling, and grain size distribution modify the optical depths and band profiles by amounts comparable to changes in the abundance itself. Within a band, the absorption and scattering cross sections of icy grains peak at different wavelengths, leading to distortions in the major band profiles for higher inclinations and larger grains \citep{dartois2022influencegraingrowth, martinien2025roleabsorptionscattering}. Ice abundances, their vertical distribution, and the grain properties therefore have to be obtained by modeling the observations with radiative transfer that includes multiple anisotropic scattering \citep{sturm2023edgeonprotoplanetarydiska, martinien2025roleabsorptionscattering}. Such dependencies mean that even with the structure of HH 48 NE constrained by HST, JWST, and ALMA imaging, its ice abundances were recovered only to within a factor of ${\sim}3$ \citep{sturm2023edgeonprotoplanetarydiska}.

This is in part due to the extreme computational expense of such forward models. Each evaluation requires a thermal Monte Carlo calculation of the dust temperature, repeated after the temperature-dependent ices are placed, a scattering Monte Carlo at every wavelength, and the ray tracing of spectra or images \citep{min2009radiativetransfervery}. A two-dimensional SED model at moderate inclination already takes minutes of CPU time \citep{kaeufer2023analysingsedsprotoplanetary}, and the edge-on case is more demanding, because the observed flux is dominated by the small fraction of photon packets that escape through the upper layers. A single three-dimensional model with anisotropic scattering takes $\sim10^1$--$10^3$ CPU hours, and we find empirically that principled Bayesian disk parameter estimation requires, at minimum, $\sim10^5$--$10^6$ likelihood evaluations. 

The Monte Carlo noise also makes gradient-based optimization difficult \citep{kaeufer2023analysingsedsprotoplanetary, sturm2023edgeonprotoplanetarydisk, dittmann2024notespracticalapplication}, so inferring parameters of single disks is prohibitively expensive, even more so for large samples of disks. The JWST analyses of HH 48 NE and Tau 042021 were therefore based on grids constructed around hand-tuned fiducial models and provide best-fit parameters without posterior distributions \citep{sturm2024jwstmirianalysis, dartois2025edgeondisctau, dartois2026nearinfraredsilhouetteflying}. Furthermore, the known edge-on sample is biased toward large, massive disks, underrepresents hosts with $M_\star \lesssim 0.5\,M_\odot$, and is predicted to be incomplete by several dozen systems in nearby star-forming regions \citep{stapelfeldt2014hstimagingnew, angelo2023demographicsprotoplanetarydisks, martinien2026observationshighlyinclined}. Quantifying these biases, comparing the known systems, and potentially identifying the missing disks all require that disks are modeled homogeneously, which multiplies the per-source cost by the sample size.

\label{sec:background_nn_emulator}

Machine-learning emulators save time by approximating the mapping between model parameters and observables with a surrogate trained on a precomputed grid of radiative transfer models. Once trained, the surrogate replaces the simulator during inference at negligible cost per evaluation \citep{alsing2020speculatoremulatingstellar}. Emulators have been used to predict disk spectra at low to moderate inclination \citep{ribas2020modelingprotoplanetarydisk, kaeufer2023analysingsedsprotoplanetary}, scattered-light images \citep{telkamp2022machinelearningframework, williams2025radiativetransfermodeling}, and, most recently, millimeter continuum images jointly with SEDs \citep{harries2026fundamentalpropertiesprotoplanetary}. However, the case of ice-bearing, edge-on disk spectra falls between these two cases, precisely where current SED emulators should fail. The SED analysis of \citet{kaeufer2023analysingsedsprotoplanetary}, for example, explicitly excludes self-obscured systems.

This failure reflects the radiative-transfer effects that complicate observational interpretation, namely multiple anisotropic scattering and strong optical-depth gradients, which also make the mapping that an emulator must learn highly non-linear and multi-scale. Changing a disk structure parameter can simultaneously reshape the continuum across more than four orders of magnitude in wavelength and modify the ice band profiles superposed on it. An emulator for this regime must therefore be fast and capable of representing correlated structure across the full spectrum, rather than treating each wavelength independently.

A natural way to build in this spectral coherence is to treat the emulator as an approximation of an operator that maps disk parameters to a function of wavelength, rather than to a fixed-length vector. This motivates the neural operator (NO) architecture adopted in our work. NOs learn maps between function spaces, with universal approximation guarantees for continuous operators \citep{chen1995universalapproximationnonlinear} and predictions that remain consistent under changes in discretization. They are established surrogates for parametric solution operators, mapping coefficients and forcing functions to solutions at a small fraction of the solver cost \citep{li2021fourierneuraloperator, lu2022comprehensivefaircomparison, kovachki2023neuraloperatorlearning, azizzadenesheli2024neuraloperatorsaccelerating, liu2025architecturesvariantsperformance, duruisseaux2026fourierneuraloperators}.

We use the trained NO as the forward model in a nested-sampling analysis with an explicit likelihood, validate it on held-out simulations and in end-to-end inference on synthetic observations with known parameters, and apply it to infer the disk structure of the nearly edge-on disk ESO-H$\alpha$ 569 from archival photometry, using the well-studied disk HH 48 NE as an external benchmark.

\section{Observations}
\label{sec:observations}

We model ESO-H$\alpha$~569 using only the broadband spectral energy distribution (SED, \autoref{tab:sedphot}) assembled by \citet{wolff2017hubblespacetelescope}, extended by spatially resolved ALMA observations at 890~$\mu$m (Band~7, \citealt{villenave2020observationsedgeonprotoplanetary}), and 1330~$\mu$m (Band~6, \citealt{bosschaart2026gapsringsnearuniversal}), and shown in the inset of \autoref{fig:eso_ha_569_posterior}. 
The four Herschel PACS/SPIRE points at 160--500~$\mu$m (marked ``$\star$'' in \autoref{tab:sedphot}) are only marginal (${\lesssim}1\sigma$) detections likely contaminated by extended cloud emission \citep{winston2012herschelfarirobservations, wolff2017hubblespacetelescope}. \citet{wolff2017hubblespacetelescope} excluded them from their SED fit for this reason. We indicate their $3\sigma$ upper limits in \autoref{fig:eso_ha_569_posterior} and exclude them from the likelihood.

\begin{table}[ht]
  \centering
  \caption{Compiled photometry for ESO-H$\alpha$~569.}
  \label{tab:sedphot}
  \begin{tabular}{
    S[table-format=4.3]
    @{}l
    S[table-format=3.3]
    @{\,${}\pm{}$\,}
    S[table-format=3.3]
    l
  }
    \hline\hline
    \multicolumn{2}{c}{$\lambda$ [$\mu$m]} &
    \multicolumn{2}{c}{Flux [mJy]} &
    Source \\
    \hline
    0.551 &              & 0.030 & 0.004 & \citet{robberto2012hstimagingsurvey} \\
    0.606 &              & 0.058 & 0.001 & \citet{wolff2017hubblespacetelescope} \\
    0.814 &              & 0.21  & 0.01  & \citet{wolff2017hubblespacetelescope} \\
    1.235 &              & 0.66  & 0.05  & \citet{skrutskie2006twomicronall} \\
    1.662 &              & 0.97  & 0.08  & \citet{skrutskie2006twomicronall} \\
    2.15  &              & 0.98  & 0.09  & \citet{skrutskie2006twomicronall} \\
    3.4   &              & 0.63  & 0.02  & \citet{cutri2012explanatorysupplementwise} \\
    3.6   &              & 0.58  & 0.03  & \citet{luhman2008diskpopulationchamaeleon} \\
    4.5   &              & 0.57  & 0.05  & \citet{luhman2008diskpopulationchamaeleon} \\
    4.6   &              & 0.71  & 0.02  & \citet{cutri2012explanatorysupplementwise} \\
    5.8   &              & 0.58  & 0.05  & \citet{luhman2008diskpopulationchamaeleon} \\
    8.0   &              & 0.67  & 0.05  & \citet{luhman2008diskpopulationchamaeleon} \\
    12    &              & 0.65  & 0.07  & \citet{cutri2012explanatorysupplementwise} \\
    22    &              & 7.5   & 0.9   & \citet{cutri2012explanatorysupplementwise} \\
    24    &              & 8.36  & 0.77  & \citet{luhman2008diskpopulationchamaeleon} \\
    70    &              & 107   & 11    & \citet{luhman2008diskpopulationchamaeleon} \\
    70    &              & 200   & 100   & \citet{winston2012herschelfarirobservations} \\
    160   & $^{\star}$   & 200   & 200   & \citet{winston2012herschelfarirobservations} \\
    250   & $^{\star}$   & 100   & 100   & \citet{winston2012herschelfarirobservations} \\
    350   & $^{\star}$   & 50    & 50    & \citet{winston2012herschelfarirobservations} \\
    500   & $^{\star}$   & 50    & 50    & \citet{winston2012herschelfarirobservations} \\
    870   &              & 72    & 14    & \citet{belloche2011endstarformation} \\
    890   &              & 40.2  & 4.0   & \citet{villenave2020observationsedgeonprotoplanetary} \\
    1330  & $^{\dagger}$ & 19.0  & 1.9   & \citet{bosschaart2026gapsringsnearuniversal} \\
    2830  &              & 3.2   & 0.1   & \citet{dunham2016almasearchsubstructure} \\
    \hline\hline
  \end{tabular}
  \\[2pt]
  {\footnotesize \textbf{Notes.} Fluxes marked ``$\star$'' are marginal detections treated as upper limits and excluded from the likelihood (\autoref{sec:fitting}). The 1330~$\mu$m (ALMA Band~6) flux marked “$\dagger$” was published after the analysis was completed and was also not included in the likelihood.}
\end{table}

We use HH~48~NE here as an external benchmark (\autoref{sec:disc_comparison}): we compare against its published model parameters \citep{sturm2023edgeonprotoplanetarydisk, sturm2023edgeonprotoplanetarydiska, sturm2024jwstmirianalysis}, and use the same photometric observations as reported by \cite{sturm2023edgeonprotoplanetarydisk} (their Table 1).

\section{Method}
\label{sec:method}

\subsection{Disk model}
\label{sec:model}

Our physical model follows the edge-on disk structure of \citet{sturm2023edgeonprotoplanetarydisk, sturm2023edgeonprotoplanetarydiska} and its JWST/MIRI application \citep{sturm2024jwstmirianalysis}, implemented in \texttt{RADMC-3D} \citep{dullemond2012radmc3dmultipurposeradiative}. In short, we compute the dust temperature structure using radiative transfer, distribute the ices in the cold regions, and ray-trace synthetic observations. A complete description of the framework, parameter choices, and its validation against photometric, HST, ALMA, and JWST observations is given in \citet{sturm2023edgeonprotoplanetarydiska, sturm2024jwstmirianalysis}; we summarize the relevant parts below.

The axisymmetric surface-density profile is specified by the gas mass $M_{\rm gas}$, characteristic radius $R_{\rm c}$, and power-law index $\gamma$, with an exponential outer taper \citep{lynden-bell1974evolutionviscousdiscs}. Its vertical structure is set by the scale-height ratio $h_{\rm c}$ at $R_{\rm c}$ and the flaring index $\psi$. An inner cavity of radius $R_{\rm cav}$ is allowed through a density-depletion factor $\delta_{\rm cav}$; the innermost dust radius is otherwise set by sublimation \citep{kama2009innerrimstructures, dullemond2010innerregionsprotoplanetary}. The viewing geometry is described by the inclination $i$.
Two dust populations represent settling. A fraction $f_{\ell}$ of the dust mass is assigned to a large-grain population, with scale height $X$ times that of the gas. The remaining small-grain population is vertically well mixed. Both populations follow a power-law size distribution with slope $-3.5$ \citep[MRN;][]{mathis1977sizedistributioninterstellar} and common minimum size $a_{\min}$. The small population has maximum size $a_{\max,\mathrm{small}}$; the settled-population maximum, $a_{\max,\mathrm{large}}=1000\,\mu$m, is fixed. Dust opacities are computed with \texttt{OpTool} \citep{dominik2021optoolcommandlinedriven} using the distribution of hollow spheres \citep[DHS;][]{min2005modelingopticalproperties}.

We use the full anisotropic-scattering capability of \texttt{RADMC-3D} in the radiative transfer for the temperature and scattered-light spectrum calculations, since isotropic approximations significantly bias both the amount and the direction of scattered light in the NIR to MIR \citep{pontoppidan2007deepspitzerspectroscopy, sturm2023edgeonprotoplanetarydisk}. This directional dependence is particularly important for the ice absorption features, which arise in the lower, colder disk layers \citep{sturm2023jwstinventoryprotoplanetary, sturm2023edgeonprotoplanetarydiska, sturm2024jwstmirianalysis, bergner2024jwsticeband}.

Ice species are placed as mantles on grains at disk radii where the temperature is below their respective desorption temperatures (Table~2 of \citealt{sturm2023edgeonprotoplanetarydiska}), with ice abundances $x_{\rm H_2O}$, $x_{\rm CO_2}$, and $x_{\rm CO}$ relative to hydrogen. Following \citet{sturm2023edgeonprotoplanetarydiska}, the ice mass is split between the two grain populations in proportion to their surface area and, within a population, by keeping a constant ice mass fraction. The size distribution refers to the total core-plus-mantle grain, so a higher ice abundance implies a smaller refractory core \citep{arabhavi2022icesplanetformingdisks}. The ice vertical extent is fixed at $z/r\leq0.6$, an empirical limit set by JWST observations of HH~48~NE rather than by a photodesorption model \citep{sturm2024jwstmirianalysis, bergner2026jwstedgeondisk}. Optical constants are taken from \citet{dorschner1995stepsinterstellarsilicate, warren1986opticalconstantscarbon, warren2008opticalconstantsice, palumbo2006profile2140cm1, zubko1996opticalconstantscosmic}, and ice-coated opacities are computed with \texttt{OpTool} using DHS \citep{dominik2021optoolcommandlinedriven, min2005modelingopticalproperties}.

\label{sec:rt_synthetic}

The stellar spectrum is specified by a blackbody temperature $T_\star$ and normalized independently to $L_\star$, while $R_\star$ defines only the finite stellar size for \texttt{RADMC-3D}, i.e.\ the radius of the photon-launching sphere and hence the angular size of the illuminating source; consequently, we do not enforce $L_\star = 4\pi R_\star^2 \sigma T_\star^4$, and $R_\star$ should not be interpreted as an independent stellar-radius measurement. We sample $R_\star$ independently to keep the training prior a hypercube and to admit an illumination radius larger than the photospheric one.

The temperature structure is solved in two passes: first without ices to locate the desorption surfaces, then again with the ices in place. From the final temperature and composition, \texttt{RADMC-3D} ray-traces images from the optical to the mm, from which a spectrum is calculated. The thermal Monte Carlo radiative transfer uses $n_{\rm phot}^{\rm therm}=5\times10^{6}$ packets and the Monte Carlo ray-tracing radiative transfer $n_{\rm phot}^{\rm spec}=2$--$5\times10^{5}$ packets ($10^{6}$ for final reported spectra). At the optically thick disk mid-plane, we enable the modified random walk \citep{fleck1984randomwalkprocedure, min2009radiativetransfervery, robitaille2010modifiedrandomwalk}. For this work we use only ray-traced spectra, so PSF convolution is not applied.

At the photon counts that are tractable across a large grid, the ray-traced images and spectra of the most optically thick disks retain residual artifacts. Rather than raising $n_{\rm phot}$ for all models, which is prohibitive, we rely on the surrogate's implicit spectral regularization (\autoref{sec:neural_operators}) and discuss the resulting noise-limited regime, and its connection to the mid-plane temperature distribution of very-high-opacity disks, in \autoref{sec:disc_superiority}.

\subsection{Grid}
\label{sec:grid}

To maximize information gain per simulation and cover the high-dimensional prior uniformly, we sample the parameter prior with a scrambled Sobol low-discrepancy sequence \citep{sobol1967distributionpointscube, balandat2020botorchframeworkefficient}. Each grid point draws the disk parameters within the bounds of the prior (\autoref{tab:results_posterior}). The gas-to-dust mass ratio (100), the maximum settled grain size ($a_{\rm max,large}=1000\,\mu$m), and the vertical ice limit ($z/r=0.6$) are fixed across the grid. We enforce the constraint $\log_{10}(a_{\rm max, small}) \geq \log_{10}(a_{\rm min}) + 1$ on the concentration of the size distribution of the small grain population.

Spectra are simulated on three nested, non-uniform wavelength grids to balance wavelength coverage against the cost of resolving narrow spectral features (\autoref{tab:model_grid_model_resolution}). Higher-resolution grids preferentially sample the ice absorption bands and the $10~\mu\mathrm{m}$ silicate feature. After quality control, the final grid contains $41{,}658$ models. Building the grid required ${\approx}120$ CPU-years. Representative spectra and the wavelength sampling density are shown in \autoref{fig:representative_prediction_examples}.

\subsection{Surrogate model}
\label{sec:surrogate_model}

Computing radiative-transfer models for Bayesian parameter estimation is prohibitively expensive. We therefore train a neural operator as a cheap surrogate model that approximates the mapping from model parameters to the discretized spectrum generated by the radiative-transfer simulator, in this case \texttt{RADMC-3D}. This enables fast spectrum evaluation during inference and exploratory analysis \citep{cozad2014learningsurrogatemodels, cranmer2020frontiersimulationbasedinference}. 

\begin{figure*}[ht]
    \sidecaption
    \includegraphics[width=120mm]{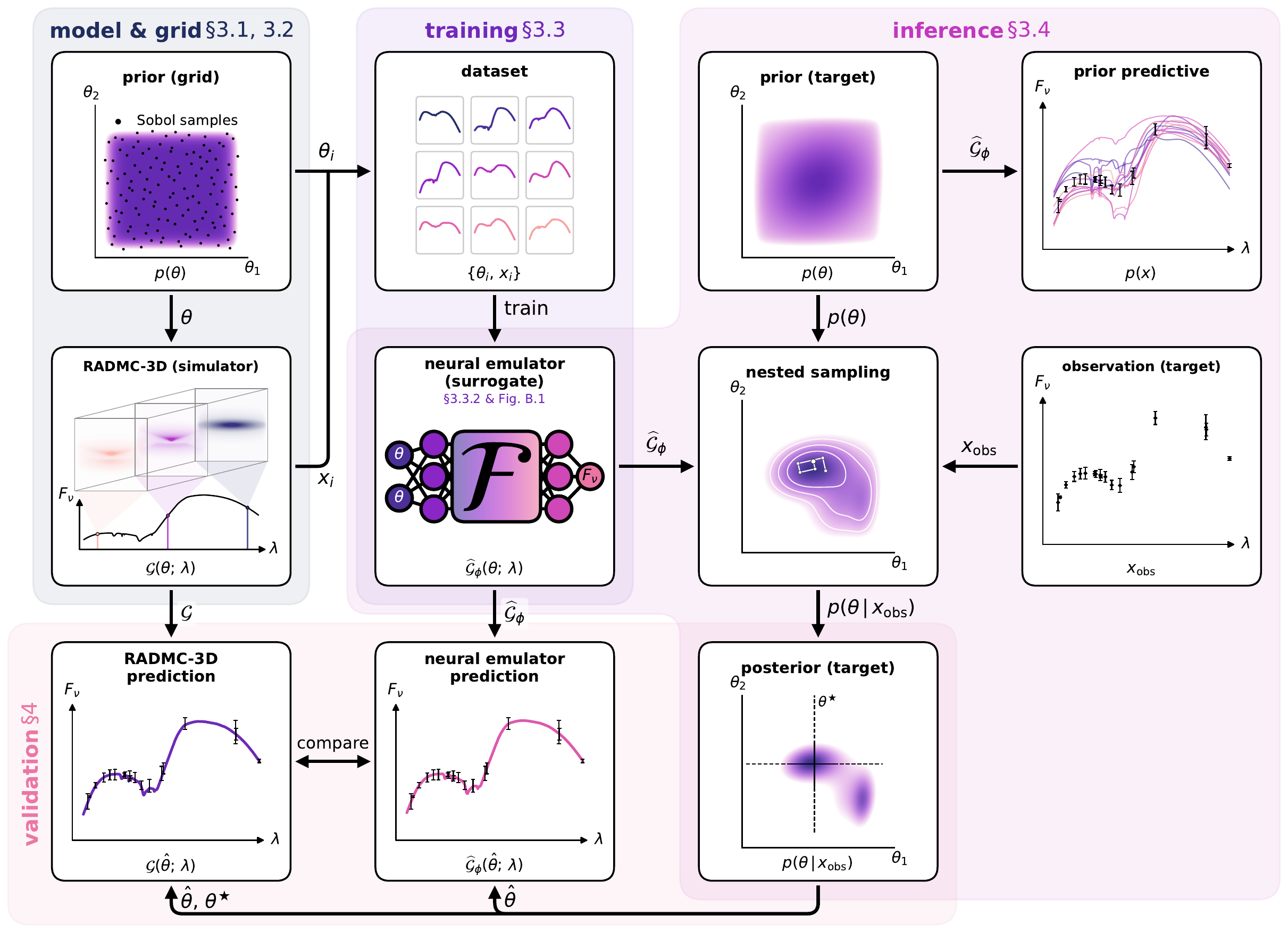}
    \caption{Overview of this work. \emph{Model \& grid} (\autoref{sec:model}, \autoref{sec:grid}): disk parameters $\boldsymbol{\theta}$ are drawn from the prior $p(\boldsymbol{\theta})$ and passed to the radiative-transfer (\texttt{RADMC-3D}) simulator $\mathcal{G}(\boldsymbol{\theta};\lambda)$, which generates the corresponding spectra. \emph{Training} (\autoref{sec:surrogate_model}): the parameter--spectrum pairs $\{\boldsymbol{\theta}_i,\,x_i\}$ form the dataset on which the neural-operator surrogate $\widehat{\mathcal{G}}_{\phi}$ is trained; its internal architecture is shown separately in \autoref{fig:nn_architecture}. \emph{Inference} (\autoref{sec:fitting}): the trained surrogate replaces the simulator, and we use traditional Bayesian inference (nested sampling) to obtain the posterior $p(\boldsymbol{\theta}\,|\,x_{\mathrm{obs}})$ from the prior $p(\boldsymbol{\theta})$, the surrogate, and an observation $x_{\mathrm{obs}}$; prior predictive samples are shown alongside. \emph{Validation} (\autoref{sec:validation}): the inferred best-fit parameters $\hat{\boldsymbol{\theta}}$ (and, for synthetic tests, the injected truth $\boldsymbol{\theta}^{\bigstar}$) are evaluated with both the surrogate, $\widehat{\mathcal{G}}_{\phi}(\hat{\boldsymbol{\theta}};\lambda)$, and the full simulator, $\mathcal{G}(\hat{\boldsymbol{\theta}};\lambda)$, and the two predictions are compared to confirm that the surrogate is accurate and does not bias the recovered parameters. Arrows trace the flow of parameters and spectra; $x$ denotes a discretized spectrum, so that $x_{\mathrm{obs}}$ is the observed spectrum ($\mathbf{F}^{\mathrm{obs}}$ in \autoref{sec:fitting}) and $\widehat{\mathcal{G}}_{\phi}$ is the neural emulator approximating the simulator $\mathcal{G}$.}
    \label{fig:methodology_overview}
\end{figure*}

The emulator is implemented as a Fourier Neural Operator \citep[FNO;][]{li2021fourierneuraloperator, kovachki2023neuraloperatorlearning} and is trained on the simulated spectra described in \autoref{sec:model}, \autoref{sec:grid}. \autoref{fig:methodology_overview} illustrates the complete workflow. The model and training grid (\autoref{sec:model}, \autoref{sec:grid}) produce the parameter--spectrum pairs on which the surrogate is trained. Once trained, it replaces the radiative-transfer simulator during inference (\autoref{sec:fitting}) and is checked against that simulator in validation (\autoref{sec:validation}). This section describes the surrogate itself, shown in \autoref{fig:nn_architecture}.

The surrogate approximates the mapping $\mathcal{G}$ of the scattered and absorbed stellar and disk radiation fields to the emergent spectrum $F_{\boldsymbol{\theta}}(\lambda)$ at the observer, $\mathcal{G}:\ \boldsymbol{\theta}\ \mapsto\ F_{\boldsymbol{\theta}}(\lambda)$, where $\boldsymbol{\theta}\in\mathbb{R}^{d}$ denotes the vector of physical disk parameters (\autoref{tab:results_posterior}) and $F_{\boldsymbol{\theta}}(\lambda)$ is the corresponding spectrum, with $\lambda$ in the simulated wavelength range. Crucially, the predicted spectrum is not a scalar but a function of wavelength. The spectrum carries correlated structure across $\lambda$, from the broad continuum spanning more than four orders of magnitude in wavelength down to the comparatively narrow ice absorption bands. As argued in \autoref{sec:neural_operators}, this functional, multi-scale character motivates an operator-learning approach rather than treating each $F_{\boldsymbol{\theta}}(\lambda_i)$ as an independent scalar regression target. We train on spectra sampled at multiple wavelength resolutions (\autoref{sec:grid}), combining broad prior coverage at low resolution with a smaller number of higher-resolution spectra that resolve the ice bands. We also aim to separate known, deterministic physics from the learned approximation. Wherever a transformation is analytic and inexpensive, we evaluate it in closed form and either supply it to the network as an input (\autoref{sec:fno_architecture}) or apply it as a fixed post-processing layer (e.g. foreground extinction; \autoref{sec:fno_architecture}), rather than having the network relearn it. This frees network capacity for the expensive part of the mapping (radiative-transfer) and keeps these steps interpretable.

Our grid is split into training ($70\%$), validation ($20\%$), and test ($10\%$) sets with a random partition. The normalization and emulator are fit on the training set, hyper-parameters are selected on the validation set, and the final performance is reported on the test set only. 

\subsubsection{Operator learning and the Fourier neural operator}
\label{sec:neural_operators}
We take the disk model plus radiative transfer solver as a physical simulator of the idealized physics that defines the mapping $\mathcal{G}~\approx~\mathcal{G}_{\rm RADMC-3D}$. Neural operators are architectures designed to approximate mappings between functions, or between finite-dimensional vectors and functions \citep{kovachki2023neuraloperatorlearning}. We adopt an FNO \citep{li2021fourierneuraloperator} that couples local nonlinear processing with global spectral convolutions in Fourier space. This combination is well matched to emulating disk spectra for three reasons.

First, disk spectra carry correlated structure across wavelength on all scales: a change in a physical parameter can reshape the continuum across orders of magnitude in wavelength, while simultaneously modifying the localized ice bands. Neighboring wavelengths vary coherently, since smooth changes in, e.g., grain size shift the opacity and continuum as a whole rather than independently per wavelength. The global spectral convolutions capture the long-range component of this correlation. In contrast, operating on the spectrum as a one-dimensional field builds the local coherence into the architecture and yields globally consistent spectra with fewer small-scale, unphysical fluctuations. This implicit smoothing is also a useful regularizer against the Monte Carlo noise (\autoref{sec:rt_synthetic}). 

Second, disk SEDs are typically composed of heterogeneous photometric observations and spectra of diverse resolutions. Because the FNO is discretization-invariant, it can be trained and evaluated on different wavelength grids. We exploit this to train on mixed-resolution spectra in a single procedure (\autoref{sec:fno_training_objective}) and to query the surrogate on arbitrary grids at inference time \citep{li2021fourierneuraloperator, kovachki2023neuraloperatorlearning, lingsch2024regulargridsfourierbased, george2024incrementalspatialspectral}. In practice the FNO invariance is only approximate \citep{bartolucci2023representationequivalentneural, lanthaler2024discretizationerrorfourier}, with consequences for our mixed-resolution grid discussed in \autoref{sec:disc_limitations}. Unlike a branch-trunk construction such as DeepONet \citep{lu2021deeponetlearningnonlinear}, the FNO does not factorize the parameter and coordinate dependence, and we found it reproduces the ice band morphology more reliably (\autoref{tab:ablation_test}).

Because standard Fourier layers assume periodicity, whereas spectra on a finite wavelength interval are non-periodic, we use the standard FNO mitigation of finite-domain padding in the spectral blocks \citep{kovachki2021universalapproximationerror, ganeshram2022fcpinohighprecision}. This reduces boundary artifacts at the shortest and longest wavelengths. We assess an alternative Fourier-continuation treatment in \autoref{sec:hpo}.

\subsubsection{FNO surrogate architecture}
\label{sec:fno_architecture}

For each simulation $i$ the training sample consists of a parameter vector $\boldsymbol{\theta}_i$ and a spectrum sampled on a wavelength grid $\Lambda_i=\{\lambda_{ij}\}_{j=1}^{n_i}$, which provides the prediction target $F_{\boldsymbol{\theta}_i}(\lambda)$. We apply the parameter transformations indicated in \autoref{tab:results_posterior} (e.g.\ $\log_{10}$ for quantities spanning several decades) and standardize the inputs using training-set statistics, $\tilde{\boldsymbol{\theta}} = (\boldsymbol{\theta}-\boldsymbol{\mu}_{\theta})/\boldsymbol{\sigma}_{\theta}$. The wavelength coordinate is represented in $\log\lambda$, both to compress the large wavelength range and to improve the resolution of short-wavelength structure, and rescaled to the unit interval, $\tilde{\lambda} = (\log\lambda - \log\lambda_{\min})/(\log\lambda_{\max} - \log\lambda_{\min})$. Fluxes are standardized globally, i.e.\ with a single mean and standard deviation of $\log_{10}F$ over all flux values of all training spectra, because per-wavelength statistics are ill-defined on the mixed grids $\Lambda_i$.

\label{sec:input_embeddings}
Rather than presenting the network with only the parameter vector, we supply it with physical quantities that are analytic and inexpensive to evaluate from $\boldsymbol{\theta}$, so that it does not need to learn known relations. In the operator picture of \autoref{sec:neural_operators}, the natural inputs to the radiative-transfer solution operator are the scalar parameters and the forcing function (the irradiating stellar spectrum) and the coefficient fields (the density, temperature, and compositional structure of the disk). Of these we supply only the closed-form functions of $\boldsymbol{\theta}$ (the stellar spectrum, the grain-size distribution, and the analytic density profiles). Computing the analytic fields in closed form and passing them as additional input channels injects a strong, physically grounded inductive bias, frees network capacity for the expensive part of the mapping, and keeps these transformations interpretable and decoupled from the trainable emulator. The embeddings fall into three families according to the domain on which each quantity is defined (left of \autoref{fig:nn_architecture}). First, the global scalar parameters $\boldsymbol{\theta}$ that are broadcast along wavelength and spectrally lifted using a sinusoidal coordinate embedding. Second, the functions that are defined on the wavelength domain, chiefly the stellar spectrum that acts as the radiative-transfer forcing term. And finally, the functions on auxiliary domains (e.g. the grain-size distribution, radial disk structure, and viewing geometry) that are mapped onto the wavelength grid by shallow FNO encoders. These channels are concatenated into a single per-wavelength input field $\mathbf{z}(\tilde{\lambda})$ (\autoref{eq:input_stack}) that is lifted into the first Fourier block. We give the detailed embeddings and their physical motivation in \autoref{sec:appendix_input_embeddings}.

We use a one-dimensional FNO with spectral-convolution layers and nonlinear mixing between layers. The network takes the concatenated field $\mathbf{z}(\tilde\lambda)$ of \autoref{eq:input_stack} (left-center of \autoref{fig:nn_architecture}) and predicts flux $\widehat{F}_{\boldsymbol{\theta}}(\lambda)$ (right of \autoref{fig:nn_architecture}) on the same normalized wavelength grid. We use domain padding in the Fourier blocks to reduce edge effects on the finite wavelength interval, and group-normalization layers \citep{wu2018groupnormalization} to improve training stability.

To match the observational forward model, we append a fixed, batched, differentiable foreground-extinction module to the FNO output. For an intrinsic spectrum $\widehat{F}_{\boldsymbol{\theta}}(\lambda)$, the attenuated spectrum is $\widehat{F}^{\mathrm{obs}}_{\boldsymbol{\theta}}(\lambda) = \widehat{F}_{\boldsymbol{\theta}}(\lambda)\,10^{-0.4\,A_V\,k(\lambda;A_V)}$, where $A_V$ is the fitted scalar line-of-sight $V$ band extinction and $k(\lambda;A_V)\equiv A_\lambda/A_V$ is the dimensionless extinction curve. The dependence of $k$ on $A_V$ denotes the use of the \citet{mathis1990interstellardustextinction, mcclure2009observational520mm} prescriptions in the $A_V$ regimes adopted by \citet{sturm2023edgeonprotoplanetarydisk}. We smooth the discontinuities between these regimes using sigmoid transition functions. This separates the adopted foreground extinction model from the learned emulator and allows alternative extinction laws to be substituted without re-training.

\subsubsection{Training objective and wavelength sampling}
\label{sec:fno_training_objective}

The simulated dataset contains spectra sampled at different wavelength resolutions (\autoref{sec:grid}). All Fourier layers operate on a fixed grid that is uniform in an internal coordinate $s\in[0,1]$. A prescribed monotonic map $\log_{10}\lambda=\Lambda(s)$ deforms this grid to the non-uniform wavelength sampling shown in \autoref{fig:representative_prediction_examples}. We train the emulator by evaluating the loss at the wavelength points available for each sample. In practice, missing wavelengths on the common internal support are represented by masked entries and excluded from the loss, while the input parameters are still broadcast and embedded over the full normalized domain.

For a mini-batch, let $\mathbf{F}$ and $\widehat{\mathbf{F}}$ denote the true and predicted $\log_{10}$-flux vectors on the valid wavelength points of a given sample. We predict and compare the $\log_{10}$-flux because of the high dynamic range of the flux across the wavelength domain. We optimize a composite objective consisting of an absolute $L_2$ term and a relative $H^1$ Sobolev term, 
\begin{equation}
\label{eq:loss}
\mathcal{L} = \mathcal{L}_{L_2} + \mathcal{L}^{\mathrm{rel}}_{H^1} = \left\|\widehat{\mathbf{F}}-\mathbf{F}\right\|_2 + \frac{\left\|\widehat{\mathbf{F}}-\mathbf{F}\right\|_{H^1}}{\left\|\mathbf{F}\right\|_{H^1}}.
\end{equation}
The $L_2$ term constrains the absolute flux level, while the relative $H^1$ term compares both flux values and their first derivatives along wavelength, enforcing the correct local shape independently of amplitude and increasing sensitivity to spectral morphology (continuum slope and ice-band curvature), following the Sobolev losses used for neural-operator training \citep{son2023sobolevtrainingphysicsinformed, kovachki2023neuraloperatorlearning}. Empirically, optimizing either term alone performs significantly worse than their combination (\autoref{tab:ablation_test}).

Because the valid wavelength points for a given spectrum may split into several contiguous segments (e.g. the continuum and the individual ice bands), we compute the Sobolev term segment-wise and combine the segment losses with weights proportional to the segment length on the $\log\lambda$ axis. This avoids introducing artificial derivative penalties across the masked gaps between segments. The two terms are combined adaptively during training with a relative loss-balancing scheme \citep[\texttt{ReLoBRaLo};][]{bischof2025multiobjectivelossbalancing}, which adjusts their weights from the optimization history, reducing the need for manual tuning when the two magnitudes differ substantially.

Mini-batches are formed from full spectra rather than from single wavelength points. This preserves the correlated spectral structure within each example and lets the Sobolev term operate on physically meaningful segments. For regularization, we add a small Gaussian perturbation ($\sigma = 10^{-5}$) to the normalized inputs during training\footnote{This $u$-space perturbation is propagated to the embedding functions (\autoref{sec:input_embeddings}) defined in physical parameter space.}; no noise is added during validation or testing. Implementation and reproducibility details are given in \autoref{sec:fno_implementation}.

\subsection{Parameter estimation}
\label{sec:fitting}

Once trained, the surrogate provides fast evaluations and enables Bayesian inference of disk model parameters from observed photometry or spectra. We denote the observed fluxes by $\mathbf{F}^{\mathrm{obs}}=\{F^{\mathrm{obs}}_m\}_{m=1}^{M}$ with uncertainties $\boldsymbol{\sigma}^{\mathrm{obs}}$, where each datum is a photometric band (or binned spectral point). We estimate the transformed physical disk and stellar parameters $\boldsymbol{\theta}\in\mathbb{R}^{d}$ (\autoref{tab:results_posterior}), and jointly fit the distance, $d$, and extinction, $A_V$, as nuisance parameters, $\boldsymbol{\phi} = \left(\boldsymbol{\theta},\, d,\, A_V\right)$. For the disk parameters we adopt uniform priors over the simulated-grid bounds (\autoref{sec:grid}), $p(\boldsymbol{\theta}) = \prod_{k=1}^{d} \mathcal{U}(\theta_k \,;\, \theta_{k,\min}, \theta_{k,\max})$. We adopt wide, physically motivated uniform priors for $d$ and $A_V\geq0$ and treat them as independent of $\boldsymbol{\theta}$, $p(\boldsymbol{\phi}) = p(\boldsymbol{\theta})\,\mathcal{U}(d)\,\mathcal{U}(A_V)$, with bounds chosen from literature or Gaia distance estimates per target, by default $d^{\mathrm{guess}}\pm 50~\mathrm{pc}$ and $0 \leq A_V \leq A_V^{\mathrm{guess}} + 5$. In the forward model, the distance enters through the usual $1/d^2$ flux scaling, and the $A_V$ correction uses the extinction law of \autoref{sec:fno_architecture}.

Given $\boldsymbol{\phi}$, the surrogate predicts fluxes $\widehat{\mathbf{F}}(\boldsymbol{\phi})=\{\widehat{F}_m(\boldsymbol{\phi})\}_{m=1}^{M}$ at the observed bands. We neglect inter-band covariance and assume independent Gaussian errors in flux space,
\begin{equation}
\ln p(\mathbf{F}^{\mathrm{obs}}\,|\,\boldsymbol{\phi}) =
-\frac{1}{2}\sum_{m=1}^{M} \left[\frac{\left(F^{\mathrm{obs}}_m-\widehat{F}_m(\boldsymbol{\phi})\right)^2}{\sigma_{m,\mathrm{tot}}^2} + \ln\left(2\pi \sigma_{m,\mathrm{tot}}^2\right) \right].
\label{eq:likelihood}
\end{equation}

To account for surrogate error, we inflate the uncertainty by a multiplicative emulator-error term calibrated on the held-out test set (\autoref{sec:validation}), $\sigma_{m,\mathrm{tot}}^2 = \left(\sigma^{\mathrm{obs}}_m\right)^2 + \left(\varepsilon_{\mathrm{emu}}\,\hat{F}_m(\phi)\right)^2$, where $\hat{F}_m(\phi)$ is the predicted flux in band $m$ and $\varepsilon_{\mathrm{emu}} = \mathrm{median}\left[\varepsilon_{\mathrm{RMS}}(\lambda)\right]$ is the median relative RMS test-set error (\autoref{sec:validation_forward}). This constant-relative discrepancy model is deliberately simple \citep[e.g.][]{kennedy2001bayesiancalibrationcomputer, bayarri2007frameworkvalidationcomputer}, and we discuss its limitations in \autoref{sec:disc_limitations}.

We sample the posterior, $p(\boldsymbol{\phi}\,|\,\mathbf{F}^{\mathrm{obs}}) \propto p(\mathbf{F}^{\mathrm{obs}}\,|\,\boldsymbol{\phi})\,p(\boldsymbol{\phi})$, using nested sampling \citep{skilling2004nestedsampling, buchner2019collaborativenestedsampling, buchner2023nestedsamplingmethods} as implemented in \texttt{UltraNest} \citep{buchner2021ultranestrobustgeneral}. We prefer nested sampling over MCMC for two reasons. First, this likelihood is expected to be degenerate and possibly multi-modal \citep{thamm1994ambiguitiesparametrizeddust, chiang1997spectralenergydistributions, robitaille2006interpretingspectralenergy, robitaille2007interpretingspectralenergy, angelo2023demographicsprotoplanetarydisks}, a regime where nested sampling is typically more robust because it compresses the prior volume through a set of live points iteratively constrained to higher likelihood \citep{buchner2023nestedsamplingmethods, dittmann2024notespracticalapplication}. Second, it simultaneously yields the Bayesian evidence, useful for comparing physical assumptions such as the presence or absence of a cavity \citep{hughes2007innerholedisk, sturm2023edgeonprotoplanetarydisk} in a consistent framework.

We use \texttt{UltraNest}'s \texttt{MLFriends}-region-based nested sampling with a population slice sampler to generate proposals within the evolving likelihood constraint. We follow the recommended calibration, doubling $n_{\mathrm{steps}}$ until the log-evidence $\ln\mathcal{Z}$ stabilizes. As a consistency check for surrogate-based inference, we re-evaluate posterior samples (including the maximum-likelihood point and the highest-posterior-density region) using the full radiative-transfer pipeline and compare the resulting spectra to the surrogate predictions. For synthetic-observation inference only, we incorporate the hot-start procedure described in \autoref{sec:ultranest_hot_start} to reduce the computational cost of the validation.

\section{Emulator validation}
\label{sec:validation}

We validate the surrogate in two complementary ways: its forward-emulation accuracy on a held-out test set (\autoref{sec:validation_forward}), and its end-to-end inference performance on synthetic observations with known ground truths (\autoref{sec:validation_inference}). The former bounds the point-wise approximation error that enters the likelihood through $\sigma_{m,\mathrm{tot}}$, while the latter checks that this error neither biases nor miscalibrates the recovered posteriors. The error metrics and the synthetic-inference setup are defined in \autoref{sec:appendix_validation_methods}.

\subsection{Forward emulation}
\label{sec:validation_forward}

We quantify forward accuracy on the emulator wavelength grid through the relative residual $r_i(\lambda_j)$ and the per-spectrum relative RMS error $\epsilon_i$ (\autoref{eq:relative_residual}, \autoref{eq:per_spectrum_relative_rmse}), applying the same validity mask for missing fluxes as in the training objective.

\begin{figure}[ht]
    \centering
    \includegraphics[width=0.99\columnwidth]{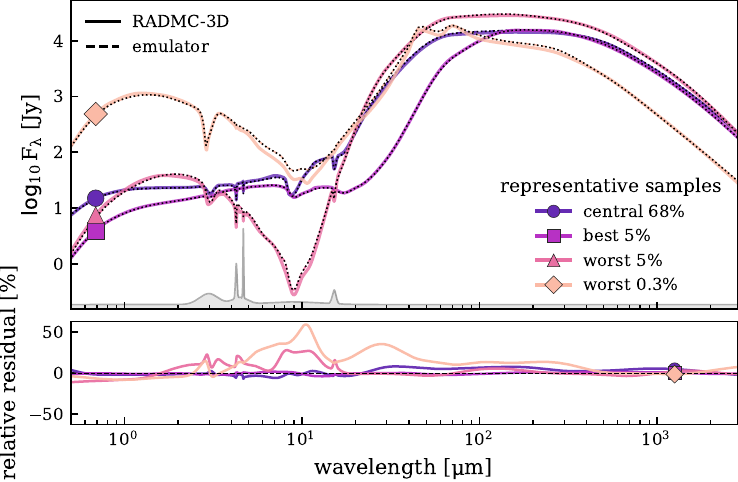}
    \caption{Representative emulations of the held-out test set. {\it Upper panel:} Comparison of simulated and emulated spectra at different quantiles of the per-spectrum relative RMS error distribution, $\epsilon_i$. The light-gray area indicates the emulator's sampling density. {\it Lower panel:} Corresponding relative residuals.}
    \label{fig:representative_prediction_examples}
\end{figure}

Representative examples spanning the best $5\%$, central $68\%$, worst $5\%$, and worst $0.3\%$ of the $\epsilon_i$ distribution (\autoref{tab:per_spectrum_and_band_errors_uniform_grid}) are shown in \autoref{fig:representative_prediction_examples}. Across the bulk of the test set the emulator reproduces both the broad continuum and the relatively narrow ice absorption bands; residuals remain at the few-percent level over most of the spectral range and grow only in localized regions and in the rare tail spectra discussed below.

\begin{figure}[ht]
    \centering
    \includegraphics[width=0.99\columnwidth]{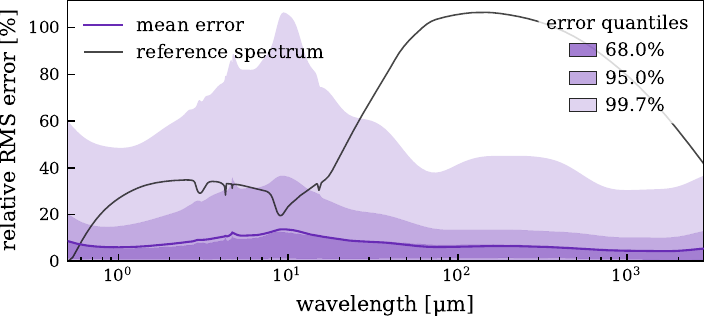}
    \caption{Wavelength dependent emulator error of the held-out test set. The relative RMS error, $\epsilon_{\rm RMS}(\lambda)$ (solid purple curve) and central intervals of the relative residual distribution (shaded bands) are compared with a representative test-set spectrum in arbitrary units  (gray curve) to highlight spectral features. The wavelength grid is non-uniform, so the curve should be interpreted together with the sampling density shown in \autoref{fig:representative_prediction_examples}.}
    \label{fig:emulator_error_by_wavelength}
\end{figure}

The wavelength-resolved RMS error $\epsilon_{\rm RMS}(\lambda)$ (\autoref{eq:wavelength_relative_rmse}) is not constant over the full wavelength grid (\autoref{fig:emulator_error_by_wavelength}). The mean error is ${\sim}\fwdRelRmseMean\%$, rising by a factor of ${\sim}1.1$-$1.5$ at the ice absorption bands, the $10\,\mu$m silicate feature, and the optical and (sub-)mm domain boundaries. The first two coincide with the sharpest curvature in the target spectra, where the relative $H^1$ term in the loss (\autoref{eq:loss}) is most stringent and the training grid is most sparsely populated in flux. The elevation at the domain edges is consistent with residual leakage from the finite-domain padding used in the Fourier blocks (\autoref{sec:neural_operators}). The median of $\epsilon_{\rm RMS}(\lambda)$ is $\fwdRelRmseMedian\%$, rising to $\fwdRelRmseQNinetyFive\%$ and $\fwdRelRmseQNineNineSeven\%$ in the worst $5\%$ and $0.3\%$ of wavelength channels. The wavelength grid is non-uniform, so these statistics should be read together with the sampling density shown in \autoref{fig:representative_prediction_examples}, and compared with the $\log_{10}\lambda$-uniform resampled statistics in \autoref{tab:per_spectrum_and_band_errors_uniform_grid}.

\begin{table}[t]
  \centering
  \caption{Forward-emulator relative RMS error on the test set.}
  \label{tab:per_spectrum_and_band_errors_uniform_grid}
  \begin{tabular}{lrrr}
    \toprule
    Quantity & Range [$\mu$m] & Mean [\%] & Median [\%] \\
    \midrule
    \multicolumn{2}{l}{\textit{Per-spectrum error $\epsilon_{i}$}} & \fwdPerSpecMean & \fwdPerSpecQFifty \\
    $Q_{68}(\epsilon_{i})$ {\scriptsize (worst 32\%)} &  & \multicolumn{2}{c}{\fwdPerSpecQSixtyEight} \\
    $Q_{95}(\epsilon_{i})$ {\scriptsize (worst 5\%)} &  & \multicolumn{2}{c}{\fwdPerSpecQNinetyFive} \\
    $Q_{99.7}(\epsilon_{i})$ {\scriptsize (worst 0.3\%)} &  & \multicolumn{2}{c}{\fwdPerSpecQNineNineSeven} \\
    \multicolumn{4}{l}{\textit{Band-resolved} $\epsilon_{\rm RMS}(\lambda)$ ($\log_{10}\lambda$-uniform)} \\
    Full support & 0.55--2830 & \fwdBandUniAllMean & \fwdBandUniAllMedian \\
    Optical & 0.5--1 & \fwdBandUniOpticalMean & \fwdBandUniOpticalMedian \\
    NIR & 1--5 & \fwdBandUniNirMean & \fwdBandUniNirMedian \\
    MIR & 5--30 & \fwdBandUniMirMean & \fwdBandUniMirMedian \\
    FIR & 30--350 & \fwdBandUniFirMean & \fwdBandUniFirMedian \\
    (sub-)mm & 350--3000 & \fwdBandUniSubMmMean & \fwdBandUniSubMmMedian \\
    Silicate (10\,$\mu$m) & 8--13 & \fwdBandUniSilicateMean & \fwdBandUniSilicateMedian \\
    Ice bands & 2.85--15.6 & \fwdBandUniIceMean & \fwdBandUniIceMedian \\
    \bottomrule
  \end{tabular}
  \par\smallskip
  {\footnotesize
  \begin{minipage}{\linewidth}
  \textbf{Notes.}~$\epsilon_{i}$: per-spectrum error (RMS over wavelength for each spectrum); the upper block is its distribution across spectra. $\epsilon_{\rm RMS}(\lambda)$: band-resolved error (RMS over spectra in each band), given as the mean and median within each band on uniform $\log_{10}\lambda$ grid. Upper quantiles are one-sided (e.g.\ $Q_{95}$ is exceeded by the worst 5\% of spectra).
  \end{minipage}}
\end{table}

\begin{figure}[ht]
    \centering
    \includegraphics[width=0.99\columnwidth]{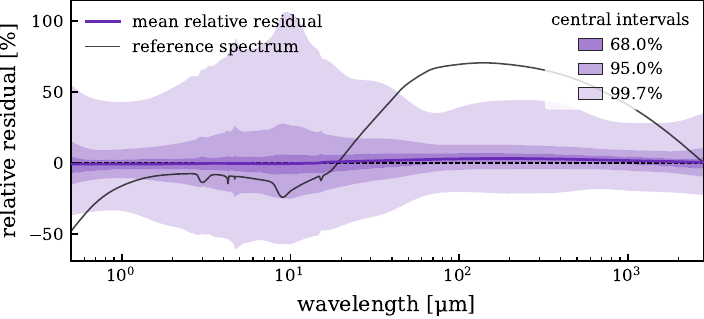}
    \caption{Mean signed residual $b(\lambda)$ on the held-out test set, showing a small ($1$-$3\%$) but coherent chromatic tilt: optical/NIR fluxes slightly low, FIR/sub-mm fluxes slightly high. The gray curve shows a representative test-set spectrum in arbitrary units for comparison with spectral features.}
    \label{fig:signed_emulator_bias_by_wavelength}
\end{figure}

The mean signed residual $b(\lambda)$ (\autoref{eq:signed_residual_bias}) is complementary to $\epsilon_{\rm RMS}(\lambda)$. The RMS error measures the amplitude of the residuals, whereas $b(\lambda)$ tests whether they cancel or whether the emulator is systematically high or low at a given wavelength (\autoref{fig:signed_emulator_bias_by_wavelength}). It reveals a small but coherent chromatic tilt: the optical/NIR fluxes are biased low and the FIR/sub-mm fluxes high, both at the $1$--$3\%$ level. This tilt is well below the photometric uncertainties modeled in \autoref{sec:results} and is absorbed by the emulator-error inflation in $\sigma_{m,\mathrm{tot}}$.

\begin{figure}[ht]
    \centering
    \includegraphics[width=0.99\columnwidth]{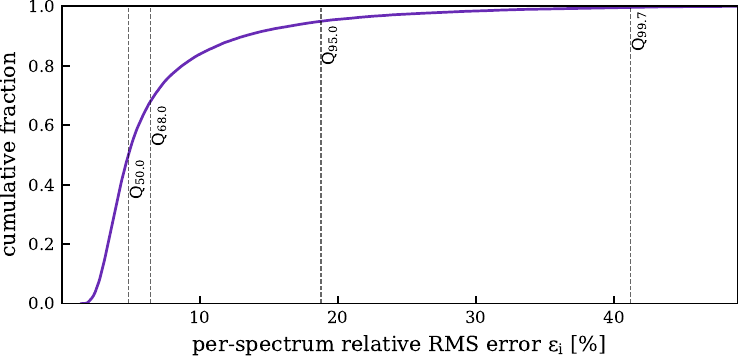}
    \caption{Cumulative distribution of the per-spectrum relative RMS error, $\epsilon_i$, on the held-out test set. The distribution is relatively compact with a median of $\epsilon_{i} = \fwdPerSpecQFifty\%$, together with a long tail that indicates rare high-error cases.}
    \label{fig:per_spectrum_relative_rmse_cdf}
\end{figure}

Finally, the cumulative distribution of $\epsilon_i$ (\autoref{fig:per_spectrum_relative_rmse_cdf}) is S-shaped and compact, with a median of $\fwdPerSpecQFifty\%$ and a $68$th percentile of $\fwdPerSpecQSixtyEight\%$, together with a pronounced heavy tail ($Q_{95}=\fwdPerSpecQNinetyFive\%$, $Q_{99.7}=\fwdPerSpecQNineNineSeven\%$). The core sets the typical emulator error for inference, while the tail is dominated by a small population of either atypical models at the corners of the parameter space, or, more importantly, high-opacity disks for which the reference simulations are noisy (see \autoref{fig:emulator_superiority_examples}). We argue in \autoref{sec:disc_superiority} that the high error of the latter is in part an artifact of imperfect targets rather than emulator failure.

\subsection{Synthetic inference}
\label{sec:validation_inference}

To verify that the surrogate supports unbiased, calibrated parameter estimation, we infer the known parameters of held-out test-set simulations with nested sampling (\autoref{sec:fitting}), using two synthetic sets of $N_{\rm synth}=512$ observations that bracket the intended use case (construction in \autoref{sec:appendix_validation_methods}). The first case, `high-$R_{\rm spec}$', uses the high-resolution test spectra at full wavelength sampling with a uniform $\sigma_{\rm synth}=10^{-2}$, probing near-ideal, information-rich conditions. The second case, `ESO-H$\alpha$~569-like', subsamples each spectrum to the wavelengths and per-band uncertainties of the ESO-H$\alpha$~569 photometry (\autoref{tab:sedphot}), probing the sparse, unevenly weighted regime representative of photometric disk observations. For both datasets we visually inspect the temperature structures computed by \texttt{RADMC-3D} and redraw any samples for which the thermal Monte Carlo clearly did not converge (9 samples rejected, see the discussion in \autoref{sec:disc_superiority}).

\begin{figure}[ht]
    \centering 
    \includegraphics[width=0.99 \columnwidth]{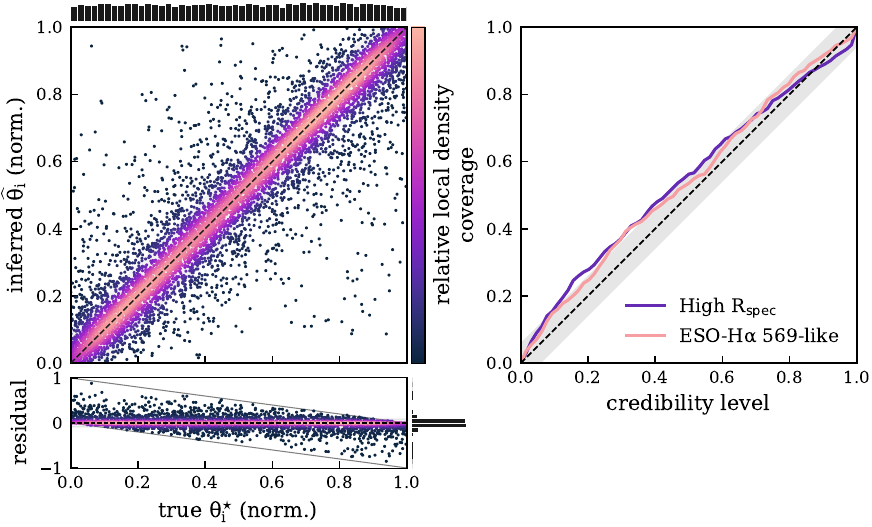}
    \caption{Synthetic-inference validation. \emph{Left:} recovered posterior medians versus the true values for the high-$R_{\rm spec}$ set. \emph{Right:} TARP coverage test for both experiments; the TARP curves follow the diagonal, with $\mathrm{ATC}=0.0111$ (high-$R_{\rm spec}$) and $\mathrm{ATC}=0.0136$ (ESO-H$\alpha$~569-like), indicating no significant miscalibration.}
    \label{fig:inferred_vs_true}
    \label{fig:tarp_plot}
\end{figure}

\autoref{fig:inferred_vs_true} compares posterior medians to the true values for the high-resolution spectra. The well-constrained structural parameters (disk mass, scale height, inclination, surface-density slope) scatter tightly about the one-to-one relation, whereas the credible intervals widen substantially for the parameters that are still weakly constrained at this spectral resolution (ice abundances and cavity properties) without indicating a systematic offset. 

We test the statistical calibration of the posteriors with the Test of Accuracy with Random Points (TARP) coverage test \citep{lemos2023samplingbasedaccuracytesting}, shown in \autoref{fig:tarp_plot} and detailed in \autoref{sec:appendix_validation_methods}, complemented by posterior-predictive checks. Both checks are consistent with no significant miscalibration: the expected-versus-credibility curve follows the diagonal with an area between the coverage curve and the diagonal of $\mathrm{ATC}=0.0111$ (high-$R_{\rm spec}$) and $\mathrm{ATC}=0.0136$ (ESO-H$\alpha$~569-like). The curve shows a shape which could be consistent with a small bias; this is the expected imprint of the few-percent optical+NIR/sub-mm broadband bias identified in \autoref{sec:validation_forward}. The posterior-predictive checks are satisfied for all synthetic observations in both sets.

We take the combination of low test-set error, unbiased parameter recovery, and no significant miscalibration as evidence that the surrogate can replace the radiative-transfer solver in the inference pipeline without introducing significant bias. We emphasize, however, that this conclusion applies to the training distribution: when applied to a real target, the surrogate's accuracy in that specific regime should always be confirmed by re-simulating the inferred models with \texttt{RADMC-3D} (\autoref{sec:fitting}, \autoref{sec:results_eso_ha_569}). We note further that broadband photometry does not carry the information content of resolved ice-band profiles, whose interpretation in turn requires an accurate disk radiative-transfer model (\autoref{sec:background_edge_on_disks}). With these caveats noted, we proceed to apply the validated emulator to a real target.

\section{Results}
\label{sec:results}

\label{sec:results_eso_ha_569}

We apply the validated surrogate and the inference procedure of \autoref{sec:fitting} to the optical-to-mm photometry of ESO-H$\alpha$~569 (\autoref{tab:sedphot}). The four marginal Herschel detections and the 1330~$\mu$m measurement are not included in the likelihood. We jointly fit the 18 physical model parameters and the nuisance parameters $d$ and $A_V$ under the priors of \autoref{tab:results_posterior}, including the held-out emulator error through $\varepsilon_{\mathrm{emu}}\,\hat{F}_m(\phi)$.
 
\begin{figure*}[ht]
    \sidecaption
    \includegraphics[width=120mm]{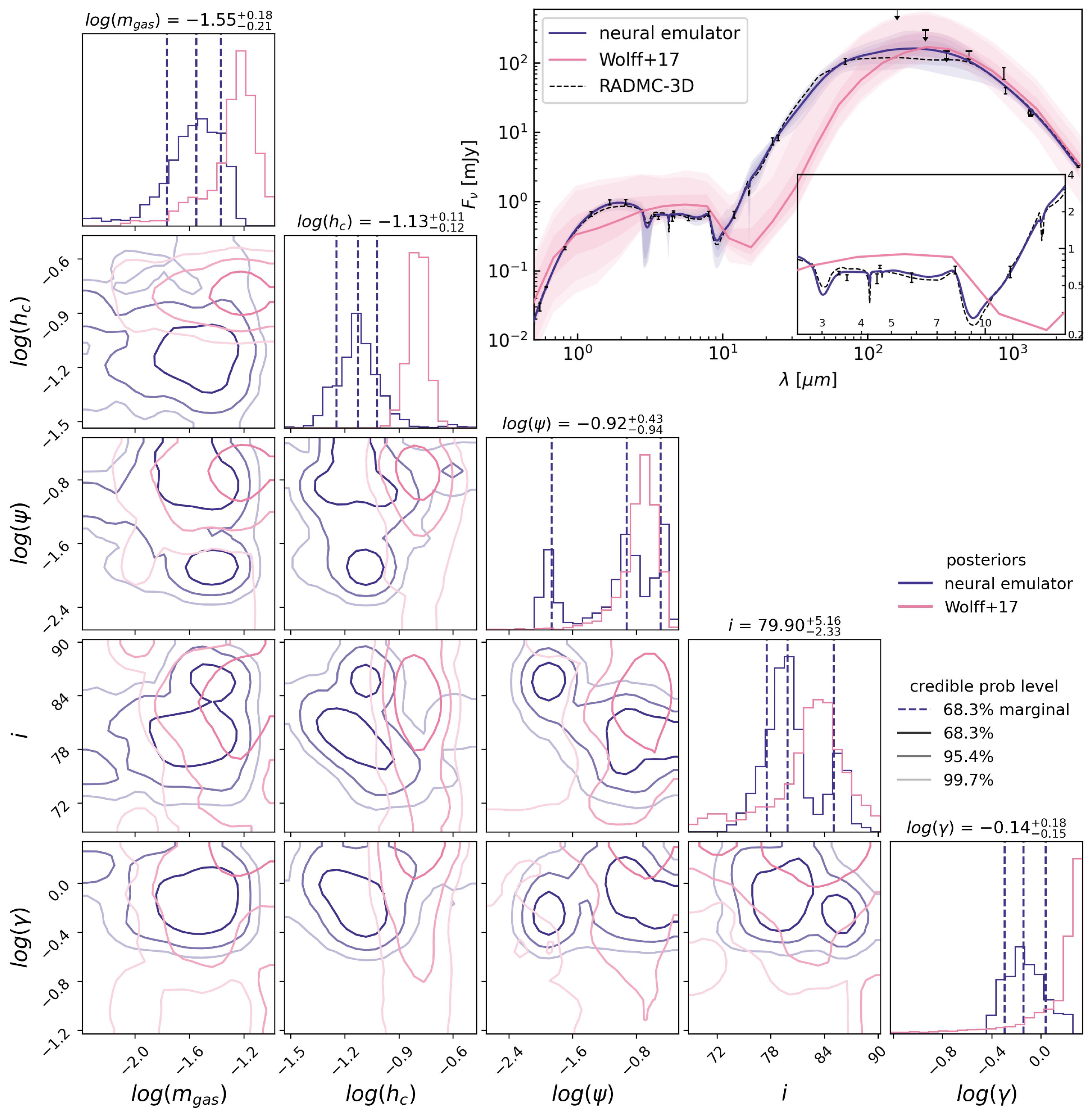}
    \caption{ESO-H$\alpha$~569 posterior and comparison with \citet{wolff2017hubblespacetelescope} for selected parameters. The upper-right panel compares the posterior-predictive spectrum (median; purple) and the \texttt{RADMC-3D} re-simulation of the maximum-likelihood parameter vector with the photometry; the Herschel $3\sigma$ upper limits and the 1330~$\mu$m measurement are shown for context but are not included in the likelihood. Because the two studies use different disk parameterizations and priors, the comparison is intended as a consistency check rather than a one-to-one parameter match.}
    \label{fig:eso_ha_569_posterior}
\end{figure*}
 
\begin{table*}[ht]
\centering
\small
\renewcommand{\arraystretch}{1.15}   %
\setlength{\tabcolsep}{4pt}
\caption{Model parameters, prior bounds, and the posteriors of ESO-H$\alpha$~569 and HH~48~NE.}
\label{tab:results_posterior}
\label{tab:prior_grid}
\label{tab:results_posterior_reduced}
\label{tab:prior_grid_reduced}
\begin{tabular}{llcccc}
\hline\hline
Parameter \hfill [unit] & Description & Prior & Prior & ESO-H$\alpha$~569 & HH~48~NE \\
                        &             & lower & upper &                   &                 \\
\hline
\multicolumn{6}{l}{\emph{Disk structure}}\\
$\log_{10}(M_{\rm gas})$ \hfill [$M_\odot$]      & total gas mass (gas-to-dust ratio 100)                & $-5.0$ & $-0.9$ & $\EsoHaLogMgasMed$ & $-2.39^{+0.59}_{-0.48}$ \\
$\log_{10}(R_{\rm c})$ \hfill [AU]               & characteristic radius of the surface-density profile  & $1.0$ & $3.0$ & $\EsoHaLogRcMed$ & $1.56^{+0.61}_{-0.35}$ \\
$\log_{10}(h_{\rm c})$ \hfill [$-$]              & scale-height ratio $H/r$ of the gas at $R_{\rm c}$     & $-2.0$ & $-0.2$ & $\EsoHaLogHcMed$ & $-0.95^{+0.28}_{-0.22}$ \\
$\log_{10}(\psi)$ \hfill [$-$]                   & flaring index, $H/r\propto r^{\psi}$                   & $-2.0$ & $-0.3$ & $\EsoHaLogPsiMed$ & $-1.12^{+0.37}_{-0.47}$ \\
$\log_{10}(\gamma)$ \hfill [$-$]                 & surface-density power-law index                       & $-0.5$ & $0.48$ & $\EsoHaLogGammaMed$ & $-0.10^{+0.22}_{-0.21}$ \\
\multicolumn{6}{l}{\emph{Dust / settling}}\\
$\log_{10}(1-f_\ell)$ \hfill [$-$]               & dust-mass fraction in the small, well-mixed population & $-4.0$ & $0.0$ & $\EsoHaLogOneMinusFlMed$ & $-1.12^{+0.50}_{-0.58}$ \\
$\log_{10}(a_{\rm min})$ \hfill [$\mu$m]         & minimum grain size (both populations)                 & $-3.0^{a}$ & $1.7$ & $\EsoHaLogAminMed$ & $-0.07^{+0.85}_{-1.55}$ \\
$\log_{10}(a_{\rm max, small})$ \hfill [$\mu$m]   & maximum grain size of the small population            & $-2.0^{a}$ & $2.7$ & $\EsoHaLogAmaxSmallMed$ & $1.17^{+0.22}_{-0.46}$ \\
$\log_{10}(X)$ \hfill [$-$]                      & scale height of the settled large grains, relative to the gas & $-2.0$ & $0.0$ & $\EsoHaLogXMed$ & $-0.39^{+0.27}_{-0.24}$ \\
\multicolumn{6}{l}{\emph{Geometry / cavity}}\\
$i$ \hfill [$^{\circ}$]                          & inclination ($90^{\circ}$: edge-on)                    & $70$ & $90$ & $\EsoHaIncMed$ & $86.1^{+2.7}_{-9.8}$ \\
$\log_{10}(\delta_{\rm cav})$ \hfill [$-$]       & density depletion factor inside $R_{\rm cav}$          & $-4.0$ & $0.0$ & $\EsoHaLogDeltaCavMed$ & $-1.80^{+1.23}_{-1.36}$ \\
$\log_{10}(R_{\rm cav})$ \hfill [AU]             & cavity radius                                          & $-2.0$ & $2.0$ & $\EsoHaLogRcavMed$ & $-0.34^{+1.52}_{-1.09}$ \\
\multicolumn{6}{l}{\emph{Ice abundances (relative to hydrogen; $x_{\rm ice} = n_{\rm ice}/n_H$)}}\\
$\log_{10}(x_{\rm CO})$ \hfill [$-$]             & CO ice abundance                                       & $-8.0$ & $-3.0$ & $\EsoHaLogXcoMed$ & $-5.43^{+1.66}_{-1.62}$ \\
$\log_{10}(x_{\rm CO_2})$ \hfill [$-$]           & CO$_2$ ice abundance                                   & $-8.0$ & $-4.0$ & $\EsoHaLogXcoTwoMed$ & $-6.10^{+1.49}_{-1.33}$ \\
$\log_{10}(x_{\rm H_2O})$ \hfill [$-$]           & H$_2$O ice abundance                                   & $-8.0$ & $-4.0$ & $\EsoHaLogXhTwoOMed$ & $-6.03^{+1.56}_{-1.34}$ \\
\multicolumn{6}{l}{\emph{Stellar}}\\
$\log_{10}(L_\star)$ \hfill [$L_\odot$]          & stellar luminosity (normalization of the blackbody)   & $-1.0$ & $0.5$ & $\EsoHaLogLstarMed$ & $0.41^{+0.22}_{-0.63}$ \\
$\log_{10}(R_\star)$$^{(b)}$ \hfill [$R_\odot$]  & radius of the photon-emitting sphere (geometry only)  & $-0.3$ & $0.5$ & $\EsoHaLogRstarMed$ & $-0.05^{+0.36}_{-0.45}$ \\
$\log_{10}(T_\star)$ \hfill [K]                  & blackbody temperature of the stellar spectrum         & $3.48$ & $3.78$ & $\EsoHaLogTstarMed$ & $3.52^{+0.07}_{-0.03}$ \\
\multicolumn{6}{l}{\emph{Observational (nuisance)}}\\
$d$ \hfill [pc]                                  & distance                                               & $150$ & $240$ & $\EsoHaDistMed$ & $186^{+14}_{-15}$ \\
$A_V$ \hfill [mag]                               & foreground $V$-band extinction                         & $0$ & $5$ & $\EsoHaAvMed$ & $2.57^{+0.97}_{-0.68}$ \\[2pt]
\hline\hline
\end{tabular}
\\[2pt]
{\footnotesize \textbf{Notes.} Posterior values are medians with $16$th/$84$th-percentile credible intervals, averaged over all posterior modes. Prior bounds are the emulator-grid bounds for the physical parameters. The nuisance parameters $d$ and $A_V$ have target-specific priors (\autoref{sec:fitting}), listed here for ESO-H$\alpha$~569. $^{(a)}$ We enforce the constraint $\log_{10}(a_{\rm max, small}) \geq \log_{10}(a_{\rm min}) + 1$. $^{(b)}$ $R_\star$ defines the finite stellar size for \texttt{RADMC-3D} and should not be interpreted as a stellar-radius measurement (\autoref{sec:rt_synthetic}).} %
\end{table*}

The posterior is bimodal in inclination: a dominant mode at $i\simeq\EsoHaIncModeA\,^{\circ}$ and a secondary mode at $i\simeq\EsoHaIncModeB\,^{\circ}$, thicker, more flared, and closer to edge-on (\autoref{fig:eso_ha_569_posterior}). The global summaries in \autoref{tab:results_posterior} average over both modes. For the adopted gas-to-dust ratio of 100, $M_{\rm gas}=\EsoHaMgasMed\,M_\odot$. The characteristic radius is $R_{\rm c}=\EsoHaRcMed\,$AU, although it remains broad and covariant with the scale height, flaring, and inclination. The recovered distance, $d=\EsoHaDistMed\,$pc, is consistent with the distance of the northern Cha~I subgroup ($d=191.4^{+0.8}_{-0.8}$\,pc, \citealt{galli2021chamaeleondancerevisiting}). The inferred foreground extinction is modest, $A_V=\EsoHaAvMed$, consistent with the foreground extinction toward the face-on disks in Cha~I \citep{manoj2011spitzerinfraredspectrograph}. The posterior-predictive spectra of both modes reproduce the scattered-light optical/NIR flux, the mid-infrared depression, the far-infrared peak, and the millimeter emission. Notably, the prediction also agrees closely with the 1330~$\mu$m measurement, which was not included in the likelihood, and the posterior-predictive spectrum, although it partially exceeds the 350 and 500~$\mu$m upper limits, remains within the range allowed by the (sub-)millimeter photometry (\autoref{sec:observations}). The Stefan--Boltzmann radius implied by $L_\star$ and $T_\star$, $\log_{10}(R_{\rm SB}/R_\odot)=0.21^{+0.32}_{-0.13}$, is consistent with the retrieved $\log_{10}(R_\star/R_\odot)=\EsoHaLogRstarMed$ and lies within the $R_\star$ prior for $76\%$ of the posterior samples (\autoref{sec:rt_synthetic}). Broadband photometry provides little information on the individual ice abundances or on the cavity parameters, whose marginals remain broad and substantially prior-driven. The fit should therefore be interpreted as a degeneracy-aware SED solution set and an end-to-end test of the inference pipeline, not as a replacement for resolved structural constraints. As a consistency check, we re-evaluated the maximum-likelihood point and posterior draws from both modes with the full \texttt{RADMC-3D} pipeline. The resulting spectra agree with the surrogate within the held-out error model, with the largest differences around the silicate and ice features (\autoref{fig:eso_ha_569_posterior}). The far-infrared to millimeter offset between the re-simulated maximum-likelihood spectrum and the posterior-predictive median reflects the averaging of the median over both inclination modes and the broad $R_{\rm c}$ posterior.

\section{Discussion}
\label{sec:discussion}

\subsection{Comparison with previous models}
\label{sec:disc_comparison}

\paragraph{ESO-H$\alpha$ 569 (Wolff et al.\ 2017).}

ESO-H$\alpha$~569 is primarily a validation target for the methodology. Despite using a different disk prescription, radiative-transfer code, and likelihood, and despite fitting no resolved imaging, our posterior is physically compatible with the image-plus-SED analysis of \citet{wolff2017hubblespacetelescope}: both favor an optically thick, massive, nearly edge-on disk without a large inner cavity. The mass agreement is not independent, because both analyses use similar (sub-)millimeter photometry and assume a gas-to-dust ratio of 100.

Fitted separately, the image and SED of \citet[their Fig.~5]{wolff2017hubblespacetelescope} prefer different inclinations; their joint MCMC settles at $i=83.0^{+2.6}_{-4.8}\,^{\circ}$. Our photometry-only posterior contains both solutions as modes along the inclination--scale-height--flaring degeneracy \citep{robitaille2006interpretingspectralenergy, robitaille2007interpretingspectralenergy, angelo2023demographicsprotoplanetarydisks}, the multimodal regime anticipated in \autoref{sec:fitting}, which nested sampling over the surrogate maps in a single run. The dominant mode coincides with the inclination preferred by the resolved imaging, data not used in our fit. The mode weights depend on the emulator-error inflation and the neglected inter-band covariance, so we do not claim that photometry alone selects the imaging solution. The modes do, however, predict different resolved morphologies (dark-lane width, nebula flux ratio), so a single resolved image discriminates between them (\autoref{sec:disc_engine}). The value of the photometry-only posterior is thus not a precise structural measurement but the recovered, weighted solution set, which resolved imaging or spectroscopy can narrow.
 
\paragraph{HH 48 NE (Sturm et al.\ 2023a,b, 2024).}

\begin{figure}[ht]
    \centering
    \includegraphics[width=0.99\columnwidth]{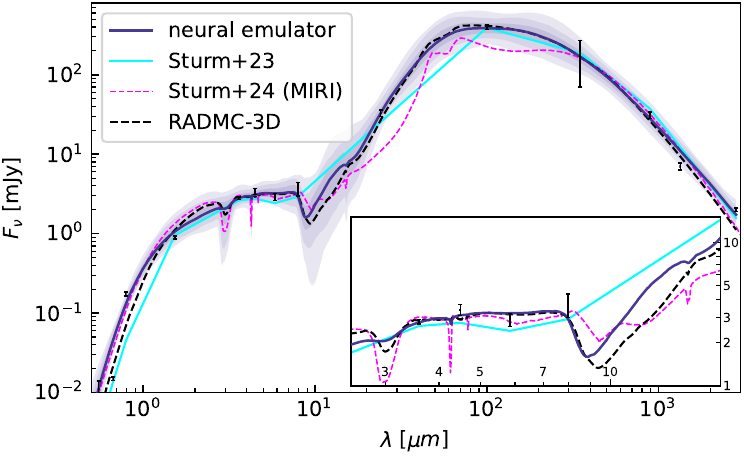}
    \caption{Posterior predictive HH 48 NE. The overplotted cyan line is the \cite{sturm2023edgeonprotoplanetarydiska} model, fit to the SED. The overplotted magenta line is the \cite{sturm2024jwstmirianalysis} model, fit to the MIRI observations of HH 48 NE.}
    \label{fig:hh_48_ne_posterior_predictive}
\end{figure}

HH 48 NE provides an external benchmark with independently constrained models from fits to the SED and resolved HST/ALMA imaging \citep{sturm2023edgeonprotoplanetarydisk, sturm2023edgeonprotoplanetarydiska} and to the JWST/MIRI spectrum \citep{sturm2024jwstmirianalysis}. We fit only the photometric compilation of \citet{sturm2023edgeonprotoplanetarydisk}, jointly sampling $d$ and $A_V$ as for ESO-H$\alpha$~569. The posterior-predictive spectra (\autoref{fig:hh_48_ne_posterior_predictive}) reproduce the photometry from the optical to the millimeter, and re-evaluated posterior draws agree with the surrogate within the held-out error model, with the largest differences at the shortest wavelengths and in the silicate and ice bands (\autoref{sec:validation_forward}). In the near- to mid-infrared, the predicted median tracks the MIRI-calibrated model of \citet{sturm2024jwstmirianalysis} more closely than the SED-based models \citep{sturm2023edgeonprotoplanetarydisk, sturm2023edgeonprotoplanetarydiska}.

The parameters constrained by our fit are consistent with \citet{sturm2023edgeonprotoplanetarydisk}, within broader credible intervals because we marginalize over $d$ and $A_V$ where the previous analyses optimized them; the inflation enters through the $1/d^{2}$ flux scaling and the $A_V$--$L_\star$--$T_\star$ degeneracy. Nuisance parameters fixed at point values propagate into the published values as unaccounted systematics; marginalizing them yields the more complete uncertainty budget required for homogeneous population comparisons.

Two differences are physically informative. First, the cavity parameters are prior-dominated. The ${\sim}55\,$au dust-depleted cavity, required by the resolved millimeter imaging and mid-infrared excess \citep{sturm2023edgeonprotoplanetarydisk}, is neither recovered nor excluded. Second, our posterior prefers $T_\star \simeq 3300\,$K, whereas \citet{sturm2023edgeonprotoplanetarydisk} adopt a spectroscopic K7 type (${\sim}4000\,$K). For an occulted star, $T_\star$ enters the observables only through the scattered-light color, which is degenerate with $A_V$ and $L_\star$, and our blackbody photosphere neglects the molecular blanketing that shapes late-K optical colors. Accordingly, the $R_\star$ marginal is nearly flat over its prior, and the Stefan--Boltzmann radius implied by the posterior, $\log_{10}(R_{\rm SB}/R_\odot)=0.63^{+0.18}_{-0.31}$, exceeds the upper prior bound of $R_\star$ for $69\%$ of the samples. The stellar parameters could be further constrained without retraining the emulator, by imposing the spectroscopic values as priors or by coupling $L_\star$, $T_\star$, and $R_\star$, which currently vary independently (\autoref{sec:rt_synthetic}), through the Stefan--Boltzmann relation or pre-main-sequence evolutionary tracks, as \citet{harries2026fundamentalpropertiesprotoplanetary} does for their training set.

\subsection{The emulator as an inference engine}
\label{sec:disc_engine}

The surrogate's principal advantage is that it amortizes the cost of radiative transfer: the training grid is expensive but paid once, after which evaluation and inference are cheap. Building the grid cost ${\sim}10^{6}$ CPU-hours, and training is negligible by comparison. Against this one-time investment, a batched forward pass evaluates $4096$ parameterizations in ${\sim}1\,$ms on a single GPU; including host--device transfers, floating-point conversions, and the slower deterministic GPU kernels used at inference, this corresponds to a conservative per-evaluation speed-up of ${\sim}10^{9}$-$10^{10}\times$ per spectrum over the simulator's ${\sim}10^{1}$-$10^{3}$ CPU hours, an indicative figure that depends on hardware and settings and sets a Monte Carlo solver at fixed photon count against a functional predictor. The ESO-H$\alpha$~569 posterior completed in less than $2\,$h of wall-clock time; a speed-up of ${\sim}4\times10^8\times$ over the ${\gtrsim}10^8$ CPU-hours it would have cost with \texttt{RADMC-3D}. The retrieval factor is smaller because it also contains the serial overhead of the nested sampler, negligible against a \texttt{RADMC-3D} model but not against a microsecond emulator call. The total emulator inference cost, including grid generation, is $<1\%$ of a comparable \texttt{RADMC-3D}-based inference; it is recovered within the first application. %

Two further properties extend the method's reach. First, the surrogate is differentiable, enabling gradient-based sensitivity analysis and uncertainty propagation that are inaccessible to a noisy Monte Carlo solver. Second, it recovers disk properties from photometry alone: the imaging-free ESO-H$\alpha$~569 fit shows that the dominant SED-shaping dependencies are captured, so the method applies to the many edge-on candidates that lack resolved data and can seed more detailed forward models once such data exist. This bears directly on the biased, incomplete edge-on sample discussed in \autoref{sec:background_edge_on_disks}: homogeneous, cheap SED modeling is a prerequisite for quantifying these selection effects and for triaging candidates for follow-up.

Several extensions follow naturally and we leave them to future work. The same neural-operator methodology can emulate images alongside or instead of spectra, complementing the scattered-light image emulators of \citet{telkamp2022machinelearningframework} and \citet{williams2025radiativetransfermodeling} and the millimeter-continuum image emulator of \citet{harries2026fundamentalpropertiesprotoplanetary}. The training grid can be grown online by refining poorly covered regions identified at inference time, or coupled to amortized SBI \citep{cranmer2020frontiersimulationbasedinference}. Our approach can be extended to high-resolution Spitzer or JWST spectroscopy, to younger (Class~0/I) disks, and to a wider diversity of structures such as ringed, non-axisymmetric, or externally irradiated disks \citep{vancapelleveen2025wideseparationplanets, kimmig2025asymmetricsignatureswarps, aru2024kaleidoscopeirradiateddisks}. Two further directions exploit the architecture itself. Because the operator is conditioned on the underlying physical fields rather than on a fixed disk prescription (\autoref{sec:input_embeddings}), it could in principle assimilate training data from different radiative-transfer codes, an aggregation that multi-fidelity and multi-physics operator learning has shown to be feasible \citep{lu2022multifidelitydeepneural, mccabe2023multiplephysicspretraining, subramanian2023foundationmodelsscientific}; and being differentiable, it could optimize the input fields directly when the data are sufficiently constraining \citep{branca2025raytraxfasttimedependent, levis2025revealingfinestructure}, both directions we consider promising but do not pursue here.

\subsection{Limitations and uncertainty}
\label{sec:disc_limitations}

The emulator inherits the assumptions of the underlying physical model, e.g. a fixed dust composition, prescribed ice distribution, and parameterized vertical structure. In particular, the retrieved ice abundances are conditional on the fixed ice prescription (\autoref{sec:model}): a lower vertical ice limit requires up to an order of magnitude higher abundance for the same band depth \citep{sturm2024jwstmirianalysis}, and because ice is added at fixed total grain size, the abundance also impacts the refractory core size and hence the continuum opacity. Its accuracy degrades where the training grid is sparse or near its boundaries. Like neural operators in general, it interpolates reliably within the training distribution but is not guaranteed to extrapolate beyond it \citep{kovachki2023neuraloperatorlearning, lehmann2025multipleinputfourierneural, azizzadenesheli2024neuraloperatorsaccelerating}. We therefore restrict inference to the prior volume and do not test extrapolation, to remain conservative.

The mixed-resolution grid also limits the emulator at high spectral frequencies. Discretization invariance holds only up to aliasing errors \citep{bartolucci2023representationequivalentneural, lanthaler2024discretizationerrorfourier} and a low-frequency bias from mode truncation \citep{qin2024betterunderstandingfourier}, and content unresolved by the training data cannot be assumed recoverable on finer grids \citep{sakarvadia2025falsepromisezeroshot}. In our grid, the majority of training spectra (\autoref{sec:grid}) do not constrain all retained Fourier modes (\autoref{sec:hpo}), and the higher modes are only constrained by moderate- and high-resolution spectra. Together with the noisy targets of \autoref{sec:disc_superiority}, this plausibly contributes to the larger ice-band test error (\autoref{sec:validation_forward}). Incremental spectral training \citep{george2024incrementalspatialspectral}, or more high-resolution spectra, would mitigate this. For the same reason, applying the emulator to Spitzer or JWST spectra ($R\gg R_{\max}\sim\fwdResHighMax$ of the training grid; \autoref{sec:appendix_validation_inference_methods}) formally requires additional training spectra sampled at instrumental resolution in the ice bands, which the mixed-resolution scheme is designed to accommodate.

Our treatment of the emulator error is approximate and deliberately conservative. We inflate the likelihood by a relative RMS error $\varepsilon_{\mathrm{emu}}$, a simplified constant-relative model that under-represents the strong wavelength dependence of the error (\autoref{sec:validation_forward}). A constant $\sigma_{\rm emu}$ overestimates the error on the well-emulated continuum and underestimates it in the ice and silicate bands, exactly the regions that determine the solid state abundances. Although the FNO already implements Monte Carlo dropout \citep{gal2016dropoutbayesianapproximation}, we do not yet use its predictive variance in the likelihood. Propagating a calibrated, wavelength-resolved emulator uncertainty, for instance with the approach of \citet{ma2024calibrateduncertaintyquantification}, would likely improve posterior calibration and is a natural next step.

\subsection{Monte Carlo radiative transfer artifacts}
\label{sec:disc_superiority}
 
Radiative transfer for highly inclined disks is dominated by the small fraction of photon packets that escape through the upper layers, forcing a trade-off between accuracy and tractable photon counts. Two distinct numerical effects degrade the reference simulations of the most optically thick disks. First, at the spectrum/scattering photon count used to build our grid ($n_{\rm phot}^{\rm spec}=2$-$5\times10^{5}$), the ray-tracer produces scattering-related artifacts, especially for optically thick inner disks, yielding Monte Carlo-noise-limited spectra. This degradation is most severe in the high-optical-depth, high-inclination regime, where dedicated benchmarks show the largest inter-code scatter and variance reduction is required to reach a fixed signal-to-noise \citep{pinte2009benchmarkproblemscontinuum, steinacker2013threedimensionaldustradiative, baes2016compositebiasingmonte, gordon2017trust3dexternally}.

Second, the thermal Monte Carlo itself can fail for these disks: the MRW (\autoref{sec:rt_synthetic}; \citealt{fleck1984randomwalkprocedure, min2009radiativetransfervery, robitaille2010modifiedrandomwalk}) is required to move packets through the very optically thick mid-plane, but the deepest, most shielded mid-plane cells of the most massive disks receive no photons, and \texttt{RADMC-3D v2.0} applies no diffusion-approximation temperature correction in those cells \citep{dullemond2012radmc3dmultipurposeradiative}. The need for a (partial-)diffusion temperature estimate where cells see few or no photons is documented throughout the disk benchmark and methods literature \citep{pascucci20042dcontinuumradiative, pinte2009benchmarkproblemscontinuum, krieger2020unbiasedmontecarlo, krieger2024improvingmontecarlo}. The resulting mid-plane temperature, and hence the ray-traced spectrum derived from it, is therefore least reliable exactly for massive disks with very high opacity regions, of which ESO-H$\alpha$~569 (with the high-opacity inner mid-plane preferred by \citealt{wolff2017hubblespacetelescope}) is a prime example. Both effects are well documented to bias Monte Carlo continuum transfer at very high optical depth \citep{min2009radiativetransfervery, robitaille2010modifiedrandomwalk, krieger2020unbiasedmontecarlo, krieger2024improvingmontecarlo}.

The implicit spectral smoothing of the FNO (\autoref{sec:neural_operators}) regularizes against the first effect \citep{rahaman2019spectralbiasneural, ulyanov2020deepimageprior, xu2020frequencyprinciplefourier}, and for such targets the emulator returns a smoother, and physically plausible spectrum compared to its own training reference; \autoref{fig:emulator_superiority_examples} shows extreme cases drawn from the test set. This connects directly to the heavy tail of the per-spectrum error distribution (\autoref{sec:validation_forward}). We posit that part of that tail reflects correct emulator predictions being scored against noisy targets \citep{lehtinen2018noise2noiselearningimage}, so the reported test-set error there overstates the true emulator error.

Re-simulating a test-set model from the high-error tail at photon counts of $10^{5}$--$5\times10^{6}$ supports this view (\autoref{fig:high_nphot_resimulation} and \autoref{fig:run2run_variance}). Relative to the $5\times10^{6}$-photon reference, the RMS residual of the \texttt{RADMC-3D} spectra over 2.5--17.5~$\mu$m falls from $23\%$ at the training photon count ($5\times10^{5}$; run-to-run scatter 20--23\%) to $11\%$ at $2\times10^{6}$, consistent with $n_{\rm phot}^{-1/2}$ scaling, whereas the emulator, trained only on spectra of this photon count and never on this model, deviates from the reference by $17\%$, on par with a model computed with twice the training photon count. An emulator that is closer to a high-fidelity reference than its own training data is what \citet{koehler2025neuralemulatorsuperiority} term emulator superiority. Here it results from the implicit spectral regularization, which removes the part of the Monte Carlo error that is uncorrelated between adjacent wavelengths.

Not all of the finite-photon error is of this kind. In the infrared of a near-edge-on disk the scattered flux is carried by thermally emitted packets that must escape the optically thick inner disk before scattering toward the observer, so that only a small fraction of the packets samples the relevant paths. The estimator is unbiased in expectation but heavy-tailed, and the typical realization under-samples these paths. The result is a coherent, wavelength-dependent deviation that decreases only slowly with photon count and, being smooth in wavelength and in the parameters, can be learned by the emulator rather than corrected \citep{mohan2024whatyousee}, a structural error in the sense of \citet{koehler2025neuralemulatorsuperiority}. For the most massive, highest-opacity disks, the thermal stage itself remains unconverged, so even a high-photon-count reference cannot be trusted, and these cases remain unconfirmed.

\section{Conclusion}
\label{sec:conclusion}

Edge-on disks provide a spatially resolved view of the vertical distribution of icy planetary building blocks, and the JWST edge-on programs, with complementary HST, Spitzer, and ALMA observations, have now assembled this view for the complete known sample. Interpreting these data, however, requires radiative transfer with multiple anisotropic scattering at ${\sim}10^{1}$--$10^{3}$ CPU-hours per model; a Bayesian analysis demands ${\sim}10^{5}$--$10^{6}$ likelihood evaluations, so fitting a single disk would cost ${\gtrsim}10^{8}$ CPU-hours, and studies have therefore relied on hand-tuned model grids without posterior distributions.

This work removes that computational barrier. Our Fourier Neural Operator surrogate reproduces \texttt{RADMC-3D} spectra of ice-bearing, edge-on disks with a median error of $\fwdPerSpecQFifty\%$ and, as the forward model of an explicit-likelihood nested sampler, recovers injected parameters without detectable bias and with nominal coverage. The surrogate predicts spectra $10^{9}$--$10^{10}$ times faster than the simulator, which allows us to compute the posterior of ESO-H$\alpha$~569, consistent with the independent image-plus-SED analysis of \citet{wolff2017hubblespacetelescope}, in less than 2\,h of wall-clock time. Including the one-time training grid, the total cost remains below $1\%$ of a single simulator-based fit.

The Bayesian analysis of the full sample of edge-on disks is thus no longer beyond computational reach. Inclined disks can now be modeled with the full radiative-transfer treatment the data demand, with complete posteriors, at negligible marginal cost. Extended to JWST spectra, this approach enables for the first time homogeneous, population-level inference of the structure and ice reservoirs of planet-forming disks, and with it an observationally calibrated view of the building blocks of exoplanets and Solar System bodies.

\section*{Data availability}
The simulator, training data, and emulator code are presently proprietary, pending specific application papers and final development of an imaging module. Current emulator-derived posteriors may be shared upon reasonable request, subject to agreement on collaboration terms.

\begin{acknowledgements}
  M.K.~McClure was financially supported by the NWO Talent Programme Vidi grant number 233.122 from the Dutch Research Council (NWO).
  This research made use of computational resources provided by both the NWO Talent Programme Vidi grant number 233.122 from the Dutch Research Council (NWO) and the European Union A-ERC grant 101019751 MOLDISK.
  This work was performed using the compute resources from the Academic Leiden Interdisciplinary Cluster Environment (ALICE) provided by Leiden University. %
  This paper makes use of the following ALMA data: ADS/JAO.ALMA\#2012.1.00031.S, ADS/JAO.ALMA\#2016.1.00460.S, ADS/JAO.ALMA\#2019.1.01792.S, and ADS/JAO.ALMA\#2022.1.01302.S. ALMA is a partnership of ESO (representing its member states), NSF (USA) and NINS (Japan), together with NRC (Canada), NSTC and ASIAA (Taiwan), and KASI (Republic of Korea), in cooperation with the Republic of Chile. The Joint ALMA Observatory is operated by ESO, AUI/NRAO and NAOJ.
  This research made use of
  \texttt{RADMC-3D},
  \texttt{OpTool},
  \texttt{UltraNest},
  \texttt{PyTorch},
  \texttt{PyTorch Lightning},
  \texttt{neuraloperator},
  \texttt{DeepXDE},
  \texttt{hydra},
  \texttt{optuna},
  \texttt{BoTorch},
  \texttt{h5py},
  \texttt{NumPy},
  \texttt{SciPy},
  \texttt{pandas},
  \texttt{Matplotlib},
  \texttt{corner},
  and \texttt{CMasher}.
\end{acknowledgements}

\bibliographystyle{bibtex/aa.bst} %
\bibliography{aa62720-26.bib}

@article{aikawa2012akariobservationsice,
  title = {{{AKARI}} Observations of Ice Absorption Bands towards Edge-on Young Stellar Objects},
  author = {Aikawa, Y. and Kamuro, D. and Sakon, I. and Itoh, Y. and Terada, H. and Noble, J. A. and Pontoppidan, K. M. and Fraser, H. J. and Tamura, M. and Kandori, R. and Kawamura, A. and Ueno, M.},
  year = 2012,
  month = feb,
  journal = {A\&A},
  volume = {538},
  pages = {A57},
  publisher = {EDP Sciences},
  issn = {0004-6361, 1432-0746},
  doi = {10.1051/0004-6361/201015999},
  urldate = {2025-03-18},
  copyright = {\copyright{} ESO, 2012},
  langid = {english}
}

@article{alsing2020speculatoremulatingstellar,
  title = {{{SPECULATOR}}: {{Emulating Stellar Population Synthesis}} for {{Fast}} and {{Accurate Galaxy Spectra}} and {{Photometry}}},
  shorttitle = {{{SPECULATOR}}},
  author = {Alsing, Justin and Peiris, Hiranya and Leja, Joel and Hahn, ChangHoon and Tojeiro, Rita and Mortlock, Daniel and Leistedt, Boris and Johnson, Benjamin D. and Conroy, Charlie},
  year = 2020,
  month = jun,
  journal = {ApJS},
  volume = {249},
  number = {1},
  pages = {5},
  publisher = {The American Astronomical Society},
  issn = {0067-0049},
  doi = {10.3847/1538-4365/ab917f},
  urldate = {2026-06-09},
  langid = {english}
}

@article{angelo2023demographicsprotoplanetarydisks,
  title = {Demographics of {{Protoplanetary Disks}}: {{A Simulated Population}} of {{Edge-on Systems}}},
  shorttitle = {Demographics of {{Protoplanetary Disks}}},
  author = {Angelo, Isabel and Duch{\^e}ne, Gaspard and Stapelfeldt, Karl and Telkamp, Zoie and M{\'e}nard, Fran{\c c}ois and Padgett, Deborah and {Van der Plas}, Gerrit and Villenave, Marion and Pinte, Christophe and Wolff, Schuyler and Fischer, William J. and Perrin, Marshall D.},
  year = 2023,
  month = mar,
  journal = {ApJ},
  volume = {945},
  number = {2},
  pages = {130},
  publisher = {The American Astronomical Society},
  issn = {0004-637X},
  doi = {10.3847/1538-4357/acbb01},
  urldate = {2026-06-09},
  langid = {english}
}

@article{arabhavi2022icesplanetformingdisks,
  title = {Ices in Planet-Forming Disks: {{Self-consistent}} Ice Opacities in Disk Models},
  shorttitle = {Ices in Planet-Forming Disks},
  author = {Arabhavi, Aditya M. and Woitke, Peter and Cazaux, St{\'e}phanie M. and Kamp, Inga and Rab, Christian and Thi, Wing-Fai},
  year = 2022,
  month = oct,
  journal = {A\&A},
  volume = {666},
  pages = {A139},
  publisher = {EDP},
  issn = {0004-6361},
  doi = {10.1051/0004-6361/202141825},
  urldate = {2026-06-15},
  copyright = {\copyright{} A. M. Arabhavi et al. 2022},
  langid = {english}
}

@article{aru2024kaleidoscopeirradiateddisks,
  title = {Kaleidoscope of Irradiated Disks: {{MUSE}} Observations of Proplyds in the {{Orion Nebula Cluster}} - {{I}}. {{Sample}} Presentation and Ionization Front Sizes},
  shorttitle = {Kaleidoscope of Irradiated Disks},
  author = {Aru, M.-L. and Mauc{\'o}, K. and Manara, C. F. and Haworth, T. J. and Facchini, S. and McLeod, A. F. and Miotello, A. and {Petr-Gotzens}, M. G. and Robberto, M. and Rosotti, G. P. and Vicente, S. and Winter, A. and Ansdell, M.},
  year = 2024,
  month = jul,
  journal = {A\&A},
  volume = {687},
  pages = {A93},
  publisher = {EDP Sciences},
  issn = {0004-6361, 1432-0746},
  doi = {10.1051/0004-6361/202349004},
  urldate = {2026-06-11},
  copyright = {\copyright{} The Authors 2024},
  langid = {english}
}

@article{arulanantham2024jwstmirimrs,
  title = {{{JWST MIRI MRS Images}} of {{Disk Winds}}, {{Water}}, and {{CO}} in an {{Edge-on Protoplanetary Disk}}},
  author = {Arulanantham, Nicole and McClure, M. K. and Pontoppidan, Klaus and Beck, Tracy L. and Sturm, J. A. and Harsono, D. and Boogert, A. C. A. and Cordiner, M. and Dartois, E. and Drozdovskaya, M. N. and Espaillat, C. and Melnick, G. J. and Noble, J. A. and Palumbo, M. E. and Pendleton, Y. J. and Terada, H. and {van Dishoeck}, E. F.},
  year = 2024,
  month = apr,
  journal = {ApJL},
  volume = {965},
  number = {1},
  pages = {L13},
  publisher = {The American Astronomical Society},
  issn = {2041-8205},
  doi = {10.3847/2041-8213/ad35c9},
  urldate = {2026-06-11},
  langid = {english}
}

@article{azizzadenesheli2024neuraloperatorsaccelerating,
  title = {Neural Operators for Accelerating Scientific Simulations and Design},
  author = {Azizzadenesheli, Kamyar and Kovachki, Nikola and Li, Zongyi and {Liu-Schiaffini}, Miguel and Kossaifi, Jean and Anandkumar, Anima},
  year = 2024,
  month = may,
  journal = {Nat. Rev. Phys.},
  volume = {6},
  number = {5},
  pages = {320--328},
  publisher = {Nature Publishing Group},
  issn = {2522-5820},
  doi = {10.1038/s42254-024-00712-5},
  urldate = {2025-03-24},
  copyright = {2024 Springer Nature Limited},
  langid = {english}
}

@article{baes2016compositebiasingmonte,
  title = {Composite Biasing in {{Monte Carlo}} Radiative Transfer},
  author = {Baes, Maarten and Gordon, Karl D. and Lunttila, Tuomas and Bianchi, Simone and Camps, Peter and Juvela, Mika and Kuiper, Rolf},
  year = 2016,
  month = jun,
  journal = {A\&A},
  volume = {590},
  pages = {A55},
  publisher = {EDP Sciences},
  issn = {0004-6361, 1432-0746},
  doi = {10.1051/0004-6361/201528063},
  urldate = {2026-07-17},
  copyright = {\copyright{} ESO, 2016},
  langid = {english}
}

@inproceedings{balandat2020botorchframeworkefficient,
  title = {{{BoTorch}}: {{A Framework}} for {{Efficient Monte-Carlo Bayesian Optimization}}},
  shorttitle = {{{BoTorch}}},
  booktitle = {Adv. {{Neural Inf}}. {{Process}}. {{Syst}}.},
  author = {Balandat, Maximilian and Karrer, Brian and Jiang, Daniel and Daulton, Samuel and Letham, Ben and Wilson, Andrew G and Bakshy, Eytan},
  year = 2020,
  volume = {33},
  pages = {21524--21538},
  urldate = {2026-07-06}
}

@article{bartolucci2023representationequivalentneural,
  title = {Representation {{Equivalent Neural Operators}}: A {{Framework}} for {{Alias}}-Free {{Operator Learning}}},
  shorttitle = {Representation {{Equivalent Neural Operators}}},
  author = {Bartolucci, Francesca and {de B{\'e}zenac}, Emmanuel and Raonic, Bogdan and Molinaro, Roberto and Mishra, Siddhartha and Alaifari, Rima},
  year = 2023,
  month = dec,
  journal = {Advances in Neural Information Processing Systems},
  volume = {36},
  pages = {69661--69672},
  doi = {10.52202/075280-3051},
  urldate = {2026-10-08},
  langid = {english}
}

@article{bayarri2007frameworkvalidationcomputer,
  title = {A {{Framework}} for {{Validation}} of {{Computer Models}}},
  author = {Bayarri, Maria J and Berger, James O and Paulo, Rui and Sacks, Jerry and Cafeo, John A and Cavendish, James and Lin, Chin-Hsu and Tu, Jian},
  year = 2007,
  month = may,
  journal = {Technometrics},
  volume = {49},
  number = {2},
  pages = {138--154},
  publisher = {Taylor \& Francis},
  issn = {0040-1706},
  doi = {10.1198/004017007000000092},
  urldate = {2026-06-15}
}

@article{belloche2011endstarformation,
  title = {The End of Star Formation in {{Chamaeleon I}}?. {{A LABOCA}} Census of Starless and Protostellar Cores},
  shorttitle = {The End of Star Formation in {{Chamaeleon I}}?},
  author = {Belloche, A. and Schuller, F. and Parise, B. and Andr{\'e}, {\relax Ph}. and Hatchell, J. and J{\o}rgensen, J. K. and Bontemps, S. and Wei{\ss}, A. and Menten, K. M. and Muders, D.},
  year = 2011,
  month = mar,
  journal = {A\&A},
  volume = {527},
  pages = {A145},
  publisher = {EDP},
  issn = {0004-6361},
  doi = {10.1051/0004-6361/201015733},
  urldate = {2026-06-16}
}

@article{bergner2024jwsticeband,
  title = {{{JWST Ice Band Profiles Reveal Mixed Ice Compositions}} in the {{HH}} 48 {{NE Disk}}},
  author = {Bergner, Jennifer B. and Sturm, J. A. and Piacentino, Elettra L. and McClure, M. K. and {\"O}berg, Karin I. and Boogert, A. C. A. and Dartois, E. and Drozdovskaya, M. N. and Fraser, H. J. and Harsono, Daniel and Ioppolo, Sergio and Law, Charles J. and Lis, Dariusz C. and McGuire, Brett A. and Melnick, Gary J. and Noble, Jennifer A. and Palumbo, M. E. and Pendleton, Yvonne J. and Perotti, Giulia and Qasim, Danna and Rocha, W. R. M. and {van Dishoeck}, E. F.},
  year = 2024,
  month = oct,
  journal = {ApJ},
  volume = {975},
  number = {2},
  pages = {166},
  publisher = {The American Astronomical Society},
  issn = {0004-637X},
  doi = {10.3847/1538-4357/ad79fc},
  urldate = {2025-03-14},
  langid = {english}
}

@article{bergner2026jwstedgeondisk,
  title = {{{JWST Edge-on Disk Ice}} ({{JEDIce}}): {{Program Overview}} and {{Ice Survey Results}}},
  shorttitle = {{{JWST Edge-on Disk Ice}} ({{JEDIce}})},
  author = {Bergner, Jennifer B. and Arulanantham, Nicole and Dartois, Emmanuel and Drozdovskaya, Maria N. and Harsono, Daniel and McClure, Melissa and Noble, Jennifer A. and {\"O}berg, Karin I. and Pontoppidan, Klaus M. and Yang, Yao-Lun and Assani, Korash and Li, Zhi-Yun and Santos, Julia C. and Thompson, Will E. and Welzel, Lukas and Yunerman, Elizabeth S. and Arabhavi, Aditya M. and Booth, Alice S. and Mentzer, Charles and Narang, Mayank and Henning, Thomas and Kamp, Inga and Perotti, Giulia and Somigliana, Alice},
  year = 2026,
  month = apr,
  journal = {ApJ},
  volume = {1002},
  number = {1},
  pages = {17},
  publisher = {The American Astronomical Society},
  issn = {0004-637X},
  doi = {10.3847/1538-4357/ae5961},
  urldate = {2026-06-09},
  langid = {english}
}

@article{bischof2025multiobjectivelossbalancing,
  title = {Multi-{{Objective Loss Balancing}} for {{Physics-Informed Deep Learning}}},
  author = {Bischof, Rafael and Kraus, Michael},
  year = 2025,
  month = may,
  journal = {Comput. Meth. Appl. Mech. Eng.},
  volume = {439},
  eprint = {2110.09813},
  primaryclass = {cs},
  pages = {117914},
  issn = {00457825},
  doi = {10.1016/j.cma.2025.117914},
  urldate = {2026-02-25},
  archiveprefix = {arXiv}
}

@article{bosschaart2026gapsringsnearuniversal,
  title = {Gaps and Rings: A near-Universal Trait of Extended Protoplanetary Discs},
  shorttitle = {Gaps and Rings},
  author = {Bosschaart, Q. and {Guerra-Alvarado}, O. M. and van der Marel, N. and Mulders, G. D.},
  year = 2026,
  month = apr,
  journal = {A\&A},
  volume = {708},
  pages = {A143},
  publisher = {EDP Sciences},
  issn = {0004-6361, 1432-0746},
  doi = {10.1051/0004-6361/202556059},
  urldate = {2026-07-17},
  copyright = {\copyright{} The Authors 2026},
  langid = {english}
}

@inproceedings{branca2025raytraxfasttimedependent,
  title = {Ray-Trax: {{Fast}}, {{Time-Dependent}}, and {{Differentiable Ray Tracing}} for {{On-the-fly Radiative Transfer}} in {{Turbulent Astrophysical Flows}}},
  booktitle = {1st {{Workshop}} on {{Differentiable Systems}} and {{Scientific Machine Learning}}, {{EurIPS}} 2025 [\eprint[arXiv]{2511.09389}]},
  author = {Branca, Lorenzo and Rost, Rune and Buck, Tobias},
  year = 2025
}

@article{buchner2019collaborativenestedsampling,
  title = {Collaborative Nested Sampling: {{Big}} Data versus Complex Physical Models},
  shorttitle = {Collaborative {{Nested Sampling}}},
  author = {Buchner, Johannes},
  year = 2019,
  month = oct,
  journal = {PASP},
  volume = {131},
  number = {1004},
  eprint = {1707.04476},
  primaryclass = {stat.CO},
  pages = {108005},
  issn = {0004-6280, 1538-3873},
  doi = {10.1088/1538-3873/aae7fc},
  urldate = {2024-05-12},
  archiveprefix = {arXiv}
}

@article{buchner2021ultranestrobustgeneral,
  title = {{{UltraNest}} - a Robust, General Purpose {{Bayesian}} Inference Engine},
  author = {Buchner, Johannes},
  year = 2021,
  month = apr,
  journal = {Journal of Open Source Software},
  volume = {6},
  number = {3001},
  eprint = {2101.09604},
  primaryclass = {stat.CO},
  pages = {3001},
  issn = {2475-9066},
  doi = {10.21105/joss.03001},
  urldate = {2026-10-08},
  archiveprefix = {arXiv},
  langid = {english}
}

@article{buchner2023nestedsamplingmethods,
  title = {Nested Sampling Methods},
  author = {Buchner, Johannes},
  year = 2023,
  month = jan,
  journal = {Statist. Surv.},
  volume = {17},
  number = {none},
  eprint = {2101.09675},
  primaryclass = {stat},
  pages = {169--215},
  publisher = {{Amer. Statist. Assoc., the Bernoulli Soc., the Inst. Math. Statist., and the Statist. Soc. Canada}},
  issn = {1935-7516},
  doi = {10.1214/23-SS144},
  urldate = {2026-06-15},
  archiveprefix = {arXiv},
  langid = {english}
}

@article{burrows1996hubblespacetelescope,
  title = {Hubble {{Space Telescope Observations}} of the {{Disk}} and {{Jet}} of {{HH}} 30},
  author = {Burrows, Christopher J. and Stapelfeldt, Karl R. and Watson, Alan M. and Krist, John E. and Ballester, Gilda E. and Clarke, John T. and Crisp, David and Gallagher III, John S. and Griffiths, Richard E. and Jeff Hester, J. and Hoessel, John G. and Holtzman, Jon A. and Mould, Jeremy R. and Scowen, Paul A. and Trauger, John T. and Westphal, James A.},
  year = 1996,
  month = dec,
  journal = {ApJ},
  volume = {473},
  number = {1},
  pages = {437},
  issn = {0004-637X},
  doi = {10.1086/178156},
  urldate = {2026-06-08},
  langid = {english}
}

@article{chen1995universalapproximationnonlinear,
  title = {Universal Approximation to Nonlinear Operators by Neural Networks with Arbitrary Activation Functions and Its Application to Dynamical Systems},
  author = {Chen, Tianping and Chen, Hong},
  year = 1995,
  month = jul,
  journal = {IEEE Trans. Neural Netw.},
  volume = {6},
  number = {4},
  pages = {911--917},
  issn = {1941-0093},
  doi = {10.1109/72.392253},
  urldate = {2026-07-09}
}

@article{chiang1997spectralenergydistributions,
  title = {Spectral {{Energy Distributions}} of {{T Tauri Stars}} with {{Passive Circumstellar Disks}}},
  author = {Chiang, E. I. and Goldreich, P.},
  year = 1997,
  month = nov,
  journal = {ApJ},
  volume = {490},
  number = {1},
  pages = {368},
  publisher = {IOP Publishing},
  issn = {0004-637X},
  doi = {10.1086/304869},
  urldate = {2026-06-11},
  langid = {english}
}

@article{collings2004laboratorysurveythermal,
  title = {A Laboratory Survey of the Thermal Desorption of Astrophysically Relevant Molecules},
  author = {Collings, Mark P. and Anderson, Mark A. and Chen, Rui and Dever, John W. and Viti, Serena and Williams, David A. and McCoustra, Martin R. S.},
  year = 2004,
  month = nov,
  journal = {MNRAS},
  volume = {354},
  number = {4},
  pages = {1133--1140},
  issn = {0035-8711},
  doi = {10.1111/j.1365-2966.2004.08272.x},
  urldate = {2026-07-09}
}

@article{cozad2014learningsurrogatemodels,
  title = {Learning Surrogate Models for Simulation-Based Optimization},
  author = {Cozad, Alison and Sahinidis, Nikolaos V. and Miller, David C.},
  year = 2014,
  journal = {AIChE J.},
  volume = {60},
  number = {6},
  pages = {2211--2227},
  issn = {1547-5905},
  doi = {10.1002/aic.14418},
  urldate = {2025-09-02},
  copyright = {\copyright{} 2014 American Institute of Chemical Engineers},
  langid = {english}
}

@article{cranmer2020frontiersimulationbasedinference,
  title = {The Frontier of Simulation-Based Inference},
  author = {Cranmer, Kyle and Brehmer, Johann and Louppe, Gilles},
  year = 2020,
  month = dec,
  journal = {Proc. Natl. Acad. Sci.},
  volume = {117},
  number = {48},
  pages = {30055--30062},
  publisher = {Proceedings of the National Academy of Sciences},
  doi = {10.1073/pnas.1912789117},
  urldate = {2025-09-02}
}

@techreport{cutri2012explanatorysupplementwise,
  title = {Explanatory {{Supplement}} to the {{WISE All-Sky Data Release Products}}},
  author = {Cutri, R. M. and Wright, E. L. and Conrow, T. and Bauer, J. and Benford, D. and Brandenburg, H. and Dailey, J. and Eisenhardt, P. R. M. and Evans, T. and {Fajardo-Acosta}, S. and Fowler, J. and Gelino, C. and Grillmair, C. and Harbut, M. and Hoffman, D. and Jarrett, T. and Kirkpatrick, J. D. and Leisawitz, D. and Liu, W. and Mainzer, A. and Marsh, K. and Masci, F. and McCallon, H. and Padgett, D. and Ressler, M. E. and Royer, D. and Skrutskie, M. F. and Stanford, S. A. and Wyatt, P. L. and Tholen, D. and Tsai, C. W. and Wachter, S. and Wheelock, S. L. and Yan, L. and Alles, R. and Beck, R. and Grav, T. and Masiero, J. and McCollum, B. and McGehee, P. and Papin, M. and Wittman, M.},
  year = 2012,
  month = mar,
  journal = {Explanatory Supplement to the WISE All-Sky Data Release Products},
  pages = {1},
  urldate = {2026-06-16}
}

@article{dartois2022influencegraingrowth,
  title = {Influence of Grain Growth on {{CO2}} Ice Spectroscopic Profiles. {{Modelling}} for Dense Cores and Disks},
  author = {Dartois, E. and Noble, J. A. and Ysard, N. and Demyk, K. and Chabot, M.},
  year = 2022,
  month = oct,
  journal = {A\&A},
  volume = {666},
  pages = {A153},
  publisher = {EDP},
  issn = {0004-6361},
  doi = {10.1051/0004-6361/202243929},
  urldate = {2025-10-09}
}

@article{dartois2024spectroscopicsizinginterstellar,
  title = {Spectroscopic Sizing of Interstellar Icy Grains with {{JWST}}},
  author = {Dartois, E. and Noble, J. A. and Caselli, P. and Fraser, H. J. and {Jim{\'e}nez-Serra}, I. and Mat{\'e}, B. and McClure, M. K. and Melnick, G. J. and Pendleton, Y. J. and Shimonishi, T. and Smith, Z. L. and Sturm, J. A. and Taillard, A. and Wakelam, V. and Boogert, A. C. A. and Drozdovskaya, M. N. and Erkal, J. and Harsono, D. and Herrero, V. J. and Ioppolo, S. and Linnartz, H. and McGuire, B. A. and Perotti, G. and Qasim, D. and Rocha, W. R. M.},
  year = 2024,
  month = mar,
  journal = {Nat. Astron.},
  volume = {8},
  number = {3},
  pages = {359--367},
  publisher = {Nature Publishing Group},
  issn = {2397-3366},
  doi = {10.1038/s41550-023-02155-x},
  urldate = {2025-03-31},
  copyright = {2024 The Author(s), under exclusive licence to Springer Nature Limited},
  langid = {english}
}

@article{dartois2025edgeondisctau,
  title = {The Edge-on Disc {{Tau}} 042021: {{Icy}} Grains at High Altitudes and a Wind Containing Astronomical Polycyclic Aromatic Hydrocarbons},
  shorttitle = {The Edge-on Disc {{Tau}} 042021},
  author = {Dartois, E. and Noble, J. A. and McClure, M. K. and Sturm, J. A. and Beck, T. L. and Arulanantham, N. and Drozdovskaya, M. N. and Espaillat, C. C. and Harsono, D. and Palumbo, M.-E. and Pendleton, Y. J. and Pontoppidan, K. M.},
  year = 2025,
  month = jun,
  journal = {A\&A},
  volume = {698},
  pages = {A8},
  publisher = {EDP Sciences},
  issn = {0004-6361, 1432-0746},
  doi = {10.1051/0004-6361/202452966},
  urldate = {2025-09-01},
  copyright = {\copyright{} The Authors 2025},
  langid = {english}
}

@article{dartois2026nearinfraredsilhouetteflying,
  title = {The Near-Infrared Silhouette of the {{Flying Saucer}} Edge-on Disc Revealed by the {{JWST JEDIce}} Program},
  author = {Dartois, Emmanuel and Noble, Jennifer A. and Bergner, Jennifer B. and Pontoppidan, Klaus M. and Assani, Korash and Harsono, Daniel and McClure, Melissa K. and Santos, Julia C. and Thompson, Will E. and Welzel, Lukas and Arulanantham, Nicole and Booth, Alice S. and Drozdovskaya, Maria N. and Li, Zhi-Yun and Ma, Jie and Martinien, Laurine and M{\'e}nard, Fran{\c c}ois and Oberg, Karin and Stapelfeldt, Karl and Yang, Yao-Lun},
  year = 2026,
  month = jan,
  journal = {A\&A},
  volume = {705},
  pages = {L14},
  publisher = {EDP Sciences},
  issn = {0004-6361, 1432-0746},
  doi = {10.1051/0004-6361/202557177},
  urldate = {2026-06-09},
  copyright = {\copyright{} The Authors 2026},
  langid = {english}
}

@article{dittmann2024notespracticalapplication,
  title = {Notes on the {{Practical Application}} of {{Nested Sampling}}: {{MultiNest}}, ({{Non}})Convergence, and {{Rectification}}},
  shorttitle = {Notes on the {{Practical Application}} of {{Nested Sampling}}},
  author = {Dittmann, Alexander},
  year = 2024,
  month = sep,
  journal = {OJAp},
  volume = {7},
  pages = {79},
  issn = {2565-6120},
  doi = {10.33232/001c.123872},
  urldate = {2026-06-15}
}

@article{dominik2021optoolcommandlinedriven,
  title = {{{OpTool}}: {{Command-line}} Driven Tool for Creating Complex Dust Opacities},
  shorttitle = {{{OpTool}}},
  author = {Dominik, Carsten and Min, Michiel and Tazaki, Ryo},
  year = 2021,
  month = apr,
  journal = {ASCL},
  pages = {ascl:2104.010},
  urldate = {2025-09-02}
}

@article{dorschner1995stepsinterstellarsilicate,
  title = {Steps toward Interstellar Silicate Mineralogy. {{II}}. {{Study}} of {{Mg-Fe-silicate}} Glasses of Variable Composition.},
  author = {Dorschner, J. and Begemann, B. and Henning, T. and Jaeger, C. and Mutschke, H.},
  year = 1995,
  month = aug,
  journal = {A\&A},
  volume = {300},
  pages = {503},
  publisher = {EDP},
  issn = {0004-6361},
  urldate = {2026-06-16}
}

@article{duchene2024jwstimagingedgeon,
  title = {{{JWST Imaging}} of {{Edge-on Protoplanetary Disks}}. {{I}}. {{Fully Vertically Mixed}} 10 {$\mu$}m {{Grains}} in the {{Outer Regions}} of a 1000 Au {{Disk}}},
  author = {Duch{\^e}ne, Gaspard and M{\'e}nard, Fran{\c c}ois and Stapelfeldt, Karl R. and Villenave, Marion and Wolff, Schuyler G. and Perrin, Marshall D. and Pinte, Christophe and Tazaki, Ryo and Padgett, Deborah L.},
  year = 2024,
  month = jan,
  journal = {AJ},
  volume = {167},
  number = {2},
  pages = {77},
  publisher = {The American Astronomical Society},
  issn = {1538-3881},
  doi = {10.3847/1538-3881/acf9a7},
  urldate = {2026-06-09},
  langid = {english}
}

@article{dullemond2004effectdustsettling,
  title = {The Effect of Dust Settling on the Appearance of Protoplanetary Disks},
  author = {Dullemond, C. P. and Dominik, C.},
  year = 2004,
  month = jul,
  journal = {A\&A},
  volume = {421},
  number = {3},
  pages = {1075--1086},
  publisher = {EDP Sciences},
  issn = {0004-6361, 1432-0746},
  doi = {10.1051/0004-6361:20040284},
  urldate = {2026-07-09},
  copyright = {\copyright{} ESO, 2004},
  langid = {english}
}

@article{dullemond2010innerregionsprotoplanetary,
  title = {The {{Inner Regions}} of {{Protoplanetary Disks}}},
  author = {Dullemond, C. P. and Monnier, J. D.},
  year = 2010,
  month = sep,
  journal = {ARA\&A},
  volume = {48},
  pages = {205--239},
  issn = {0066-4146},
  doi = {10.1146/annurev-astro-081309-130932},
  urldate = {2026-06-15}
}

@article{dullemond2012radmc3dmultipurposeradiative,
  title = {{{RADMC-3D}}: {{A}} Multi-Purpose Radiative Transfer Tool},
  shorttitle = {{{RADMC-3D}}},
  author = {Dullemond, C. P. and Juhasz, A. and Pohl, A. and Sereshti, F. and Shetty, R. and Peters, T. and Commercon, B. and Flock, M.},
  year = 2012,
  month = feb,
  journal = {ASCL},
  pages = {ascl:1202.015},
  urldate = {2025-09-02}
}

@article{dunham2016almasearchsubstructure,
  title = {An {{ALMA Search}} for {{Substructure}}, {{Fragmentation}}, and {{Hidden Protostars}} in {{Starless Cores}} in {{Chamaeleon I}}},
  author = {Dunham, Michael M. and Offner, Stella S. R. and Pineda, Jaime E. and Bourke, Tyler L. and Tobin, John J. and Arce, H{\'e}ctor G. and Chen, Xuepeng and Di Francesco, James and Johnstone, Doug and Lee, Katherine I. and Myers, Philip C. and Price, Daniel and Sadavoy, Sarah I. and Schnee, Scott},
  year = 2016,
  month = jun,
  journal = {ApJ},
  volume = {823},
  pages = {160},
  publisher = {IOP},
  issn = {0004-637X},
  doi = {10.3847/0004-637X/823/2/160},
  urldate = {2026-06-16}
}

@article{duruisseaux2026fourierneuraloperators,
  title = {Fourier {{Neural Operators Explained}}: {{A Practical Perspective}}},
  shorttitle = {Fourier {{Neural Operators Explained}}},
  author = {Duruisseaux, Valentin and Kossaifi, Jean and Anandkumar, Anima},
  year = 2026,
  month = jan,
  number = {arXiv:2512.01421},
  eprint = {2512.01421},
  primaryclass = {cs},
  publisher = {arXiv},
  doi = {10.48550/arXiv.2512.01421},
  urldate = {2026-02-09},
  archiveprefix = {arXiv}
}

@misc{falcon2025pytorchlightning,
  title = {{{PyTorch Lightning}}},
  author = {Falcon, William and {The PyTorch Lightning team}},
  year = 2025,
  month = sep,
  doi = {10.5281/zenodo.17064382},
  urldate = {2026-06-16},
  howpublished = {Zenodo}
}

@article{fayolle2011coicephotodesorption,
  title = {{{CO Ice Photodesorption}}: {{A Wavelength-dependent Study}}},
  shorttitle = {{{CO Ice Photodesorption}}},
  author = {Fayolle, Edith C. and Bertin, Mathieu and Romanzin, Claire and Michaut, Xavier and {\"O}berg, Karin I. and Linnartz, Harold and Fillion, Jean-Hugues},
  year = 2011,
  month = oct,
  journal = {ApJ},
  volume = {739},
  pages = {L36},
  publisher = {IOP},
  issn = {0004-637X},
  doi = {10.1088/2041-8205/739/2/L36},
  urldate = {2026-06-15}
}

@article{fleck1984randomwalkprocedure,
  title = {A Random Walk Procedure for Improving the Computational Efficiency of the Implicit {{Monte Carlo}} Method for Nonlinear Radiation Transport},
  author = {Fleck, J. A and Canfield, E. H},
  year = 1984,
  month = jun,
  journal = {J. Comput. Phys.},
  volume = {54},
  number = {3},
  pages = {508--523},
  issn = {0021-9991},
  doi = {10.1016/0021-9991(84)90130-X},
  urldate = {2026-06-16}
}

@inproceedings{gal2016dropoutbayesianapproximation,
  title = {Dropout as a {{Bayesian Approximation}}: {{Representing Model Uncertainty}} in {{Deep Learning}}},
  shorttitle = {Dropout as a {{Bayesian Approximation}}},
  booktitle = {Proc. 33rd {{Int}}. {{Conf}}. {{Mach}}. {{Learn}}.},
  author = {Gal, Yarin and Ghahramani, Zoubin},
  year = 2016,
  month = jun,
  pages = {1050--1059},
  issn = {1938-7228},
  urldate = {2026-06-15},
  langid = {english}
}

@article{galli2021chamaeleondancerevisiting,
  title = {Chamaeleon {{DANCe}}. {{Revisiting}} the Stellar Populations of {{Chamaeleon I}} and {{Chamaeleon II}} with {{Gaia-DR2}} Data},
  author = {Galli, P. A. B. and Bouy, H. and Olivares, J. and {Miret-Roig}, N. and Sarro, L. M. and Barrado, D. and Berihuete, A. and Bertin, E. and Cuillandre, J.-C.},
  year = 2021,
  month = feb,
  journal = {A\&A},
  volume = {646},
  pages = {A46},
  publisher = {EDP},
  issn = {0004-6361},
  doi = {10.1051/0004-6361/202039395},
  urldate = {2026-06-15}
}

@article{ganeshram2022fcpinohighprecision,
  title = {{{FC-PINO}}: {{High Precision Physics-Informed Neural Operators}} via {{Fourier Continuation}}},
  shorttitle = {{{FC-PINO}}},
  author = {Ganeshram, Adarsh and Maust, Haydn and Duruisseaux, Valentin and Li, Zongyi and Wang, Yixuan and Leibovici, Daniel and Bruno, Oscar and Hou, Thomas and Anandkumar, Anima},
  year = 2022,
  month = nov,
  number = {arXiv:2211.15960},
  eprint = {2211.15960},
  publisher = {arXiv},
  doi = {10.48550/arXiv.2211.15960},
  urldate = {2026-06-15},
  archiveprefix = {arXiv}
}

@article{george2024incrementalspatialspectral,
  title = {Incremental Spatial and Spectral Learning of Neural Operators for Solving Large-Scale {{PDEs}}},
  author = {George, Robert Joseph and Zhao, Jiawei and Kossaifi, Jean and Li, Zongyi and Anandkumar, Anima},
  year = 2024,
  journal = {Transactions on Machine Learning Research},
  issn = {2835-8856}
}

@article{gordon2017trust3dexternally,
  title = {{{TRUST}} - {{I}}. {{A 3D}} Externally Illuminated Slab Benchmark for Dust Radiative Transfer},
  author = {Gordon, K. D. and Baes, M. and Bianchi, S. and Camps, P. and Juvela, M. and Kuiper, R. and Lunttila, T. and Misselt, K. A. and Natale, G. and Robitaille, T. and Steinacker, J.},
  year = 2017,
  month = jul,
  journal = {A\&A},
  volume = {603},
  pages = {A114},
  publisher = {EDP Sciences},
  issn = {0004-6361, 1432-0746},
  doi = {10.1051/0004-6361/201629976},
  urldate = {2026-07-09},
  copyright = {\copyright{} ESO, 2017},
  langid = {english}
}

@article{henning2024mindsjwstmiri,
  title = {{{MINDS}}: {{The JWST MIRI Mid-INfrared Disk Survey}}},
  shorttitle = {{{MINDS}}},
  author = {Henning, Thomas and Kamp, Inga and Samland, Matthias and Arabhavi, Aditya M. and Kanwar, Jayatee and {van Dishoeck}, Ewine F. and G{\"u}del, Manuel and Lagage, Pierre-Olivier and Waelkens, Christoffel and Abergel, Alain and Absil, Olivier and Barrado, David and Boccaletti, Anthony and Bouwman, Jeroen and {Caratti o Garatti}, Alessio and Geers, Vincent and Glauser, Adrian M. and Lahuis, Fred and Mueller, Michael and Nehm{\'e}, Cyrine and Olofsson, G{\"o}ran and Pantin, Eric and Ray, Tom P. and Scheithauer, Silvia and Vandenbussche, Bart and Waters, L. B. F. M. and Wright, Gillian and Argyriou, Ioannis and Christiaens, Valentin and Franceschi, Riccardo and Gasman, Danny and Grant, Sierra L. and Guadarrama, Rodrigo and Jang, Hyerin and {Morales-Calder{\'o}n}, Maria and Pawellek, Nicole and Perotti, Giulia and {Rodgers-Lee}, Donna and Schreiber, J{\"u}rgen and Schwarz, Kamber and Tabone, Beno{\^i}t and Temmink, Milou and Vlasblom, Marissa and Colina, Luis and Greve, Thomas R. and {\"O}stlin, G{\"o}ran},
  year = 2024,
  month = may,
  journal = {PASP},
  volume = {136},
  number = {5},
  pages = {054302},
  publisher = {The Astronomical Society of the Pacific},
  issn = {1538-3873},
  doi = {10.1088/1538-3873/ad3455},
  urldate = {2025-03-14},
  langid = {english}
}

@article{honda2009detectionwaterice,
  title = {Detection of {{Water Ice Grains}} on the {{Surface}} of the {{Circumstellar Disk Around HD}} 142527},
  author = {Honda, M. and Inoue, A. K. and Fukagawa, M. and Oka, A. and Nakamoto, T. and Ishii, M. and Terada, H. and Takato, N. and Kawakita, H. and Okamoto, Y. K. and Shibai, H. and Tamura, M. and Kudo, T. and Itoh, Y.},
  year = 2009,
  month = jan,
  journal = {ApJ},
  volume = {690},
  pages = {L110-L113},
  publisher = {IOP},
  issn = {0004-637X},
  doi = {10.1088/0004-637X/690/2/L110},
  urldate = {2025-06-25}
}

@article{honda2016watericesurface,
  title = {Water {{Ice}} at the {{Surface}} of the {{HD}} 100546 {{Disk}}},
  author = {Honda, M. and Kudo, T. and Takatsuki, S. and Inoue, A. K. and Nakamoto, T. and Fukagawa, M. and Tamura, M. and Terada, H. and Takato, N.},
  year = 2016,
  month = apr,
  journal = {ApJ},
  volume = {821},
  pages = {2},
  publisher = {IOP},
  issn = {0004-637X},
  doi = {10.3847/0004-637X/821/1/2},
  urldate = {2025-06-25}
}

@article{hornik1989multilayerfeedforwardnetworks,
  title = {Multilayer Feedforward Networks Are Universal Approximators},
  author = {Hornik, Kurt and Stinchcombe, Maxwell and White, Halbert},
  year = 1989,
  month = jan,
  journal = {Neural Networks},
  volume = {2},
  number = {5},
  pages = {359--366},
  issn = {0893-6080},
  doi = {10.1016/0893-6080(89)90020-8},
  urldate = {2026-07-08}
}

@article{hughes2007innerholedisk,
  title = {An {{Inner Hole}} in the {{Disk}} around {{TW Hydrae Resolved}} in 7 Mm {{Dust Emission}}},
  author = {Hughes, A. M. and Wilner, D. J. and Calvet, N. and D'Alessio, P. and Claussen, M. J. and Hogerheijde, M. R.},
  year = 2007,
  month = jul,
  journal = {ApJ},
  volume = {664},
  number = {1},
  pages = {536},
  publisher = {IOP Publishing},
  issn = {0004-637X},
  doi = {10.1086/518885},
  urldate = {2026-06-14},
  langid = {english}
}

@article{jin2025neuralnetworkemulator,
  title = {Neural Network Emulator to Constrain the High-z {{IGM}} Thermal State from {{Lyman-$\alpha$}} Forest Flux Autocorrelation Function},
  author = {Jin, Zhenyu and Wolfson, Molly and Hennawi, Joseph F. and {Gonz{\'a}lez-Hern{\'a}ndez}, Diego},
  year = 2025,
  month = jan,
  journal = {MNRAS},
  volume = {536},
  pages = {2277--2293},
  publisher = {OUP},
  issn = {0035-8711},
  doi = {10.1093/mnras/stae2741},
  urldate = {2026-06-15}
}

@article{kaeufer2023analysingsedsprotoplanetary,
  title = {Analysing the {{SEDs}} of Protoplanetary Disks with Machine Learning},
  author = {Kaeufer, T. and Woitke, P. and Min, M. and Kamp, I. and Pinte, C.},
  year = 2023,
  month = apr,
  journal = {A\&A},
  volume = {672},
  pages = {A30},
  publisher = {EDP},
  issn = {0004-6361},
  doi = {10.1051/0004-6361/202245461},
  urldate = {2026-06-15},
  copyright = {\copyright{} The Authors 2023},
  langid = {english}
}

@article{kama2009innerrimstructures,
  title = {The Inner Rim Structures of Protoplanetary Discs},
  author = {Kama, M. and Min, M. and Dominik, C.},
  year = 2009,
  month = nov,
  journal = {A\&A},
  volume = {506},
  pages = {1199--1213},
  publisher = {EDP},
  issn = {0004-6361},
  doi = {10.1051/0004-6361/200912068},
  urldate = {2026-06-15}
}

@article{kennedy2001bayesiancalibrationcomputer,
  title = {Bayesian Calibration of Computer Models},
  author = {Kennedy, Marc C. and O'Hagan, Anthony},
  year = 2001,
  journal = {J. R. Stat. Soc. Ser. B Stat. Methodol.},
  volume = {63},
  number = {3},
  pages = {425--464},
  issn = {1467-9868},
  doi = {10.1111/1467-9868.00294},
  urldate = {2026-06-15},
  copyright = {2001 Royal Statistical Society},
  langid = {english}
}

@article{kimmig2025asymmetricsignatureswarps,
  title = {Asymmetric Signatures of Warps in Edge-on Disks},
  author = {Kimmig, Carolin N. and Villenave, Marion},
  year = 2025,
  month = jun,
  journal = {A\&A},
  volume = {698},
  pages = {A146},
  publisher = {EDP Sciences},
  issn = {0004-6361, 1432-0746},
  doi = {10.1051/0004-6361/202453313},
  urldate = {2025-11-05},
  copyright = {\copyright{} The Authors 2025},
  langid = {english}
}

@article{kossaifi2026librarylearningneural,
  title = {A {{Library}} for {{Learning Neural Operators}}},
  author = {Kossaifi, Jean and Kovachki, Nikola and Li, Zongyi and Pitt, David and {Liu-Schiaffini}, Miguel and George, Robert J. and Bonev, Boris and Azizzadenesheli, Kamyar and Berner, Julius and Duruisseaux, Valentin and Anandkumar, Anima},
  year = 2026,
  journal = {Journal of Machine Learning Research},
  volume = {27},
  number = {205},
  pages = {1--6},
  issn = {1533-7928},
  urldate = {2026-10-08}
}

@article{kovachki2021universalapproximationerror,
  title = {On Universal Approximation and Error Bounds for {{Fourier}} Neural Operators},
  author = {Kovachki, Nikola and Lanthaler, Samuel and Mishra, Siddhartha},
  year = 2021,
  month = jan,
  journal = {J. Mach. Learn. Res.},
  volume = {22},
  number = {1},
  pages = {13237--13312},
  issn = {1532-4435},
  urldate = {2026-07-06}
}

@article{kovachki2023neuraloperatorlearning,
  title = {Neural {{Operator}}: {{Learning Maps Between Function Spaces With Applications}} to {{PDEs}}},
  shorttitle = {Neural {{Operator}}},
  author = {Kovachki, Nikola and Li, Zongyi and Liu, Burigede and Azizzadenesheli, Kamyar and Bhattacharya, Kaushik and Stuart, Andrew and Anandkumar, Anima},
  year = 2023,
  journal = {J. Mach. Learn. Res.},
  volume = {24},
  number = {89},
  pages = {1--97},
  issn = {1533-7928},
  urldate = {2025-03-24}
}

@article{krieger2020unbiasedmontecarlo,
  title = {Unbiased {{Monte Carlo}} Continuum Radiative Transfer in Optically Thick Regions},
  author = {Krieger, A. and Wolf, S.},
  year = 2020,
  month = mar,
  journal = {A\&A},
  volume = {635},
  pages = {A148},
  publisher = {EDP Sciences},
  issn = {0004-6361, 1432-0746},
  doi = {10.1051/0004-6361/201937355},
  urldate = {2026-06-16},
  copyright = {\copyright{} ESO 2020},
  langid = {english}
}

@article{krieger2024improvingmontecarlo,
  title = {Improving {{Monte Carlo}} Radiative Transfer in the Regime of High Optical Depths: {{The}} Minimum Scattering Order},
  shorttitle = {Improving {{Monte Carlo}} Radiative Transfer in the Regime of High Optical Depths},
  author = {Krieger, A. and Wolf, S.},
  year = 2024,
  month = feb,
  journal = {A\&A},
  volume = {682},
  pages = {A99},
  publisher = {EDP Sciences},
  issn = {0004-6361, 1432-0746},
  doi = {10.1051/0004-6361/202346195},
  urldate = {2026-06-16},
  copyright = {\copyright{} The Authors 2024},
  langid = {english}
}

@article{lanthaler2024discretizationerrorfourier,
  title = {Discretization {{Error}} of {{Fourier Neural Operators}}},
  author = {Lanthaler, Samuel and Stuart, Andrew M. and Trautner, Margaret},
  year = 2024,
  month = may,
  number = {arXiv:2405.02221},
  eprint = {2405.02221},
  publisher = {arXiv},
  doi = {10.48550/arXiv.2405.02221},
  urldate = {2026-07-07},
  archiveprefix = {arXiv}
}

@article{lehmann2025multipleinputfourierneural,
  title = {Multiple-Input {{Fourier Neural Operator}} ({{MIFNO}}) for Source-Dependent {{3D}} Elastodynamics},
  author = {Lehmann, Fanny and Gatti, Filippo and Clouteau, Didier},
  year = 2025,
  month = apr,
  journal = {J. Comput. Phys.},
  volume = {527},
  pages = {113813},
  publisher = {Elsevier},
  issn = {0021-9991},
  doi = {10.1016/j.jcp.2025.113813},
  urldate = {2026-06-17}
}

@inproceedings{lehtinen2018noise2noiselearningimage,
  title = {{{Noise2Noise}}: {{Learning Image Restoration}} without {{Clean Data}}},
  shorttitle = {{{Noise2Noise}}},
  booktitle = {Proc. 35th {{Int}}. {{Conf}}. {{Mach}}. {{Learn}}.},
  author = {Lehtinen, Jaakko and Munkberg, Jacob and Hasselgren, Jon and Laine, Samuli and Karras, Tero and Aittala, Miika and Aila, Timo},
  year = 2018,
  month = jul,
  pages = {2965--2974},
  publisher = {PMLR},
  issn = {2640-3498},
  urldate = {2026-07-06},
  langid = {english}
}

@inproceedings{lemos2023samplingbasedaccuracytesting,
  title = {Sampling-{{Based Accuracy Testing}} of {{Posterior Estimators}} for {{General Inference}}},
  booktitle = {Proc. 40th {{Int}}. {{Conf}}. {{Mach}}. {{Learn}}.},
  author = {Lemos, Pablo and Coogan, Adam and Hezaveh, Yashar and {Perreault-Levasseur}, Laurence},
  year = 2023,
  month = jul,
  pages = {19256--19273},
  publisher = {PMLR},
  issn = {2640-3498},
  urldate = {2026-06-15},
  langid = {english}
}

@article{levis2025revealingfinestructure,
  title = {Revealing {{Fine Structure}} in {{Protoplanetary Disks}} with {{Physics Constrained Neural Fields}}},
  author = {Levis, Aviad and Luong, Nhan and Teague, Richard and Bouman, Katherine L. and {Barraza-Alfaro}, Marcelo and Flaherty, Kevin},
  year = 2025,
  month = sep,
  number = {arXiv:2509.03623},
  eprint = {2509.03623},
  primaryclass = {astro-ph.EP},
  publisher = {arXiv},
  doi = {10.48550/arXiv.2509.03623},
  urldate = {2026-06-16},
  archiveprefix = {arXiv}
}

@inproceedings{li2021fourierneuraloperator,
  title = {Fourier {{Neural Operator}} for {{Parametric Partial Differential Equations}}},
  booktitle = {Proc. 9th {{Int}}. {{Conf}}. {{Learn}}. {{Represent}}.},
  author = {Li, Zongyi and Kovachki, Nikola and Azizzadenesheli, Kamyar and Liu, Burigede and Bhattacharya, Kaushik and Stuart, Andrew and Anandkumar, Anima},
  year = 2021
}

@article{ligterink2024mindtrapnonnegligible,
  title = {Mind the Trap - {{Non-negligible}} Effect of Volatile Trapping in Ice on {{C}}/{{O}} Ratios in Protoplanetary Disks and Exoplanetary Atmospheres},
  author = {Ligterink, N. F. W. and Kipfer, K. A. and Gavino, S.},
  year = 2024,
  month = jul,
  journal = {A\&A},
  volume = {687},
  pages = {A224},
  publisher = {EDP Sciences},
  issn = {0004-6361, 1432-0746},
  doi = {10.1051/0004-6361/202450405},
  urldate = {2026-06-10},
  copyright = {\copyright{} The Authors 2024},
  langid = {english}
}

@article{ligterink2024rapidformationmacromolecules,
  title = {The Rapid Formation of Macromolecules in Irradiated Ice of Protoplanetary Disk Dust Traps},
  author = {Ligterink, Niels F. W. and Pinilla, Paola and {van der Marel}, Nienke and {van Scheltinga}, Jeroen Terwisscha and Booth, Alice S. and Alexander, Conel M. O'D and Riebe, My E. I.},
  year = 2024,
  month = oct,
  journal = {Nat. Astron.},
  volume = {8},
  number = {10},
  pages = {1257--1263},
  publisher = {Nature Publishing Group},
  issn = {2397-3366},
  doi = {10.1038/s41550-024-02334-4},
  urldate = {2025-03-14},
  copyright = {2024 The Author(s), under exclusive licence to Springer Nature Limited},
  langid = {english}
}

@inproceedings{lingsch2024regulargridsfourierbased,
  title = {Beyond {{Regular Grids}}: {{Fourier-Based Neural Operators}} on {{Arbitrary Domains}}},
  shorttitle = {Beyond {{Regular Grids}}},
  booktitle = {Proceedings of the 41st {{International Conference}} on {{Machine Learning}}},
  author = {Lingsch, Levi E. and Michelis, Mike Yan and Bezenac, Emmanuel De and Perera, Sirani M. and Katzschmann, Robert K. and Mishra, Siddhartha},
  year = 2024,
  month = jul,
  pages = {30610--30629},
  publisher = {PMLR},
  issn = {2640-3498},
  urldate = {2026-10-08},
  langid = {english}
}

@article{liu2025architecturesvariantsperformance,
  title = {Architectures, Variants, and Performance of Neural Operators: {{A}} Comparative Review},
  shorttitle = {Architectures, Variants, and Performance of Neural Operators},
  author = {Liu, Shengjun and Yu, Yu and Zhang, Ting and Liu, Hanchao and Liu, Xinru and Meng, Deyu},
  year = 2025,
  month = oct,
  journal = {Neurocomputing},
  volume = {648},
  pages = {130518},
  issn = {0925-2312},
  doi = {10.1016/j.neucom.2025.130518},
  urldate = {2025-11-14}
}

@inproceedings{loshchilov2019decoupledweightdecay,
  title = {Decoupled {{Weight Decay Regularization}}},
  booktitle = {Proc. 7th {{Int}}. {{Conf}}. {{Learn}}. {{Represent}}.},
  author = {Loshchilov, Ilya and Hutter, Frank},
  year = 2019
}

@article{lu2021deeponetlearningnonlinear,
  title = {{{DeepONet}}: {{Learning}} Nonlinear Operators for Identifying Differential Equations Based on the Universal Approximation Theorem of Operators},
  shorttitle = {{{DeepONet}}},
  author = {Lu, Lu and Jin, Pengzhan and Karniadakis, George Em},
  year = 2021,
  month = mar,
  journal = {Nat. Mach. Intell.},
  volume = {3},
  number = {3},
  eprint = {1910.03193},
  primaryclass = {cs},
  pages = {218--229},
  issn = {2522-5839},
  doi = {10.1038/s42256-021-00302-5},
  urldate = {2025-11-12},
  archiveprefix = {arXiv}
}

@article{lu2021deepxdedeeplearning,
  title = {{{DeepXDE}}: A Deep Learning Library for Solving Differential Equations},
  author = {Lu, Lu and Meng, Xuhui and Mao, Zhiping and Karniadakis, George Em},
  year = 2021,
  journal = {SIAM Rev.},
  volume = {63},
  number = {1},
  eprint = {https://doi.org/10.1137/19M1274067},
  pages = {208--228},
  doi = {10.1137/19M1274067}
}

@article{lu2022comprehensivefaircomparison,
  title = {A Comprehensive and Fair Comparison of Two Neural Operators (with Practical Extensions) Based on {{FAIR}} Data},
  author = {Lu, Lu and Meng, Xuhui and Cai, Shengze and Mao, Zhiping and Goswami, Somdatta and Zhang, Zhongqiang and Karniadakis, George Em},
  year = 2022,
  month = apr,
  journal = {Comput. Meth. Appl. Mech. Eng.},
  volume = {393},
  pages = {114778},
  issn = {0045-7825},
  doi = {10.1016/j.cma.2022.114778},
  urldate = {2026-06-11}
}

@article{lu2022multifidelitydeepneural,
  title = {Multifidelity Deep Neural Operators for Efficient Learning of Partial Differential Equations with Application to Fast Inverse Design of Nanoscale Heat Transport},
  author = {Lu, Lu and Pestourie, Rapha{\"e}l and Johnson, Steven G. and Romano, Giuseppe},
  year = 2022,
  month = jun,
  journal = {Phys. Rev. Res.},
  volume = {4},
  number = {2},
  pages = {023210},
  publisher = {American Physical Society},
  doi = {10.1103/PhysRevResearch.4.023210},
  urldate = {2026-06-16}
}

@article{luhman2008diskpopulationchamaeleon,
  title = {The {{Disk Population}} of the {{Chamaeleon I Star-forming Region}}},
  author = {Luhman, K. L. and Allen, L. E. and Allen, P. R. and Gutermuth, R. A. and Hartmann, L. and Mamajek, E. E. and Megeath, S. T. and Myers, P. C. and Fazio, G. G.},
  year = 2008,
  month = mar,
  journal = {ApJ},
  volume = {675},
  pages = {1375--1406},
  publisher = {IOP},
  issn = {0004-637X},
  doi = {10.1086/527347},
  urldate = {2026-06-16}
}

@article{lynden-bell1974evolutionviscousdiscs,
  title = {The Evolution of Viscous Discs and the Origin of the Nebular Variables.},
  author = {{Lynden-Bell}, D. and Pringle, J. E.},
  year = 1974,
  month = sep,
  journal = {MNRAS},
  volume = {168},
  pages = {603--637},
  publisher = {OUP},
  issn = {0035-8711},
  doi = {10.1093/mnras/168.3.603},
  urldate = {2026-06-14}
}

@article{ma2024calibrateduncertaintyquantification,
  title = {Calibrated {{Uncertainty Quantification}} for {{Operator Learning}} via {{Conformal Prediction}}},
  author = {Ma, Ziqi and Pitt, David and Azizzadenesheli, Kamyar and Anandkumar, Anima},
  year = 2024,
  month = feb,
  journal = {Transactions on Machine Learning Research},
  eprint = {2402.01960},
  primaryclass = {cs.LG},
  publisher = {arXiv},
  doi = {10.48550/arXiv.2402.01960},
  urldate = {2026-08-17},
  archiveprefix = {arXiv}
}

@article{manoj2011spitzerinfraredspectrograph,
  title = {{{SPITZER INFRARED SPECTROGRAPH SURVEY OF YOUNG STARS IN THE CHAMAELEON I STAR-FORMING REGION}}},
  author = {Manoj, P. and Kim, K. H. and Furlan, E. and McClure, M. K. and Luhman, K. L. and Watson, Dan M. and Espaillat, C. and Calvet, N. and Najita, J. R. and D'Alessio, P. and Adame, L. and Sargent, B. A. and Forrest, W. J. and Bohac, C. and Green, J. D. and Arnold, L. A.},
  year = 2011,
  month = feb,
  journal = {ApJS},
  volume = {193},
  number = {1},
  pages = {11},
  publisher = {The American Astronomical Society},
  issn = {0067-0049},
  doi = {10.1088/0067-0049/193/1/11},
  urldate = {2026-08-17},
  langid = {english}
}

@article{martinien2025roleabsorptionscattering,
  title = {The Role of Absorption and Scattering in Shaping Ice Bands - {{Spatially}} Resolved Spectroscopy of Protoplanetary Disks},
  author = {Martinien, L. and Duch{\^e}ne, G. and M{\'e}nard, F. and Tazaki, R. and Stapelfeldt, K. R.},
  year = 2025,
  month = apr,
  journal = {A\&A},
  volume = {696},
  pages = {L6},
  publisher = {EDP Sciences},
  issn = {0004-6361, 1432-0746},
  doi = {10.1051/0004-6361/202554360},
  urldate = {2025-11-04},
  copyright = {\copyright{} The Authors 2025},
  langid = {english}
}

@article{martinien2026observationshighlyinclined,
  title = {Observations of Highly Inclined Disks with {{ALMA}} - {{Results}} from {{12CO}} Gas and Continuum Observations},
  author = {Martinien, L. and Duch{\^e}ne, G. and Ribas, {\'A} and Villenave, M. and M{\'e}nard, F. and Stapelfeldt, K. R. and Pinte, C.},
  year = 2026,
  month = jun,
  journal = {A\&A},
  volume = {710},
  pages = {A140},
  publisher = {EDP Sciences},
  issn = {0004-6361, 1432-0746},
  doi = {10.1051/0004-6361/202659107},
  urldate = {2026-07-14},
  copyright = {\copyright{} The Authors 2026},
  langid = {english}
}

@article{mathis1977sizedistributioninterstellar,
  title = {The Size Distribution of Interstellar Grains.},
  author = {Mathis, J. S. and Rumpl, W. and Nordsieck, K. H.},
  year = 1977,
  month = oct,
  journal = {ApJ},
  volume = {217},
  pages = {425--433},
  publisher = {IOP},
  issn = {0004-637X},
  doi = {10.1086/155591},
  urldate = {2026-06-15}
}

@article{mathis1990interstellardustextinction,
  title = {Interstellar Dust and Extinction.},
  author = {Mathis, John S.},
  year = 1990,
  month = jan,
  journal = {ARA\&A},
  volume = {28},
  pages = {37--70},
  issn = {0066-4146},
  doi = {10.1146/annurev.aa.28.090190.000345},
  urldate = {2026-06-16}
}

@inproceedings{mccabe2023multiplephysicspretraining,
  title = {Multiple {{Physics Pretraining}} for {{Spatiotemporal Surrogate Models}}},
  booktitle = {Advances in {{Neural Information Processing Systems}}},
  author = {McCabe, Michael and {R{\'e}galdo-Saint Blancard}, Bruno and Parker, Liam and Ohana, Ruben and Cranmer, Miles and Bietti, Alberto and Eickenberg, Michael and Golkar, Siavash and Krawezik, Geraud and Lanusse, Francois and Pettee, Mariel and Tesileanu, Tiberiu and Cho, Kyunghyun and Ho, Shirley},
  year = 2024,
  volume = {37},
  pages = {119301--119335},
  publisher = {Curran Associates, Inc.},
  doi = {10.52202/079017-3791},
  urldate = {2026-10-08}
}

@article{mcclure2009observational520mm,
  title = {Observational 5-20 {$\mu$}m {{Interstellar Extinction Curves Toward Star-Forming Regions Derived From Spitzer IRS Spectra}}},
  author = {McClure, M.},
  year = 2009,
  month = feb,
  journal = {ApJ},
  volume = {693},
  number = {2},
  pages = {L81},
  publisher = {The American Astronomical Society},
  issn = {0004-637X},
  doi = {10.1088/0004-637X/693/2/L81},
  urldate = {2025-04-02},
  langid = {english}
}

@article{mcclure2023iceagejwst,
  title = {An {{Ice Age JWST}} Inventory of Dense Molecular Cloud Ices},
  author = {McClure, M. K. and Rocha, W. R. M. and Pontoppidan, K. M. and Crouzet, N. and Chu, L. E. U. and Dartois, E. and Lamberts, T. and Noble, J. A. and Pendleton, Y. J. and Perotti, G. and Qasim, D. and Rachid, M. G. and Smith, Z. L. and Sun, Fengwu and Beck, Tracy L. and Boogert, A. C. A. and Brown, W. A. and Caselli, P. and Charnley, S. B. and Cuppen, Herma M. and Dickinson, H. and Drozdovskaya, M. N. and Egami, E. and Erkal, J. and Fraser, H. and Garrod, R. T. and Harsono, D. and Ioppolo, S. and {Jim{\'e}nez-Serra}, I. and Jin, M. and J{\o}rgensen, J. K. and Kristensen, L. E. and Lis, D. C. and McCoustra, M. R. S. and McGuire, Brett A. and Melnick, G. J. and {\"O}berg, Karin I. and Palumbo, M. E. and Shimonishi, T. and Sturm, J. A. and {van Dishoeck}, E. F. and Linnartz, H.},
  year = 2023,
  month = apr,
  journal = {Nat. Astron.},
  volume = {7},
  number = {4},
  pages = {431--443},
  publisher = {Nature Publishing Group},
  issn = {2397-3366},
  doi = {10.1038/s41550-022-01875-w},
  urldate = {2024-09-18},
  copyright = {2023 The Author(s), under exclusive licence to Springer Nature Limited},
  langid = {english}
}

@article{mildenhall2021nerfrepresentingscenes,
  title = {{{NeRF}}: Representing Scenes as Neural Radiance Fields for View Synthesis},
  shorttitle = {{{NeRF}}},
  author = {Mildenhall, Ben and Srinivasan, Pratul P. and Tancik, Matthew and Barron, Jonathan T. and Ramamoorthi, Ravi and Ng, Ren},
  year = 2021,
  month = dec,
  journal = {Commun. ACM},
  volume = {65},
  number = {1},
  pages = {99--106},
  issn = {0001-0782},
  doi = {10.1145/3503250},
  urldate = {2026-06-08}
}

@article{min2005modelingopticalproperties,
  title = {Modeling Optical Properties of Cosmic Dust Grains Using a Distribution of Hollow Spheres},
  author = {Min, M. and Hovenier, J. W. and de Koter, A.},
  year = 2005,
  month = mar,
  journal = {A\&A},
  volume = {432},
  number = {3},
  pages = {909--920},
  publisher = {EDP Sciences},
  issn = {0004-6361, 1432-0746},
  doi = {10.1051/0004-6361:20041920},
  urldate = {2025-11-07},
  copyright = {\copyright{} ESO, 2005},
  langid = {english}
}

@article{min2009radiativetransfervery,
  title = {Radiative Transfer in Very Optically Thick Circumstellar Disks},
  author = {Min, M. and Dullemond, C. P. and Dominik, C. and de Koter, A. and Hovenier, J. W.},
  year = 2009,
  month = apr,
  journal = {A\&A},
  volume = {497},
  number = {1},
  pages = {155--166},
  publisher = {EDP Sciences},
  issn = {0004-6361, 1432-0746},
  doi = {10.1051/0004-6361/200811470},
  urldate = {2026-06-10},
  copyright = {\copyright{} ESO, 2009},
  langid = {english}
}

@article{oberg2007photodesorptioncoice,
  title = {Photodesorption of {{CO Ice}}},
  author = {{\"O}berg, Karin I. and Fuchs, Guido W. and Awad, Zainab and Fraser, Helen J. and Schlemmer, Stephan and van Dishoeck, Ewine F. and Linnartz, Harold},
  year = 2007,
  month = may,
  journal = {ApJ},
  volume = {662},
  number = {1},
  pages = {L23},
  publisher = {IOP Publishing},
  issn = {0004-637X},
  doi = {10.1086/519281},
  urldate = {2026-06-15},
  langid = {english}
}

@article{oberg2009photodesorptionicesco,
  title = {Photodesorption of Ices {{I}}: {{CO}}, {{N2}}, and {{CO2}}},
  shorttitle = {Photodesorption of Ices {{I}}},
  author = {{\"O}berg, K. I. and {van Dishoeck}, E. F. and Linnartz, H.},
  year = 2009,
  month = mar,
  journal = {A\&A},
  volume = {496},
  pages = {281--293},
  publisher = {EDP},
  issn = {0004-6361},
  doi = {10.1051/0004-6361/200810207},
  urldate = {2026-06-15}
}

@article{oberg2009photodesorptionicesii,
  title = {Photodesorption of {{Ices}}. {{II}}. {{H2O}} and {{D2O}}},
  author = {{\"O}berg, Karin I. and Linnartz, Harold and Visser, Ruud and {van Dishoeck}, Ewine F.},
  year = 2009,
  month = mar,
  journal = {ApJ},
  volume = {693},
  pages = {1209--1218},
  publisher = {IOP},
  issn = {0004-637X},
  doi = {10.1088/0004-637X/693/2/1209},
  urldate = {2026-06-15}
}

@article{oberg2011effectssnowlinesplanetary,
  title = {{{THE EFFECTS OF SNOWLINES ON C}}/{{O IN PLANETARY ATMOSPHERES}}},
  author = {{\"O}berg, Karin I. and {Murray-Clay}, Ruth and Bergin, Edwin A.},
  year = 2011,
  month = nov,
  journal = {ApJL},
  volume = {743},
  number = {1},
  pages = {L16},
  publisher = {The American Astronomical Society},
  issn = {2041-8205},
  doi = {10.1088/2041-8205/743/1/L16},
  urldate = {2026-06-08},
  langid = {english}
}

@article{oberg2023protoplanetarydiskchemistry,
  title = {Protoplanetary {{Disk Chemistry}}},
  author = {{\"O}berg, Karin I. and Facchini, Stefano and Anderson, Dana E.},
  year = 2023,
  month = aug,
  journal = {ARA\&A},
  volume = {61},
  number = {Volume 61, 2023},
  pages = {287--328},
  publisher = {Annual Reviews},
  issn = {0066-4146, 1545-4282},
  doi = {10.1146/annurev-astro-022823-040820},
  urldate = {2024-09-02},
  langid = {english}
}

@article{palumbo2006profile2140cm1,
  title = {The Profile of the 2140 Cm-1 Solid {{CO}} Band on Different Substrates},
  author = {Palumbo, M. E. and Baratta, G. A. and Collings, M. P. and McCoustra, M. R. S.},
  year = 2006,
  month = jan,
  journal = {Phys. Chem. Chem. Phys.},
  volume = {8},
  pages = {279--284},
  issn = {1463-9076},
  doi = {10.1039/B509279E},
  urldate = {2026-06-16}
}

@article{pascucci20042dcontinuumradiative,
  title = {The {{2D}} Continuum Radiative Transfer Problem - {{Benchmark}} Results for Disk Configurations},
  author = {Pascucci, I. and Wolf, S. and Steinacker, J. and Dullemond, C. P. and Henning, Th and Niccolini, G. and Woitke, P. and Lopez, B.},
  year = 2004,
  month = apr,
  journal = {A\&A},
  volume = {417},
  number = {3},
  pages = {793--805},
  publisher = {EDP Sciences},
  issn = {0004-6361, 1432-0746},
  doi = {10.1051/0004-6361:20040017},
  urldate = {2026-06-16},
  copyright = {\copyright{} ESO, 2004},
  langid = {english}
}

@inproceedings{paszke2019pytorchimperativestyle,
  title = {{{PyTorch}}: {{An Imperative Style}}, {{High-Performance Deep Learning Library}}},
  shorttitle = {{{PyTorch}}},
  booktitle = {Adv. {{Neural Inf}}. {{Process}}. {{Syst}}.},
  author = {Paszke, Adam and Gross, Sam and Massa, Francisco and Lerer, Adam and Bradbury, James and Chanan, Gregory and Killeen, Trevor and Lin, Zeming and Gimelshein, Natalia and Antiga, Luca and Desmaison, Alban and K{\"o}pf, Andreas and Yang, Edward and DeVito, Zach and Raison, Martin and Tejani, Alykhan and Chilamkurthy, Sasank and Steiner, Benoit and Fang, Lu and Bai, Junjie and Chintala, Soumith},
  year = 2019,
  volume = {32},
  pages = {8026--8037}
}

@article{pinte2009benchmarkproblemscontinuum,
  title = {Benchmark Problems for Continuum Radiative Transfer - {{High}} Optical Depths, Anisotropic Scattering, and Polarisation},
  author = {Pinte, C. and Harries, T. J. and Min, M. and Watson, A. M. and Dullemond, C. P. and Woitke, P. and M{\'e}nard, F. and {Dur{\'a}n-Rojas}, M. C.},
  year = 2009,
  month = may,
  journal = {A\&A},
  volume = {498},
  number = {3},
  pages = {967--980},
  publisher = {EDP Sciences},
  issn = {0004-6361, 1432-0746},
  doi = {10.1051/0004-6361/200811555},
  urldate = {2026-03-09},
  copyright = {\copyright{} ESO, 2009},
  langid = {english}
}

@article{pontoppidan2005icesedgeondisk,
  title = {Ices in the {{Edge-on Disk CRBR}} 2422.8-3423: {{Spitzer Spectroscopy}} and {{Monte Carlo Radiative Transfer Modeling}}},
  shorttitle = {Ices in the {{Edge-on Disk CRBR}} 2422.8-3423},
  author = {Pontoppidan, Klaus M. and Dullemond, Cornelis P. and {van Dishoeck}, Ewine F. and Blake, Geoffrey A. and Boogert, Adwin C. A. and Evans II, Neal J. and {Kessler-Silacci}, Jacqueline E. and Lahuis, Fred},
  year = 2005,
  month = mar,
  journal = {ApJ},
  volume = {622},
  number = {1},
  pages = {463},
  issn = {0004-637X},
  doi = {10.1086/427688},
  urldate = {2025-03-18},
  langid = {english}
}

@article{pontoppidan2007deepspitzerspectroscopy,
  title = {Deep {{Spitzer Spectroscopy}} of the ``{{Flying Saucer}}'' {{Edge-on Disk}}: {{Large Grains}} beyond 50 {{AU}}},
  shorttitle = {Deep {{Spitzer Spectroscopy}} of the ``{{Flying Saucer}}'' {{Edge-on Disk}}},
  author = {Pontoppidan, Klaus M. and Stapelfeldt, Karl R. and Blake, Geoffrey A. and van Dishoeck, Ewine F. and Dullemond, Cornelis P.},
  year = 2007,
  month = feb,
  journal = {ApJ},
  volume = {658},
  number = {2},
  pages = {L111},
  publisher = {IOP Publishing},
  issn = {0004-637X},
  doi = {10.1086/514817},
  urldate = {2026-06-14},
  langid = {english}
}

@article{potapov2025simplemoleculescomplex,
  title = {Simple Molecules and Complex Chemistry in a Protoplanetary Disk - {{A JWST}} Investigation of the Highly Inclined Disk D216-0939},
  shorttitle = {Simple Molecules and Complex Chemistry in a Protoplanetary Disk},
  author = {Potapov, A. and Linz, H. and Bouwman, J. and Rocha, W. R. M. and Martin, J. S. and Wolf, S. and Henning, Th and Terada, H.},
  year = 2025,
  month = may,
  journal = {A\&A},
  volume = {697},
  pages = {A53},
  publisher = {EDP Sciences},
  issn = {0004-6361, 1432-0746},
  doi = {10.1051/0004-6361/202453385},
  urldate = {2026-06-11},
  copyright = {\copyright{} The Authors 2025},
  langid = {english}
}

@article{potapov2026firstdetectionhdo,
  title = {First Detection of {{HDO}} Ice in a Protoplanetary Disk},
  author = {Potapov, Alexey and Kalambkar, Piyush and Bouwman, Jeroen and Boersma, Christiaan and Terada, Hiroshi and Rocha, Will R. and Linz, Hendrik},
  year = 2026,
  month = jul,
  journal = {Astronomy \& Astrophysics},
  volume = {711},
  pages = {A135},
  publisher = {EDP Sciences},
  issn = {0004-6361, 1432-0746},
  doi = {10.1051/0004-6361/202659525},
  urldate = {2026-10-08},
  copyright = {\copyright{} The Authors 2026},
  langid = {english}
}

@article{qin2024betterunderstandingfourier,
  title = {Toward a {{Better Understanding}} of {{Fourier Neural Operators}} from a {{Spectral Perspective}}},
  author = {Qin, Shaoxiang and Lyu, Fuyuan and Peng, Wenhui and Geng, Dingyang and Wang, Ju and Tang, Xing and Leroyer, Sylvie and Gao, Naiping and Liu, Xue and Wang, Liangzhu Leon},
  year = 2024,
  month = apr,
  publisher = {arXiv},
  doi = {10.48550/arXiv.2404.07200},
  urldate = {2026-06-12},
  number = {arXiv:2404.07200},
  eprint = {2404.07200},
  archiveprefix = {arXiv}
}

@inproceedings{rahaman2019spectralbiasneural,
  title = {On the {{Spectral Bias}} of {{Neural Networks}}},
  booktitle = {Proc. 36th {{Int}}. {{Conf}}. {{Mach}}. {{Learn}}.},
  author = {Rahaman, Nasim and Baratin, Aristide and Arpit, Devansh and Draxler, Felix and Lin, Min and Hamprecht, Fred and Bengio, Yoshua and Courville, Aaron},
  year = 2019,
  month = may,
  pages = {5301--5310},
  publisher = {PMLR},
  issn = {2640-3498},
  urldate = {2026-07-06},
  langid = {english}
}

@article{ran2026planetformationconstraints,
  title = {Planet Formation Constraints from Exoplanet Population Synthesis and Simulation Based Inference},
  author = {Ran, Jason and Kama, Mihkel and {Sommerville-Thomas}, Anna},
  year = 2026,
  month = apr,
  journal = {Front. Astron. Space Sci.},
  volume = {13},
  pages = {1736938},
  publisher = {Frontiers},
  issn = {2296-987X},
  doi = {10.3389/fspas.2026.1736938},
  urldate = {2026-06-15},
  langid = {english}
}

@article{ribas2020modelingprotoplanetarydisk,
  title = {Modeling Protoplanetary Disk {{SEDs}} with Artificial Neural Networks - {{Revisiting}} the Viscous Disk Model and Updated Disk Masses},
  author = {Ribas, {\'A} and Espaillat, C. C. and Mac{\'i}as, E. and Sarro, L. M.},
  year = 2020,
  month = oct,
  journal = {A\&A},
  volume = {642},
  pages = {A171},
  publisher = {EDP Sciences},
  issn = {0004-6361, 1432-0746},
  doi = {10.1051/0004-6361/202038352},
  urldate = {2024-09-02},
  copyright = {\copyright{} ESO 2020},
  langid = {english}
}

@article{robberto2012hstimagingsurvey,
  title = {An {{HST Imaging Survey}} of {{Low-mass Stars}} in the {{Chamaeleon I Star-forming Region}}},
  author = {Robberto, M. and Spina, L. and Da Rio, N. and Apai, D. and Pascucci, I. and Ricci, L. and Goddi, C. and Testi, L. and Palla, F. and Bacciotti, F.},
  year = 2012,
  month = sep,
  journal = {AJ},
  volume = {144},
  pages = {83},
  publisher = {IOP},
  issn = {0004-6256},
  doi = {10.1088/0004-6256/144/3/83},
  urldate = {2026-06-16}
}

@article{robitaille2006interpretingspectralenergy,
  title = {Interpreting {{Spectral Energy Distributions}} from {{Young Stellar Objects}}. {{I}}. {{A Grid}} of 200,000 {{YSO Model SEDs}}},
  author = {Robitaille, Thomas P. and Whitney, Barbara A. and Indebetouw, Remy and Wood, Kenneth and Denzmore, Pia},
  year = 2006,
  month = dec,
  journal = {ApJS},
  volume = {167},
  number = {2},
  pages = {256},
  publisher = {IOP Publishing},
  issn = {0067-0049},
  doi = {10.1086/508424},
  urldate = {2026-06-11},
  langid = {english}
}

@article{robitaille2007interpretingspectralenergy,
  title = {Interpreting {{Spectral Energy Distributions}} from {{Young Stellar Objects}}. {{II}}. {{Fitting Observed SEDs Using}} a {{Large Grid}} of {{Precomputed Models}}},
  author = {Robitaille, Thomas P. and Whitney, Barbara A. and Indebetouw, Remy and Wood, Kenneth},
  year = 2007,
  month = apr,
  journal = {ApJS},
  volume = {169},
  number = {2},
  pages = {328},
  publisher = {IOP Publishing},
  issn = {0067-0049},
  doi = {10.1086/512039},
  urldate = {2026-06-14},
  langid = {english}
}

@article{robitaille2010modifiedrandomwalk,
  title = {On the Modified Random Walk Algorithm for {{Monte-Carlo}} Radiation Transfer},
  author = {Robitaille, T. P.},
  year = 2010,
  month = sep,
  journal = {A\&A},
  volume = {520},
  pages = {A70},
  publisher = {EDP Sciences},
  issn = {0004-6361, 1432-0746},
  doi = {10.1051/0004-6361/201015025},
  urldate = {2026-06-16},
  copyright = {\copyright{} ESO, 2010},
  langid = {english}
}

@inproceedings{sakarvadia2025falsepromisezeroshot,
  title = {The {{False Promise}} of {{Zero-Shot Super-Resolution}} in {{Machine-Learned Operators}}},
  booktitle = {International {{Conference}} on {{Learning Representations}}},
  author = {Sakarvadia, Mansi and Hegazy, Kareem and Totounferoush, Amin and Chard, Kyle and Yang, Yaoqing and Foster, Ian and Mahoney, Michael W.},
  year = 2026,
  urldate = {2026-10-08},
  langid = {english},
}

@article{simon2019entrapmentcoco2,
  title = {Entrapment of {{CO}} in {{CO2 Ice}}},
  author = {Simon, Alexia and {\"O}berg, Karin I. and Rajappan, Mahesh and Maksiutenko, Pavlo},
  year = 2019,
  month = sep,
  journal = {ApJ},
  volume = {883},
  pages = {21},
  publisher = {IOP},
  issn = {0004-637X},
  doi = {10.3847/1538-4357/ab32e5},
  urldate = {2026-06-30}
}

@inproceedings{skilling2004nestedsampling,
  title = {Nested {{Sampling}}},
  booktitle = {Bayesian {{Inference}} and {{Maximum Entropy Methods}} in {{Science}} and {{Engineering}}: 24th {{International Workshop}} on {{Bayesian Inference}} and {{Maximum Entropy Methods}} in {{Science}} and {{Engineering}}},
  author = {Skilling, John},
  year = 2004,
  month = nov,
  volume = {735},
  pages = {395--405},
  publisher = {AIP},
  doi = {10.1063/1.1835238},
  urldate = {2026-07-06}
}

@article{skrutskie2006twomicronall,
  title = {The {{Two Micron All Sky Survey}} ({{2MASS}})},
  author = {Skrutskie, M. F. and Cutri, R. M. and Stiening, R. and Weinberg, M. D. and Schneider, S. and Carpenter, J. M. and Beichman, C. and Capps, R. and Chester, T. and Elias, J. and Huchra, J. and Liebert, J. and Lonsdale, C. and Monet, D. G. and Price, S. and Seitzer, P. and Jarrett, T. and Kirkpatrick, J. D. and Gizis, J. E. and Howard, E. and Evans, T. and Fowler, J. and Fullmer, L. and Hurt, R. and Light, R. and Kopan, E. L. and Marsh, K. A. and McCallon, H. L. and Tam, R. and Van Dyk, S. and Wheelock, S.},
  year = 2006,
  month = feb,
  journal = {AJ},
  volume = {131},
  pages = {1163--1183},
  publisher = {IOP},
  issn = {0004-6256},
  doi = {10.1086/498708},
  urldate = {2026-06-16}
}

@article{smith2025cospatialicemapping,
  title = {Cospatial Ice Mapping of {{H2O}} with {{CO2}} and {{CO}} across a Molecular Cloud with {{JWST}}/{{NIRCam}}},
  author = {Smith, Z. L. and Dickinson, H. J. and Fraser, H. J. and McClure, M. K. and Noble, J. A. and Boogert, A. C. A. and Sun, F. and Egami, E. and Dartois, E. and Erkal, J. and Shimonishi, T. and Beck, T. L. and Bergner, J. B. and Caselli, P. and Charnley, S. B. and Chu, L. and Drozdovskaya, M. N. and Garrod, R. and Harsono, D. and Ioppolo, S. and {Jimenez-Serra}, I. and J{\o}rgensen, J. K. and Melnick, G. J. and {\"O}berg, K. I. and Palumbo, M. E. and Pendleton, Y. J. and Perotti, G. and Pontoppidan, K. M. and Qasim, D. and Rocha, W. R. M. and Sturm, J. A. and Taillard, A. and Urso, R. G. and {van Dishoeck}, E. F.},
  year = 2025,
  month = mar,
  journal = {Nat. Astron.},
  volume = {9},
  number = {6},
  pages = {883--894},
  publisher = {Nature Publishing Group},
  issn = {2397-3366},
  doi = {10.1038/s41550-025-02511-z},
  urldate = {2025-03-31},
  copyright = {2025 The Author(s)},
  langid = {english}
}

@article{sobol1967distributionpointscube,
  title = {On the Distribution of Points in a Cube and the Approximate Evaluation of Integrals},
  author = {Sobol', I. M},
  year = 1967,
  month = jan,
  journal = {USSR Computational Mathematics and Mathematical Physics},
  volume = {7},
  number = {4},
  pages = {86--112},
  issn = {0041-5553},
  doi = {10.1016/0041-5553(67)90144-9},
  urldate = {2026-07-09}
}

@article{son2023sobolevtrainingphysicsinformed,
  title = {Sobolev Training for Physics-Informed Neural Networks},
  author = {Son, Hwijae and Jang, Jin Woo and Han, Woo Jin and Hwang, Hyung Ju},
  year = 2023,
  month = sep,
  journal = {Comm. Math. Sci.},
  volume = {21},
  number = {6},
  pages = {1679--1705},
  publisher = {International Press of Boston},
  issn = {1945-0796, 1539-6746},
  doi = {10.4310/CMS.2023.v21.n6.a11},
  urldate = {2026-07-06},
  langid = {english}
}

@inproceedings{stapelfeldt2014hstimagingnew,
  title = {{{HST Imaging}} of {{New Edge-on Circumstellar Disks}} in {{Nearby Star-forming Regions}}},
  booktitle = {Exploring the {{Formation}} and {{Evolution}} of {{Planetary Systems}}},
  author = {Stapelfeldt, K. R. and Duch{\^e}ne, G. and Perrin, M. and Wolff, S. and Krist, J. E. and Padgett, D. L. and M{\'e}nard, F. and Pinte, C.},
  year = 2014,
  month = jan,
  volume = {299},
  pages = {99--103},
  doi = {10.1017/S1743921313008004},
  urldate = {2026-07-20}
}

@article{steinacker2013threedimensionaldustradiative,
  title = {Three-{{Dimensional Dust Radiative Transfer}}*},
  author = {Steinacker, J{\"u}rgen and Baes, Maarten and Gordon, Karl D.},
  year = 2013,
  month = aug,
  journal = {ARA\&A},
  volume = {51},
  pages = {63--104},
  issn = {0066-4146},
  doi = {10.1146/annurev-astro-082812-141042},
  urldate = {2026-06-16}
}

@article{sturm2023edgeonprotoplanetarydisk,
  title = {The Edge-on Protoplanetary Disk {{HH}} 48 {{NE}} - {{I}}. {{Modeling}} the Geometry and Stellar Parameters},
  author = {Sturm, J. A. and McClure, M. K. and Law, C. J. and Harsono, D. and Bergner, J. B. and Dartois, E. and Drozdovskaya, M. N. and Ioppolo, S. and {\"O}berg, K. I. and Palumbo, M. E. and Pendleton, Y. J. and Rocha, W. R. M. and Terada, H. and Urso, R. G.},
  year = 2023,
  month = sep,
  journal = {A\&A},
  volume = {677},
  pages = {A17},
  publisher = {EDP Sciences},
  issn = {0004-6361, 1432-0746},
  doi = {10.1051/0004-6361/202346052},
  urldate = {2024-09-02},
  copyright = {\copyright{} The Authors 2023},
  langid = {english}
}

@article{sturm2023edgeonprotoplanetarydiska,
  title = {The Edge-on Protoplanetary Disk {{HH}} 48 {{NE}} - {{II}}. {{Modeling}} Ices and Silicates},
  author = {Sturm, J. A. and McClure, M. K. and Bergner, J. B. and Harsono, D. and Dartois, E. and Drozdovskaya, M. N. and Ioppolo, S. and {\"O}berg, K. I. and Law, C. J. and Palumbo, M. E. and Pendleton, Y. J. and Rocha, W. R. M. and Terada, H. and Urso, R. G.},
  year = 2023,
  month = sep,
  journal = {A\&A},
  volume = {677},
  pages = {A18},
  publisher = {EDP Sciences},
  issn = {0004-6361, 1432-0746},
  doi = {10.1051/0004-6361/202346053},
  urldate = {2024-09-02},
  copyright = {\copyright{} The Authors 2023},
  langid = {english}
}

@article{sturm2023jwstinventoryprotoplanetary,
  title = {A {{JWST}} Inventory of Protoplanetary Disk Ices - {{The}} Edge-on Protoplanetary Disk {{HH}} 48 {{NE}}, Seen with the {{Ice Age ERS}} Program},
  author = {Sturm, J. A. and McClure, M. K. and Beck, T. L. and Harsono, D. and Bergner, J. B. and Dartois, E. and Boogert, A. C. A. and Chiar, J. E. and Cordiner, M. A. and Drozdovskaya, M. N. and Ioppolo, S. and Law, C. J. and Linnartz, H. and Lis, D. C. and Melnick, G. J. and McGuire, B. A. and Noble, J. A. and {\"O}berg, K. I. and Palumbo, M. E. and Pendleton, Y. J. and Perotti, G. and Pontoppidan, K. M. and Qasim, D. and Rocha, W. R. M. and Terada, H. and Urso, R. G. and van Dishoeck, E. F.},
  year = 2023,
  month = nov,
  journal = {A\&A},
  volume = {679},
  pages = {A138},
  publisher = {EDP Sciences},
  issn = {0004-6361, 1432-0746},
  doi = {10.1051/0004-6361/202347512},
  urldate = {2024-09-02},
  copyright = {\copyright{} The Authors 2023},
  langid = {english}
}

@article{sturm2024jwstmirianalysis,
  title = {A {{JWST}}/{{MIRI}} Analysis of the Ice Distribution and Polycyclic Aromatic Hydrocarbon Emission in the Protoplanetary Disk {{HH}} 48 {{NE}}},
  author = {Sturm, J. A. and McClure, M. K. and Harsono, D. and Bergner, J. B. and Dartois, E. and Boogert, A. C. A. and Cordiner, M. A. and Drozdovskaya, M. N. and Ioppolo, S. and Law, C. J. and Lis, D. C. and McGuire, B. A. and Melnick, G. J. and Noble, J. A. and {\"O}berg, K. I. and Palumbo, M. E. and Pendleton, Y. J. and Perotti, G. and Rocha, W. R. M. and Urso, R. G. and van Dishoeck, E. F.},
  year = 2024,
  month = sep,
  journal = {A\&A},
  volume = {689},
  pages = {A92},
  publisher = {EDP Sciences},
  issn = {0004-6361, 1432-0746},
  doi = {10.1051/0004-6361/202450865},
  urldate = {2026-06-09},
  copyright = {\copyright{} The Authors 2024},
  langid = {english}
}

@inproceedings{subramanian2023foundationmodelsscientific,
  title = {Towards Foundation Models for Scientific Machine Learning: Characterizing Scaling and Transfer Behavior},
  shorttitle = {Towards Foundation Models for Scientific Machine Learning},
  booktitle = {Proc. 37th {{Int}}. {{Conf}}. {{Neural Inf}}. {{Process}}. {{Syst}}.},
  author = {Subramanian, Shashank and Harrington, Peter and Keutzer, Kurt and Bhimji, Wahid and Morozov, Dmitriy and Mahoney, Michael W. and Gholami, Amir},
  year = 2023,
  month = dec,
  pages = {71242--71262},
  urldate = {2026-07-06}
}

@inproceedings{tancik2020fourierfeatureslet,
  title = {Fourier {{Features Let Networks Learn High Frequency Functions}} in {{Low Dimensional Domains}}},
  booktitle = {Proc. 34th {{Int}}. {{Conf}}. {{Neural Inf}}. {{Process}}. {{Syst}}.},
  author = {Tancik, Matthew and Srinivasan, Pratul and Mildenhall, Ben and {Fridovich-Keil}, Sara and Raghavan, Nithin and Singhal, Utkarsh and Ramamoorthi, Ravi and Barron, Jonathan and Ng, Ren},
  year = 2020,
  volume = {33},
  pages = {7537--7547},
  urldate = {2026-06-08}
}

@article{telkamp2022machinelearningframework,
  title = {A {{Machine Learning Framework}} to {{Predict Images}} of {{Edge-on Protoplanetary Disks}}},
  author = {Telkamp, Zoie and {Mart{\'i}nez-Palomera}, Jorge and Duch{\^e}ne, Gaspard and Ashimbekova, Aishabibi and Wolfe, Edward and Angelo, Isabel and Pinte, Christophe},
  year = 2022,
  month = nov,
  journal = {ApJ},
  volume = {939},
  number = {2},
  pages = {73},
  publisher = {The American Astronomical Society},
  issn = {0004-637X},
  doi = {10.3847/1538-4357/ac96f1},
  urldate = {2026-06-09},
  langid = {english}
}

@article{terada2007detectionwaterice,
  title = {Detection of {{Water Ice}} in {{Edge-on Protoplanetary Disks}}: {{HK Tauri B}} and {{HV Tauri C}}},
  shorttitle = {Detection of {{Water Ice}} in {{Edge-on Protoplanetary Disks}}},
  author = {Terada, Hiroshi and Tokunaga, Alan T. and Kobayashi, Naoto and Takato, Naruhisa and Hayano, Yutaka and Takami, Hideki},
  year = 2007,
  month = sep,
  journal = {ApJ},
  volume = {667},
  number = {1},
  pages = {303},
  issn = {0004-637X},
  doi = {10.1086/520951},
  urldate = {2025-03-18},
  langid = {english}
}

@article{thamm1994ambiguitiesparametrizeddust,
  title = {Ambiguities of Parametrized Dust Disk Models for Young Stellar Objects.},
  author = {Thamm, E. and Steinacker, J. and Henning, {\relax Th}.},
  year = 1994,
  month = jul,
  journal = {A\&A},
  volume = {287},
  pages = {493--502},
  publisher = {EDP},
  issn = {0004-6361},
  urldate = {2026-06-14}
}

@article{ulyanov2020deepimageprior,
  title = {Deep {{Image Prior}}},
  author = {Ulyanov, Dmitry and Vedaldi, Andrea and Lempitsky, Victor},
  year = 2020,
  month = jul,
  journal = {Int. J. Comput. Vis.},
  volume = {128},
  number = {7},
  eprint = {1711.10925},
  primaryclass = {cs.CV},
  pages = {1867--1888},
  issn = {0920-5691, 1573-1405},
  doi = {10.1007/s11263-020-01303-4},
  urldate = {2026-06-16},
  archiveprefix = {arXiv}
}

@article{vancapelleveen2025wideseparationplanets,
  title = {{{WIde Separation Planets In Time}} ({{WISPIT}}): {{Two}} Directly Imaged Exoplanets around the {{Sun-like}} Stellar Binary {{WISPIT}} 1},
  shorttitle = {{{WIde Separation Planets In Time}} ({{WISPIT}})},
  author = {{van Capelleveen}, Richelle F. and Kenworthy, Matthew A. and Ginski, Christian and Mamajek, Eric E. and Bohn, Alexander J. and Landman, Rico and Stolker, Tomas and Zhang, Yapeng and {van der Marel}, Nienke and Snellen, Ignas},
  year = 2025,
  month = dec,
  journal = {A\&A},
  volume = {704},
  pages = {A221},
  publisher = {EDP},
  issn = {0004-6361},
  doi = {10.1051/0004-6361/202556584},
  urldate = {2026-07-01}
}

@article{villenave2020observationsedgeonprotoplanetary,
  title = {Observations of Edge-on Protoplanetary Disks with {{ALMA}} - {{I}}. {{Results}} from Continuum Data},
  author = {Villenave, M. and M{\'e}nard, F. and Dent, W. R. F. and Duch{\^e}ne, G. and Stapelfeldt, K. R. and Benisty, M. and Boehler, Y. and van der Plas, G. and Pinte, C. and Telkamp, Z. and Wolff, S. and Flores, C. and Lesur, G. and Louvet, F. and Riols, A. and Dougados, C. and Williams, H. and Padgett, D.},
  year = 2020,
  month = oct,
  journal = {A\&A},
  volume = {642},
  pages = {A164},
  publisher = {EDP Sciences},
  issn = {0004-6361, 1432-0746},
  doi = {10.1051/0004-6361/202038087},
  urldate = {2024-09-02},
  copyright = {\copyright{} M. Villenave et al. 2020},
  langid = {english}
}

@article{villenave2023modestdustsettling,
  title = {Modest {{Dust Settling}} in the {{IRAS04302}}+2247 {{Class I Protoplanetary Disk}}},
  author = {Villenave, M. and Podio, L. and Duch{\^e}ne, G. and Stapelfeldt, K. R. and Melis, C. and {Carrasco-Gonzalez}, C. and Le Gouellec, V. J. M. and M{\'e}nard, F. and {de Simone}, M. and Chandler, C. and Garufi, A. and Pinte, C. and Bianchi, E. and Codella, C.},
  year = 2023,
  month = mar,
  journal = {ApJ},
  volume = {946},
  number = {2},
  pages = {70},
  publisher = {The American Astronomical Society},
  issn = {0004-637X},
  doi = {10.3847/1538-4357/acb92e},
  urldate = {2026-06-09},
  langid = {english}
}

@article{warren1986opticalconstantscarbon,
  title = {Optical Constants of Carbon Dioxide Ice},
  author = {Warren, S. G.},
  year = 1986,
  month = aug,
  journal = {Appl. Opt.},
  volume = {25},
  pages = {2650--2674},
  publisher = {OSA},
  issn = {0003-6935},
  doi = {10.1364/AO.25.002650},
  urldate = {2026-06-16}
}

@article{warren2008opticalconstantsice,
  title = {Optical Constants of Ice from the Ultraviolet to the Microwave: {{A}} Revised Compilation},
  shorttitle = {Optical Constants of Ice from the Ultraviolet to the Microwave},
  author = {Warren, Stephen G. and Brandt, Richard E.},
  year = 2008,
  month = jul,
  journal = {J. Geophys. Res. Atmos.},
  volume = {113},
  pages = {D14220},
  publisher = {Wiley},
  issn = {0148-0227},
  doi = {10.1029/2007JD009744},
  urldate = {2026-06-16}
}

@article{williams2025lockedicehow,
  title = {Locked in Ice: How Pebble Drift and Volatile Entrapment Can Significantly Impact Carbon and Oxygen Ratios in Evolving Protoplanetary Discs},
  shorttitle = {Locked in Ice},
  author = {Williams, Joe and Krijt, Sebastiaan and Bitsch, Bertram and Houge, Adrien and Bergner, Jennifer},
  year = 2025,
  month = dec,
  journal = {MNRAS},
  volume = {544},
  number = {4},
  pages = {3562--3578},
  issn = {0035-8711},
  doi = {10.1093/mnras/staf1839},
  urldate = {2026-06-10}
}

@article{williams2025radiativetransfermodeling,
  title = {Radiative {{Transfer Modeling}} of a {{Shadowed Protoplanetary Disk Assisted}} by a {{Neural Network}}},
  author = {Williams, Jonathan P. and Benisty, Myriam and Ginski, Christian and Lodato, Giuseppe and Vincent, Maria},
  year = 2025,
  month = sep,
  journal = {ApJ},
  volume = {991},
  number = {2},
  pages = {176},
  publisher = {The American Astronomical Society},
  issn = {0004-637X},
  doi = {10.3847/1538-4357/adfa15},
  urldate = {2026-06-08},
  langid = {english}
}

@article{winston2012herschelfarirobservations,
  title = {Herschel Far-{{IR}} Observations of the {{Chamaeleon}} Molecular Cloud Complex . {{Chamaeleon I}}: {{A}} First View of Young Stellar Objects in the Cloud},
  shorttitle = {Herschel Far-{{IR}} Observations of the {{Chamaeleon}} Molecular Cloud Complex . {{Chamaeleon I}}},
  author = {Winston, E. and Cox, N. L. J. and Prusti, T. and Mer{\'i}n, B. and Ribas, A. and Royer, P. and Vavrek, R. and Puga, E. and Andr{\'e}, {\relax Ph}. and Men'shchikov, A. and K{\"o}nyves, V. and K{\'o}sp{\'a}l, {\'A}. and {Alves de Oliveira}, C. and Pilbratt, G. L. and Waelkens, C.},
  year = 2012,
  month = sep,
  journal = {A\&A},
  volume = {545},
  pages = {A145},
  publisher = {EDP},
  issn = {0004-6361},
  doi = {10.1051/0004-6361/201118665},
  urldate = {2026-06-16}
}

@article{wolff2017hubblespacetelescope,
  title = {Hubble {{Space Telescope Scattered-light Imaging}} and {{Modeling}} of the {{Edge-on Protoplanetary Disk ESO-H$\alpha$}} 569},
  author = {Wolff, Schuyler G. and Perrin, Marshall D. and Stapelfeldt, Karl and Duch{\^e}ne, Gaspard and M{\'e}nard, Francois and Padgett, Deborah and Pinte, Christophe and Pueyo, Laurent and Fischer, William J.},
  year = 2017,
  month = dec,
  journal = {ApJ},
  volume = {851},
  number = {1},
  pages = {56},
  publisher = {The American Astronomical Society},
  issn = {0004-637X},
  doi = {10.3847/1538-4357/aa9981},
  urldate = {2024-09-02},
  langid = {english}
}

@inproceedings{wu2018groupnormalization,
  title = {Group {{Normalization}}},
  booktitle = {Proc. {{Eur}}. {{Conf}}. {{Comput}}. {{Vis}}.},
  author = {Wu, Yuxin and He, Kaiming},
  year = 2018,
  pages = {3--19},
  urldate = {2026-06-08}
}

@article{xu2020frequencyprinciplefourier,
  title = {Frequency {{Principle}}: {{Fourier Analysis Sheds Light}} on {{Deep Neural Networks}}},
  shorttitle = {Frequency {{Principle}}},
  author = {Xu, Zhi-Qin John and Zhang, Yaoyu and Luo, Tao and Xiao, Yanyang and Ma, Zheng},
  year = 2020,
  month = jun,
  journal = {CiCP},
  volume = {28},
  number = {5},
  eprint = {1901.06523},
  primaryclass = {cs.LG},
  pages = {1746--1767},
  issn = {1815-2406, 1991-7120},
  doi = {10.4208/cicp.OA-2020-0085},
  urldate = {2026-06-16},
  archiveprefix = {arXiv}
}

@article{zubko1996opticalconstantscosmic,
  title = {Optical Constants of Cosmic Carbon Analogue Grains - {{I}}. {{Simulation}} of Clustering by a Modified Continuous Distribution of Ellipsoids},
  author = {Zubko, V. G. and Mennella, V. and Colangeli, L. and Bussoletti, E.},
  year = 1996,
  month = oct,
  journal = {MNRAS},
  volume = {282},
  pages = {1321--1329},
  publisher = {OUP},
  issn = {0035-8711},
  doi = {10.1093/mnras/282.4.1321},
  urldate = {2026-06-16}
}

@article{harries2026fundamentalpropertiesprotoplanetary,
  title = {Fundamental Properties of Protoplanetary Discs Determined from Simultaneous Fits to Thermal Dust Images and Spectral Energy Distributions},
  author = {Harries, Tim J},
  year = 2026,
  month = may,
  journal = {Mon. Not. R. Astron. Soc.},
  volume = {548},
  number = {2},
  pages = {stag628},
  issn = {0035-8711},
  doi = {10.1093/mnras/stag628},
  urldate = {2026-09-16}
}

@article{koehler2025neuralemulatorsuperiority,
    title={Neural Emulator Superiority: When Machine Learning for {PDE}s Surpasses its Training Data},
    author={Felix Koehler and Nils Thuerey},
    journal={Advances in Neural Information Processing Systems (NeurIPS)},
    volume={38},
    year={2025}
}

@article{mohan2024whatyousee,
  title = {What {{You See}} Is {{Not What You Get}}: {{Neural Partial Differential Equations}} and {{The Illusion}} of {{Learning}}},
  shorttitle = {What {{You See}} Is {{Not What You Get}}},
  author = {Mohan, Arvind and Chattopadhyay, Ashesh and Miller, Jonah},
  year = 2024,
  month = nov,
  number = {arXiv:2411.15101},
  eprint = {2411.15101},
  primaryclass = {cs.LG},
  publisher = {arXiv},
  doi = {10.48550/arXiv.2411.15101},
  urldate = {2026-06-16},
  archiveprefix = {arXiv}
}

\begin{appendix}
\nolinenumbers

\section{Model error, grid \& wavelength resolution} %
\label{sec:grid_app}
\FloatBarrier

\autoref{tab:model_grid_model_resolution} summarizes the allocation of models across wavelength resolutions and the total submitted and accepted model count.

\begin{table}[ht]
\renewcommand{\arraystretch}{1.2}   %
\caption{Overview of \texttt{RADMC-3D} model grid \& resolutions.}
\label{tab:model_grid_model_resolution}
\centering
\begin{tabular}{l|c|c|c}
\hline
Resolution & Models & Fraction & Wavelengths \\
\hline
Low & $2^{15}=32{,}768$ & $76.19\%$ & $21$ \\
Moderate & $2^{13}=8{,}192$ & $19.05\%$ & $42$ \\
High & $2^{11}=2{,}048$ & $4.76\%$ & $103$ \\
\hline
Total (submitted) & $43{,}008$ & $100\%$ & \\
\hline
Failed/rejected & $1{,}350$ & $3.14\%$ &  \\
Total (final) & $41{,}658$ & $96.86\%$ & \\
\hline
\end{tabular}
\\[2pt]
{\footnotesize \textbf{Notes.} All wavelength grids span $0.55$--$2830~\mu\mathrm{m}$. The grids are nested: each higher-resolution grid contains the wavelengths of all lower-resolution grids.}
\end{table}

\begin{figure}[ht]
    \centering
    \includegraphics[width=0.99\columnwidth]{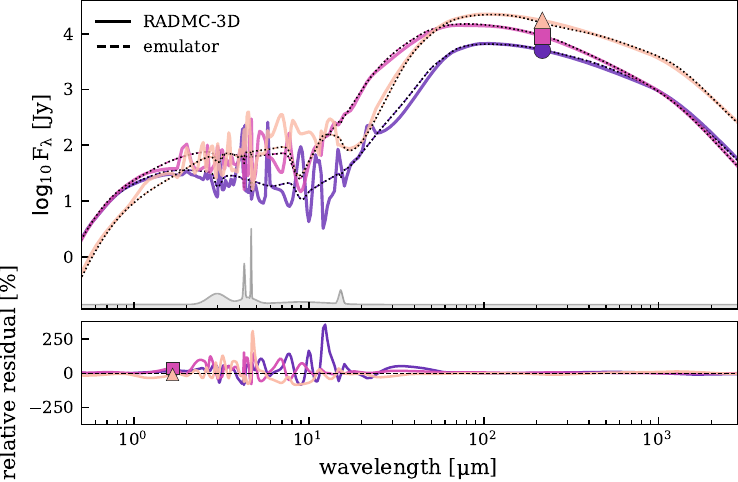}
    \caption{Examples from the high-error tail in the high-opacity regime. The solid curves show the emulator predictions and the points the finite-photon \texttt{RADMC-3D} references, the lower panel shows their relative residuals. The smoother emulator spectra and irregular reference spectra illustrate the confounding of emulator error with Monte Carlo artifacts but are not, by themselves, evidence that either prediction is closer to the converged radiative-transfer solution. The light-gray area indicates the wavelength-sampling density.}
    \label{fig:emulator_superiority_examples}
\end{figure}

\begin{figure}[ht]
    \centering
    \includegraphics[width=0.99\columnwidth]{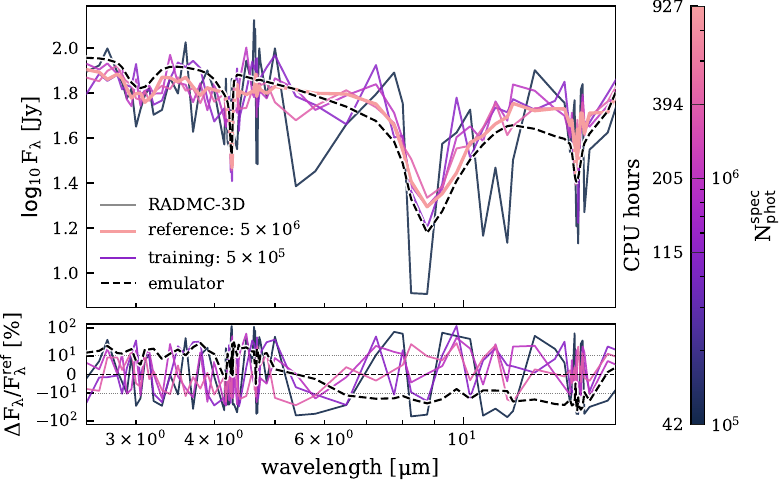}
    \caption{Photon-count convergence of a randomly selected high-resolution test-set model from the high-error tail ($\epsilon_i$ more than one standard deviation above the median; $M_{\rm gas}=0.037\,M_\odot$, $i=81.7^{\circ}$, converged thermal stage), re-simulated with \texttt{RADMC-3D} at $n_{\rm phot}^{\rm spec}=10^{5}$--$5\times10^{6}$ with the same $n_{\rm phot}^{\rm therm}$. The $5\times10^{6}$-photon run (927 CPU h) serves as reference and the $5\times10^{5}$ run is the (unseen) target. The emulator (dashed) was never trained on spectra computed with $n_{\rm phot}^{\rm spec}>5\times10^{5}$. The lower panel shows relative residuals with respect to the reference. Over 2.5--17.5~$\mu$m the RMS residuals are 42, 23, 17, and 11\% for $10^{5}$, $5\times10^{5}$, $10^{6}$, and $2\times10^{6}$ photons, and 17\% for the emulator, whose residual is a smooth offset rather than noise.}
    \label{fig:high_nphot_resimulation}
\end{figure}

\begin{figure}[ht]
    \centering
    \includegraphics[width=0.99\columnwidth]{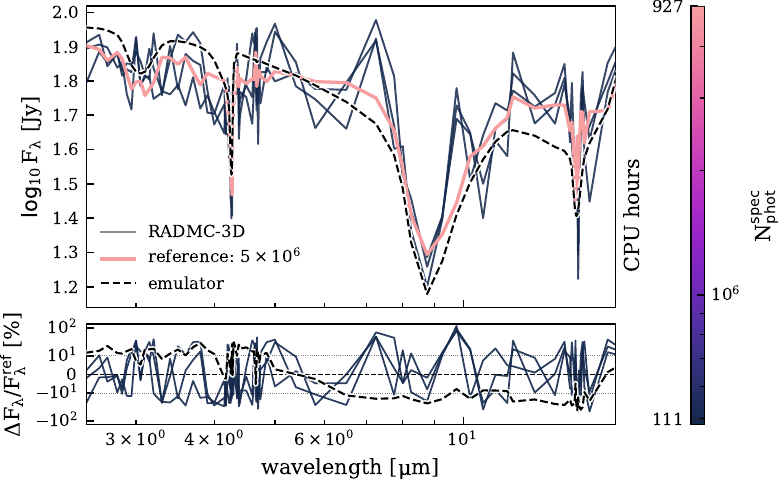}
    \caption{Run-to-run scatter at the training photon count. Three independent $5\times10^{5}$-photon realizations of the model of \autoref{fig:high_nphot_resimulation} (RMS residuals of 20--23\% relative to the $5\times10^{6}$-photon reference) together with the emulator prediction. All three realizations lie above the reference near 7~$\mu$m (\autoref{sec:disc_superiority}).}
    \label{fig:run2run_variance}
\end{figure}

The $3.14\%$ of submitted models that failed or were rejected (\autoref{tab:model_grid_model_resolution}) are of three kinds: models exceeding the wall-clock limit of the compute queue in the thermal or the ray-tracing stage, models terminated by the consistency check of \texttt{RADMC-3D} between the scattering matrices and the \texttt{OpTool} optical constants, resolved during grid construction by raising the lower bound of the opacity wavelength grid from $0.01$ to $0.1~\mu$m (adequate for $T_\star<10^{4}$~K), and models rejected by our quality control, i.e.\ spectra whose mean absolute point-to-point difference of $\log_{10}F$ between adjacent near- and mid-infrared samples exceeds $0.68$ times the total range of $\log_{10}F$, so that Monte Carlo noise dominates any spectral structure (\autoref{sec:disc_superiority}). In \autoref{fig:emulator_superiority_examples} we show examples of irregular spectra predicted by \texttt{RADMC-3D}, and in \autoref{fig:high_nphot_resimulation} and \autoref{fig:run2run_variance} the photon-count convergence and the run-to-run scatter of a test-set model from the high-error tail (\autoref{sec:disc_superiority}).

\FloatBarrier

\section{Physically motivated input embeddings}
\label{sec:appendix_input_embeddings}

\begin{figure*}[ht]
    \centering
    \includegraphics[width=1.99\columnwidth]{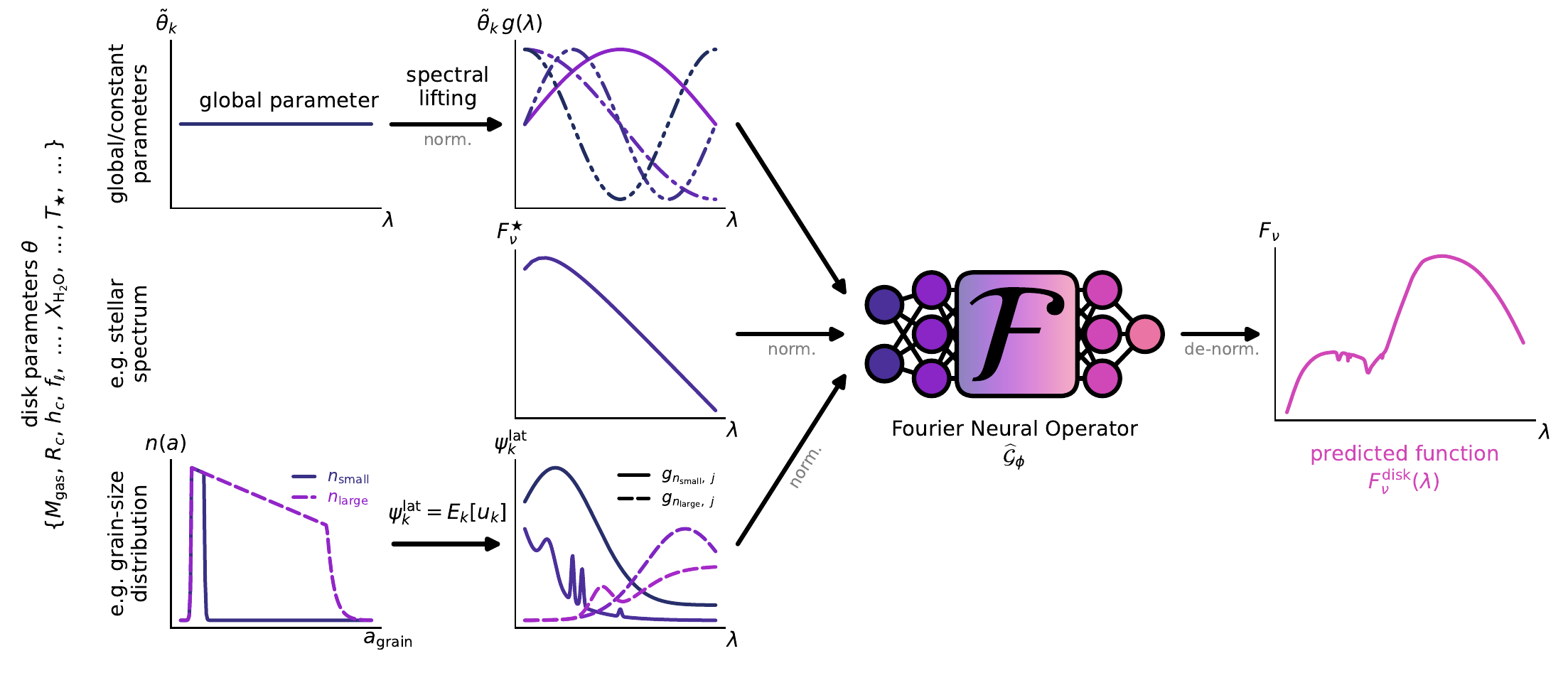}
    \caption{Emulator architecture. The neural operator predicts the flux at arbitrary wavelength discretizations from the disk parameters and the analytic input fields derived from them. In addition to the neural network, the full emulator includes a non-trainable, differentiable module for foreground extinction.}
    \label{fig:nn_architecture}
\end{figure*}

This appendix expands the physically motivated input embeddings summarized in \autoref{sec:input_embeddings}. The simulator we emulate solves the radiative-transfer equation for a specific disk model: the observed spectrum is the stellar radiation field reprocessed by a dusty disk whose opacity and density structure are themselves functions of $\boldsymbol{\theta}$ through our disk prescription. In the operator picture of \autoref{sec:neural_operators}, the natural inputs to such a solution operator (both the radiative-transfer model and its approximation using a neural operator) are therefore the scalar parameters together with the forcing function (the irradiating stellar spectrum) and the coefficient fields (e.g.\ the density, temperature and compositional structure of the disk). Only the stellar spectrum, the grain-size distribution, the density profiles, and the viewing geometry are closed-form functions of $\boldsymbol{\theta}$ and are supplied as input channels. The temperature structure and the ice placement are outputs of the thermal Monte Carlo, which is not computed during training or at inference. This injects a strong, physically grounded inductive bias, frees network capacity to model the expensive part of the mapping rather than relearning known functions, and keeps these transformations interpretable and decoupled from the trainable emulator, in the same spirit as the non-trainable foreground-extinction layer of \autoref{sec:fno_architecture}. Strictly, because all of these fields are deterministic functions of $\boldsymbol{\theta}$, they add no information beyond $\boldsymbol{\theta}$ itself and act purely as inductive bias. The resulting embeddings are shown on the left of \autoref{fig:nn_architecture} and fall into three families according to the domain on which each quantity is defined.

\emph{Global parameters.} %
The disk model parameters $\boldsymbol{\theta}$ are scalars and carry no wavelength dependence, whereas the FNO operates along the wavelength axis. We therefore broadcast the normalized vector $\tilde{\boldsymbol{\theta}}$ across the wavelength grid, so that the global conditioning is available at every point, and combine it with a Neural Radiance Fields (NeRF)-style sinusoidal embedding $g(\tilde{\lambda})$ of the normalized coordinate \citep{tancik2020fourierfeatureslet, mildenhall2021nerfrepresentingscenes}. Including parameter-modulated coordinate features $\tilde{\theta}_k\,g(\tilde{\lambda})$ (a spectral lifting of the constant parameters) gives each parameter an explicit, multi-frequency wavelength basis through which to imprint wavelength-localized structure, rather than only a constant offset (top-left row of \autoref{fig:nn_architecture}).

\emph{Functions on the wavelength domain.} %
Some physical quantities are themselves functions of wavelength. If they can be adequately described and resolved on the target grid\footnote{For the stellar temperatures we adopt in this study, this is the case for the stellar spectrum; however, for hotter stars, the chosen wavelength support might not be sufficient. Conversely, if one would include the absorption and scattering coefficients (or the scattering matrices), they might necessitate a much higher resolution wavelength grid.}, they can be concatenated directly as input channels $\psi^{\mathrm{wl}}_m(\tilde\lambda)$ (left-middle row of \autoref{fig:nn_architecture}). The most important is the stellar spectrum, supplied in $\log_{10}$ space for numerical stability,
\begin{equation}
\label{eq:stellar_forcing}
\psi^{\mathrm{wl}}_\star(\tilde{\lambda}) = \log_{10} F_{\nu}^{\star}\!\left(\lambda;\,L_\star, T_\star\right),
\end{equation}
which we evaluate analytically as a luminosity-scaled blackbody. Physically, $F_{\nu}^{\star}$ is the forcing function of the radiative-transfer problem: the disk is illuminated and heated by the central star, and the shape of the stellar spectrum sets the source term for both the scattered-light continuum at short wavelengths and the thermal reprocessing that dominates at longer wavelengths. Providing $\psi^{\mathrm{wl}}_\star$ as an input field therefore supplies the source term and the short-wavelength continuum anchor directly, making the analogy with a PDE forcing term explicit and motivating the operator formulation.

\emph{Functions on auxiliary domains.} %
Other physically informative quantities are functions of an independent coordinate other than wavelength, and so should not be concatenated directly. The first is the grain-size distribution $n(a)$ on a $\log_{10}$ grain-size grid, which conditions the dust opacity. The second is the radial disk structure on a $\log_{10}$ radius grid: the scale height $H(r)$ and aspect ratio $H/R$, the gas and dust surface densities and their partition between the small- and large-grain populations, the vertical density at a set of scale heights, and the cavity depletion and location. These fields control the temperature structure and emitting area as a function of radius, the radial range over which the dust is cold enough for each species to freeze out (and therefore the strength of the ice bands), and, for the near-edge-on geometries considered here, the line-of-sight optical depth and self-occultation that shape the observed spectrum. A third group encodes the viewing geometry (inclination only), represented by adopting a geometrically thin, flat-disk projection and evaluating inclination-conditioned projection factors on a cyclic azimuthal domain. We treat the viewing geometry separately on purpose: for different inclinations of the same disk the intrinsic physical disk structure is unchanged; however, the observed spectrum differs. We conjecture that easing the separation of a latent observer-independent disk model from a latent observer-dependent projection will help condition the neural operator. Empirically, removing the viewing-geometry channels, with an otherwise unchanged architecture, yields worse performance on the test set \autoref{tab:ablation_test}.
 
Because these functions are defined on domains with no defined correspondence to wavelength, we map each group on its own domain independently to latent functions on the wavelength coordinate using shallow, narrow FNO encoders $E_k$ (\autoref{tab:embedding_fno}),
\begin{equation}
\label{eq:latent_encoder}
\psi^{\mathrm{lat}}_k(\tilde{\lambda}) = E_k\!\left[u_k\right](\tilde{\lambda}),
\end{equation}
where $u_k$ denotes the auxiliary function on its native coordinate $\xi_k$ (grain size, radius, or viewing angle). The latent channels $\psi^{\mathrm{lat}}_k$ are concatenated with the other embeddings and processed by the main network (left-bottom row of \autoref{fig:nn_architecture}). An FNO is a natural choice for this projection: the underlying physical map, for example from $n(a)$ to the opacity $\kappa(\lambda)$ \citep{min2005modelingopticalproperties, dominik2021optoolcommandlinedriven}, is itself an operator between function spaces with both global and localized structure, exactly the regime FNOs are designed for. Computing these maps exactly (e.g.\ a full DHS opacity calculation) is the expense the surrogate is built to avoid. The shallow encoders instead learn a cheap, differentiable approximation that retains the wavelength-dependent imprint of each field while adding relatively few parameters. These cross-domain encoders are the only trainable component of the embedding stack and are optimized jointly with the rest of the emulator.

Collecting the three families, the field lifted to the first Fourier block is the per-wavelength concatenation
\begin{equation}
\label{eq:input_stack}
\mathbf{z}(\tilde{\lambda}) =
\Big[\,
\tilde{\lambda},\;
g(\tilde{\lambda}),\;
\tilde{\boldsymbol{\theta}},\;
\{\tilde{\theta}_k\,g(\tilde{\lambda})\},\;
\{\psi^{\mathrm{wl}}_m(\tilde{\lambda})\},\;
\{\psi^{\mathrm{lat}}_k(\tilde{\lambda})\}
\,\Big],
\end{equation}
combining the global conditioning, the wavelength-domain forcing functions $\psi^{\mathrm{wl}}_m$ (e.g.\ $\psi^{\mathrm{wl}}_\star$), and the projected auxiliary fields $\psi^{\mathrm{lat}}_k$. With the exception of the cross-domain encoders, all of these transforms are fixed and evaluated on the fly from $\boldsymbol{\theta}$; they add no trainable parameters and, like the extinction layer, keep known deterministic physics separate from the learned radiative-transfer operator. 

\FloatBarrier

\section{Hyper-parameter optimization}
\label{sec:hpo}
 
We select the emulator hyper-parameters with a grid-search over the FNO architecture, optimization settings, and regularization strength. The search space is given in \autoref{tab:hpo_grid}. In addition, we tested whether a Fourier continuation \citep{ganeshram2022fcpinohighprecision} of the non-periodic targets improves performance over simple domain padding; we find that it does not improve performance appreciably, so we adopt domain padding throughout. Furthermore, because the spectra are represented as flux densities, $F_\nu$, in Jy, we tested whether predicting the rescaled quantity $\nu F_\nu$ ($\log_{10}\left(\nu F_\nu\right)$) instead of $F_\nu$ ($\log_{10}\left(F_\nu\right)$) improves emulator performance. We found that it does not. Hyper-parameter selection used a $25\%$ hold-out subset of the validation set, which was not used for the final estimate of emulator generalization performance.

\begin{table}[ht]
\caption{Hyper-parameter optimization search space for the FNO surrogate.}
\label{tab:hpo_grid}
\centering
\begin{tabular}{l|l}
\hline
Hyper-parameter & Values \\
\hline
learning rate${}^{(a)}$ & $10^{-2},\,\mathbf{10^{-3}},\,10^{-4}$ \\
weight decay & $10^{-4},\,\mathbf{10^{-5}},\,10^{-6}$ \\
noise level ($\sigma_{\epsilon}$) & $10^{-4},\,\mathbf{10^{-5}},\,10^{-6}$ \\
number of Fourier modes ($n_{\mathrm{modes}}$) & $8,\,\mathbf{12},\,16,\,24,\,32$ \\
hidden channels ($n_{\mathrm{hidden}}$) & $16,\,32,\,\mathbf{64},\,128$ \\
number of FNO layers ($n_{\mathrm{layers}}$) & $4,\,6,\,\mathbf{8}$ \\
coordinate embedding type & \textbf{\texttt{nerf}}, \texttt{transformer} \\
number of coordinate frequencies & $4,\,\mathbf{5},\,6$ \\
\hline
\end{tabular}
\\[2pt]
{\footnotesize \textbf{Notes.} The best-performing hyper-parameters are in bold. $^{(a)}$ We optimize for the initial learning rate; the learning rate scheduler decays the learning rate to $10^{-7}$ (at minimum) during training.}
\end{table}
 
\begin{table}[ht]
\caption{Embedding-FNO hyper-parameters.}
\label{tab:embedding_fno}
\centering
\begin{tabular}{l|l}
\hline
Hyper-parameter & Values$^{(a)}$ \\
\hline
number of Fourier modes ($n_{\mathrm{modes}}$) & same as main FNO \\
hidden channels ($n_{\mathrm{hidden}}$) & $2 \times k$ \\
number of FNO layers ($n_{\mathrm{layers}}$) & $2$ \\
output channels ($n_{\mathrm{out}}$) & $2 \times k$ \\
\hline
\end{tabular}
\\[2pt]
{\footnotesize \textbf{Notes.} Applies only to the learned embeddings defined on auxiliary domains. $^{(a)}$ Values are reported per encoder group $E_k$, where $k$ is the number of input functions of that group (\autoref{sec:input_embeddings}).}
\end{table}

\autoref{tab:ablation_test} summarizes the architecture ablations and baselines, evaluated as the median per-spectrum relative RMS error (\autoref{eq:per_spectrum_relative_rmse}) on the held-out test set. The gap over the multi-layer perceptron (MLP; \citealt{hornik1989multilayerfeedforwardnetworks}) indicates that the operator formulation, rather than network capacity alone, carries most of the performance. MLP and DeepONet are implemented using \texttt{deepxde} \citep{lu2021deepxdedeeplearning}; to isolate the effect of architecture they match the FNO data split, normalization, composite loss, optimizer, and approximate trainable-parameter count.

\begin{table}[ht]
\caption{Ablation study.}
\label{tab:ablation_test}
\centering
\begin{tabular}{l|c}
\hline
Architecture & Median $\epsilon_{i}$ [\%] \\
\hline
FNO & $\mathbf{\fwdPerSpecQFifty}$ \\
FNO (no viewing angle) & $5.35$ \\
FNO (spectral lifting only) & $6.09$ \\
FNO (no $H^1$ loss) & $5.22$ \\
K-Nearest Neighbors${}^{(a)}$  & $22.98$ \\
Radial Basis Function${}^{(a)}$  & $24.92$ \\
MLP${}^{(b)}$ & $18.77$ \\
DeepONet${}^{(b),(c)}$ & $12.86$ \\
\hline
\end{tabular}
\\[2pt]
{\footnotesize \textbf{Notes.} $\epsilon_{i}$: per-spectrum error (RMS over wavelength for each spectrum). MLP and DeepONet predict on the high-resolution wavelength grid and are evaluated using the same masked metrics as the FNO. $^{(a)}$ We perform a simple grid-search hyper-parameter optimization for K-Nearest Neighbors and Radial Basis Function. $^{(b)}$ No dedicated hyper-parameter optimization was performed; capacity and training pipeline were matched to the FNO. $^{(c)}$ DeepONet shows notably worse performance in ice bands and appears more affected by the bad targets discussed in \autoref{sec:disc_superiority}.}
\end{table}

\section{Emulator implementation}
\label{sec:fno_implementation}

The emulator has $\sim4\times10^5$ trainable parameters, is implemented in \texttt{PyTorch} \citep{paszke2019pytorchimperativestyle}, and trained with \texttt{PyTorch Lightning} \citep{falcon2025pytorchlightning} on a single NVIDIA GeForce RTX 5090 GPU in $\sim\!8$ hours. We use the AdamW optimizer with decoupled weight decay \citep{loshchilov2019decoupledweightdecay} and a \texttt{ReduceLROnPlateau} scheduler driven by the validation loss to limit overfitting. The FNO layers and the sinusoidal embedding use the \texttt{neuraloperator} library \citep{kossaifi2026librarylearningneural}. The main architectural hyper-parameters are selected following the recommendations of \citet{duruisseaux2026fourierneuraloperators}, using the grid-search described in \autoref{sec:hpo}, with embedding FNO hyper-parameters derived from the number of input channels and main FNO hyper-parameters. 

To support reproducible downstream use, the trained emulator stores (i) the wavelength support, (ii) the parameter names, and (iii) the parameter bounds inferred from the training set. The normalization transform is fitted on the training subset and serialized with the model so that identical transformations are applied at inference. Inference can be performed on an explicitly supplied wavelength grid or on the default stored support; the emulator returns spectra in units of $\mathrm{mJy}$ after applying the inverse normalization transform and, when enabled, the deterministic post-processing layers. This makes the surrogate a drop-in replacement for the expensive radiative-transfer model (\autoref{sec:fitting}).

\section{Emulator validation: methodology}
\label{sec:appendix_validation_methods}

This appendix details the error metrics and the synthetic-inference setup underlying the validation results of \autoref{sec:validation}.

\subsection{Forward-emulation metrics}
\label{sec:appendix_validation_forward_methods}

For each test-set spectrum $i$ we compare the prediction $\widehat{\mathbf{F}}_i$ to the simulation $\mathbf{F}_i$ on the emulator wavelength grid, applying the same validity mask for missing fluxes as in the training objective. To assess the relative accuracy we define the relative residual
\begin{equation}
\label{eq:relative_residual}
r_i(\lambda_j) = \frac{\widehat{F}_i(\lambda_j)-F_i(\lambda_j)}{F_i(\lambda_j)},
\end{equation}
and summarize it per spectrum by the relative RMS error
\begin{equation}
\label{eq:per_spectrum_relative_rmse}
\epsilon_i = \left[ \frac{1}{N_i} \sum_{j\in\mathcal{V}_i} r_i^2(\lambda_j) \right]^{1/2},
\end{equation}
where $\mathcal{V}_i$ is the set of valid wavelength points for spectrum $i$ and $N_i = |\mathcal{V}_i|$. We summarize the wavelength-dependent error by the relative RMS flux error at each wavelength,
\begin{equation}
\label{eq:wavelength_relative_rmse}
\epsilon_{\rm RMS}(\lambda_j) = \left[ \frac{1}{N_j} \sum_{i\in\mathcal{V}_j} r_i^2(\lambda_j) \right]^{1/2},
\end{equation}
where $\mathcal{V}_j$ is the set of valid test-set spectra at wavelength $\lambda_j$ and $N_j = |\mathcal{V}_j|$. Finally, to test for coherent wavelength-dependent bias we compute the mean signed residual,
\begin{equation}
\label{eq:signed_residual_bias}
b(\lambda_j) = \frac{1}{N_j} \sum_{i\in\mathcal{V}_j} r_i(\lambda_j),
\end{equation}
which tests whether the residuals cancel or whether the emulator is systematically high or low at a given wavelength.

\subsection{Synthetic-inference setup}
\label{sec:appendix_validation_inference_methods}

We draw two sets of $N_{\rm synth}=512$ parameter vectors $\{\boldsymbol{\phi}^{(n)}_{\rm true}\}_{n=1}^{N_{\rm synth}}$ from the held-out test set, where
\begin{equation}
\boldsymbol{\phi}_{\rm true} = \left(\boldsymbol{\theta}_{\rm true},\, d_{\rm true},\, A_{V,{\rm true}}\right),
\end{equation}
drawing the two additional parameters uniformly from $d_{\rm true}\sim\mathcal{U}(100~\mathrm{pc},\,240~\mathrm{pc})$ and $A_{V,{\rm true}}\sim\mathcal{U}(0,\,5)$. For synthetic inference the prior matches these distributions. For each draw we construct a synthetic observation by perturbing the simulated spectrum with Gaussian noise,
\begin{equation}
\label{eq:synth_data}
F^{(n)}_{m} = F_{m}\left(\boldsymbol{\phi}^{(n)}_{\rm true}\right) + \epsilon^{(n)}_{m},
\qquad
\epsilon^{(n)}_{m}\sim\mathcal{N}\left(0,\sigma_{m}^2\right),
\end{equation}
with $\sigma_{m}=\sigma_{\rm synth}\,F_{m}(\boldsymbol{\phi}^{(n)}_{\rm true})$, and run the full pipeline of \autoref{sec:fitting} with the same priors, likelihood (\autoref{eq:likelihood}), and emulator-error inflation, using $n_{\rm live}=400$. The two sets bracket the intended use case. The first (high-$R_{\rm spec}$) draws only from the high-resolution test spectra, uses every wavelength-flux pair, and adopts a uniform $\sigma_{\rm synth}=10^{-2}$, testing the emulator under near-ideal, information-rich conditions\footnote{Still, these data have a maximum spectral resolution of only $R_{\rm max}\sim\fwdResHighMax$, at the ice bands, compared to $R\sim\fwdResSpitzer$ for Spitzer, and $R\sim\fwdResJwstNirspecIfu$ and ${\sim}\fwdResJwstMiriMrs$ for JWST NIRSpec IFU and MIRI MRS, respectively.}. The second (ESO-H$\alpha$~569-like) draws from the full test set but subsamples each spectrum to the wavelengths and adopts the per-band uncertainties of the ESO-H$\alpha$~569 photometry (\autoref{tab:sedphot}), probing the sparse, unevenly weighted regime that is representative of photometric disk observations.

\subsection{Nested sampling hot-start procedure}
\label{sec:ultranest_hot_start}

For synthetic-observation inference we use a hot-start procedure in \texttt{UltraNest} to reduce the computational cost of the validation. Sharply localized likelihoods can otherwise force many iterations of compression from the full prior toward the posterior. We first generate a global low-discrepancy scan of the prior using $2^{22}\approx 4.2\times 10^6$ Sobol points and evaluate the likelihood at each. We retain the $512$ highest-likelihood points as an elite set. To estimate the scale of the auxiliary distribution, we draw $512$ refinement points $u_i$ per elite point from a local Gaussian proposal in unit-cube coordinates, rejecting points outside the prior bounds. The local covariance is estimated from the geometry of the refined elite set: for each elite point we identify its $K=2n_{\mathrm{dim}}$ nearest neighbors and compute their empirical covariance,
\begin{equation}
    \mathbf{C}_i = \mathrm{Cov}(\mathcal{N}_K(\mathbf{u}_i)) + \epsilon^2 I,
\end{equation}
with $\epsilon=10^{-3}$, and draw local refinement samples from $\mathcal{N}(\mathbf{u}_i,\mathbf{C}_i)$. This makes the refinement scale adaptive to the local concentration and anisotropy of the high-likelihood region. The resulting $512^2\approx 2.6\times 10^5$ local refinement samples estimate an approximate posterior center and covariance in unit-cube coordinates, which define the Student-$t$ auxiliary distribution used by \texttt{UltraNest}'s hot-start procedure to generate an auxiliary problem identical to the original inference problem.

\subsection{Coverage testing}
\label{sec:appendix_validation_tarp_methods}

We assess the statistical calibration of the posteriors with the TARP coverage test \citep{lemos2023samplingbasedaccuracytesting}. TARP was developed to assess generative posterior estimators. We apply it here to a surrogate-plus-nested-sampling posterior, treating it, together with posterior-predictive checks, as a necessary and sufficient calibration diagnostic. Coverage tests have been previously applied to emulator- and grid-based likelihood pipelines coupled to MCMC or nested sampling \citep[e.g.,][]{jin2025neuralnetworkemulator, ran2026planetformationconstraints}, which supports using TARP beyond neural posterior estimation.

\section{Posteriors}
\label{sec:posteriors}

\begin{figure*}[ht]
    \centering
    \includegraphics[width=1.99\columnwidth]{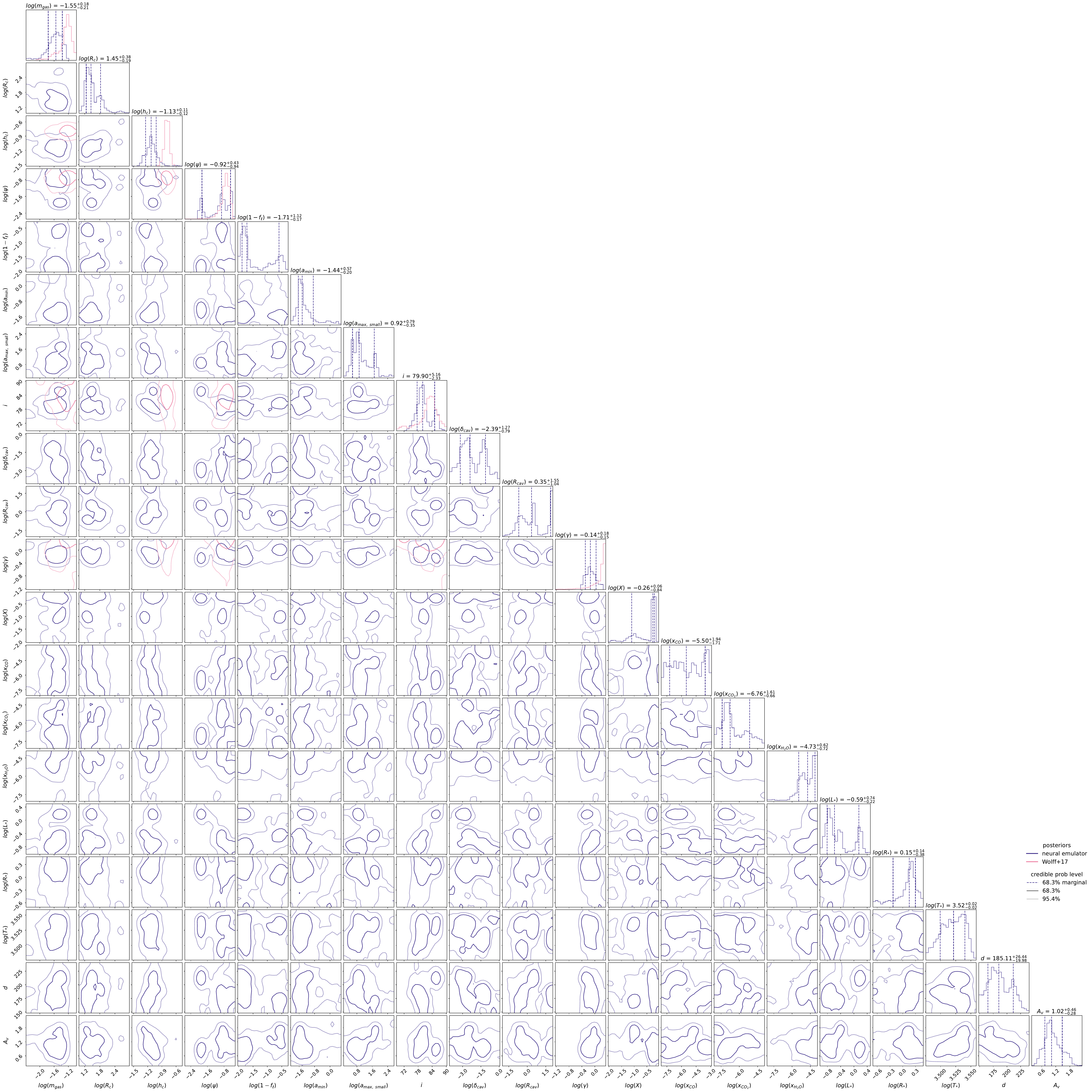}
    \caption{Full ESO-H$\alpha$~569 posterior, compared with the combined SED and HST imaging posterior of \citet{wolff2017hubblespacetelescope}.}
    \label{fig:posterior_esoha569}
\end{figure*}

\begin{figure*}[ht]
    \centering
    \includegraphics[width=1.99\columnwidth]{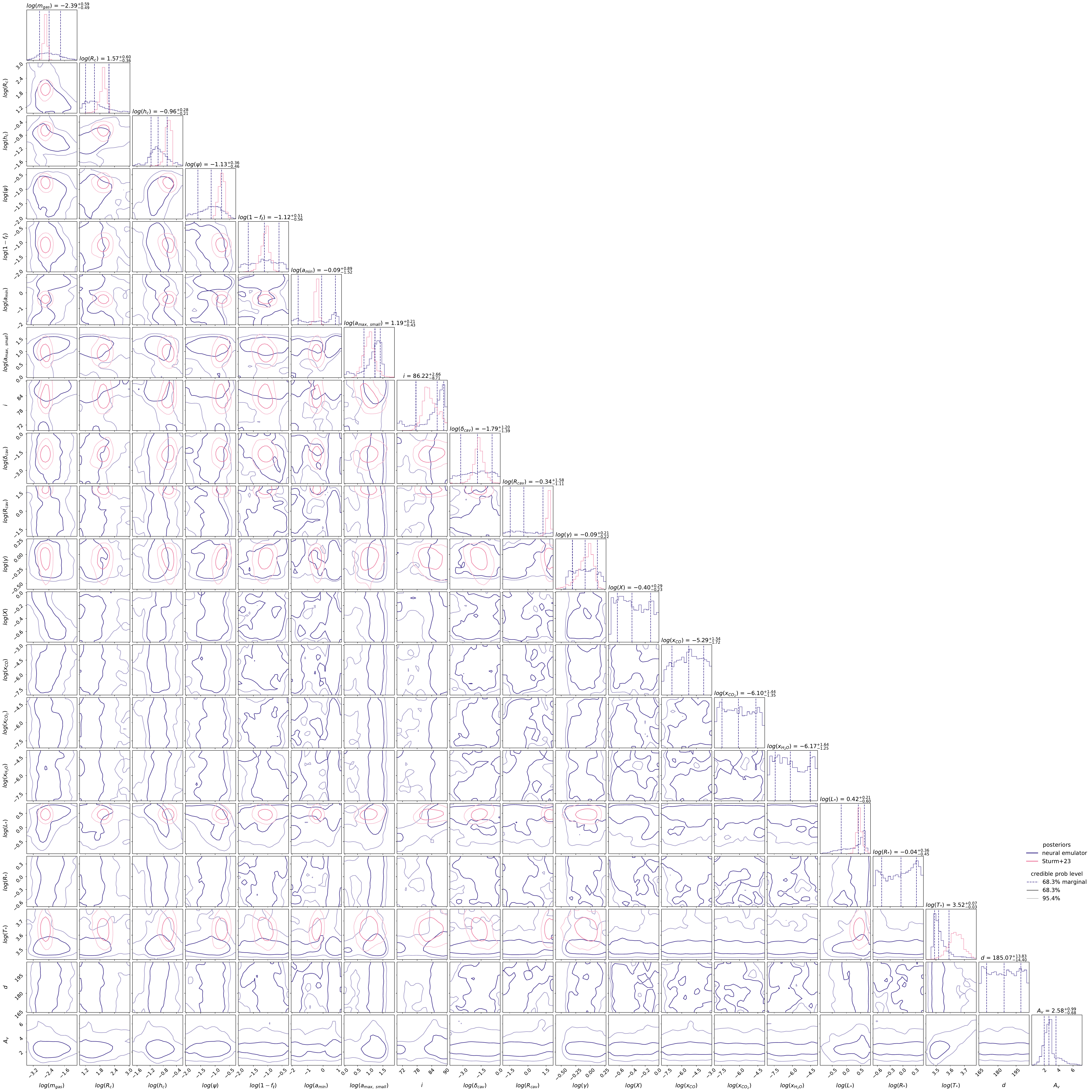}
    \caption{Full HH~48~NE posterior, compared with the SED posterior of \citet{sturm2023edgeonprotoplanetarydisk}.}
    \label{fig:posterior_hh48ne}
\end{figure*}

\autoref{fig:posterior_esoha569} and \autoref{fig:posterior_hh48ne} show the full posteriors as weighted nested-sampling samples with kernel-density contours (effective sample sizes ${\approx}\EsoHaESS$ and ${\approx}1.1\times10^{4}$). The substructure in the two-dimensional contours of the prior-dominated parameters reflects the finite effective sample size spread over 20 dimensions, not structure in the likelihood. The posteriors of the successive $n_{\rm steps}$-doubling runs (\autoref{sec:fitting}) are consistent with each other. The literature posteriors are smoother because they were obtained with fewer free parameters and with $d$ and $A_V$ optimized rather than marginalized (\autoref{sec:disc_comparison}).

\end{appendix}

\end{document}